\documentclass[12pt,preprint]{aastex}
\usepackage{comment}
\usepackage[T1]{fontenc}
\usepackage{pslatex}

\shorttitle{FIR-Radio Correlation with {\it Spitzer}}
\shortauthors{Murphy et al.}

\begin{document}

\title{An Initial Look at the Far Infrared-Radio Correlation within
  Nearby Star-forming Galaxies using the {\it Spitzer} Space
  Telescope}
\author{ 
E.J.~Murphy\altaffilmark{1}, R.~Braun\altaffilmark{2},
G.~Helou\altaffilmark{3}, L.~Armus\altaffilmark{3},
J.D.P.~Kenney\altaffilmark{1}, K.D.~Gordon\altaffilmark{4},
G.J.~Bendo\altaffilmark{4}, D.A.~Dale\altaffilmark{5},
F.~Walter\altaffilmark{6}, T.A.~Oosterloo\altaffilmark{2},
R.C.~Kennicutt\altaffilmark{4}, D.~Calzetti\altaffilmark{7}, 
J.M.~Cannon\altaffilmark{6}, B.T.~Draine\altaffilmark{8}, 
C.W.~Engelbracht\altaffilmark{4}, D.J.~Hollenbach\altaffilmark{9},
T.H.~Jarrett\altaffilmark{3}, L.J.~Kewley\altaffilmark{10},
C.~Leitherer\altaffilmark{7}, A.~Li\altaffilmark{11},
M.J.~Meyer\altaffilmark{7}, M.W.~Regan\altaffilmark{7},
G.H.~Rieke\altaffilmark{4}, M.J.~Rieke\altaffilmark{4}, 
H.~Roussel\altaffilmark{3}, K.~Sheth\altaffilmark{3}, 
J.D.T.~Smith\altaffilmark{4}, M.D.~Thornley\altaffilmark{12}}

\altaffiltext{1}{\scriptsize Department of Astronomy, Yale University,
  P.O. Box 208101, New Haven, CT 06520-8101; murphy@astro.yale.edu}
\altaffiltext{2}{\scriptsize ASTRON, P.O. Box 2, 7990 AA Dwingeloo,
  The Netherlands}
\altaffiltext{3}{\scriptsize California Institute of Technology, MC
  314-6, Pasadena, CA 91101}
\altaffiltext{4}{\scriptsize Steward Observatory, University of Arizona,
  933 North Cherry Avenue, Tucson, AZ 85721}
\altaffiltext{5}{\scriptsize Department of Physics and Astronomy,
  University of Wyoming, Laramie, WY 82071}
\altaffiltext{6}{\scriptsize Max Planck Institut f\"{u}r Astronomie,
  K\"{o}nigstuhl 17, 69117 Heidelberg, Germany}
\altaffiltext{7}{\scriptsize Space Telescope Science Institute, 3700 San
  Martin Drive, Baltimore, MD 21218}
\altaffiltext{8}{\scriptsize Princeton University Observatory, Peyton
  Hall, Princeton, NJ 08544}
\altaffiltext{9}{\scriptsize NASA/Ames Research Center, MS 245-6,
  Moffett Field, CA 94035}
\altaffiltext{10}{\scriptsize Institute for Astronomy, 2680 Woodlawn
  Drive, Honolulu, HI 96822}
\altaffiltext{11}{\scriptsize Department of Physics and Astronomy,
  University of Missouri, Columbia, MO 65211}
\altaffiltext{12}{\scriptsize Department of Physics, Bucknell
  University, Lewisburg, PA 17837}

\begin{abstract}
We present an initial look at the far infrared-radio correlation
within the star-forming disks of four nearby, nearly face-on galaxies
(NGC~2403, NGC~3031, NGC~5194, and NGC~6946).
Using {\it Spitzer} MIPS imaging, observed as part of the {\it
  Spitzer} Infrared Nearby Galaxies Survey (SINGS), and Westerbork
Synthesis Radio Telescope (WSRT) radio continuum data, taken for the
WSRT SINGS radio continuum survey, 
we are able to probe variations in the logarithmic 24~$\micron$/22~cm
($q_{24}$) and 70~$\micron$/22~cm ($q_{70}$) surface brightness
ratios across each disk at sub-kpc scales.
We find general trends of decreasing $q_{24}$ and $q_{70}$ with
declining surface brightness and with increasing radius.
We also find that the dispersion in $q_{24}$ is generally a bit larger
than what is found for $q_{70}$ within galaxies, and both are
comparable to what is measured {\it globally} among galaxies at around
0.2~dex.
The residual dispersion around the trend of $q_{24}$ and $q_{70}$ 
versus surface brightness is smaller than the residual dispersion
around the trend of  $q_{24}$ and $q_{70}$ versus radius, on average
by $\sim$0.1~dex, indicating that the distribution of star formation
sites is more important in determining the infrared/radio disk
appearance than the exponential profiles of disks.  
We have also performed preliminary phenomenological modeling of cosmic
ray electron (CR$e^{-}$) diffusion using an image-smearing technique, 
and find that smoothing the infrared maps improves their correlation 
with the radio maps.  
We find that exponential smoothing kernels work marginally better than
Gaussian kernels, independent of projection for these nearly face-on
galaxies.  
This result suggests that additional processes besides simple
random-walk diffusion in three dimensions must affect the evolution of
CR$e^{-}$s.
The best fit smoothing kernels for the two less active star-forming
galaxies (NGC~2403 and NGC~3031) have much larger scale-lengths than
those of the more active star-forming galaxies (NGC~5194 and
NGC~6946).  
This difference may be due to the relative deficit of recent CR$e^{-}$
injection into the interstellar medium (ISM) for the galaxies having
largely quiescent disks.
\end{abstract}

\section{Introduction}
A major result of the {\it Infrared Astronomical Satellite} (IRAS;
\citet{gxn84}) all-sky survey was the discovery of a correlation
between the globally measured far infrared (42-122~$\micron$, FIR) dust
emission and the optically thin radio continuum emission of normal
late-type star-forming galaxies without an active galactic
nucleus (AGN) \citep{de85,gxh85}.  
The most remarkable feature of this correlation is that it displays
such little scatter, $\sim$0.2~dex, among 
galaxies spanning 5 orders of magnitude in luminosity.
While the FIR emission is due to the thermal re-radiation of
interstellar starlight by dust grains, the radio emission is
primarily non-thermal synchrotron emission from cosmic ray electrons
(CR$e^{-}$s) that propagate in a galaxy's magnetic field after
initially being accelerated by supernova shocks or other processes.
The physics which maintains a strong correlation between  
these two quantities over such a wide range of galaxies remains
unclear.  
   
The connection between radio and infrared emission from galaxies
is that they are both powered by massive stars, as pointed out
originally for starbursts by  \citet{har75}.
Young massive stars, which heat up dust to provide the bulk of the FIR
emission, are thought to be the same stars which end as supernovae (SNe)
and bring about the synchrotron emission.
Such a simplified picture, however, cannot fully explain the small
dispersion measured among galaxies spanning ranges in magnetic field
strength, metallicity, interstellar medium (ISM) mass, dust grain
chemistry and distributions, and star formation rates (SFRs), which
all contribute to the observed FIR/radio ratio of galaxies.  
In fact, some of these parameters individually have a larger
dispersion among galaxies than what is measured for the FIR-radio
correlation.

Various physical models for the {\it global} FIR-radio
correlation have been introduced (e.g. \citet{volk89, hb93, nb97, hoer98,
  bres02}), though progress on the theoretical front has been
limited.  
Observations of the FIR-radio correlation within galaxies, using
both IRAS and ISO \citep{bg88,Xu92,mh95,mh98,hoer98} and early {\it
  Spitzer} data \citep{kdg04, hin04},  has provided tantalizing hints
of the variations in the correlation, motivating this detailed {\it
  Spitzer} follow-up.  
Prior to {\it Spitzer}, the physical scale below which the
FIR-radio  correlation breaks down, a few hundred parsecs, has only
been possible to study by looking at star-forming regions within our
own Milky Way galaxy \citep{bp88}, and the nearest galaxies (e.g. M31,
\citet{hoer98}).  
Due to the limited spatial resolution and/or sensitivity of previous
instruments compared to {\it Spitzer}, the measurement of detailed
variations within other galaxies had not been possible. 

If the general picture of the FIR-radio correlation is correct, and
massive stars are largely responsible for both the infrared and radio
emission from galaxies, 
the fact that the mean free path of UV photons ($\sim$100~pc) which
heat the dust is much less than the diffusion length for a CR$e^{-}$
($\sim$1-2~kpc) suggests that the radio image should resemble a
smeared version of the infrared image. 
This idea was first introduced by \citet{bh90}, who attempted
to model the propagation of CR$e^{-}$s by smearing IRAS scan data of
galaxies using parameterized kernels containing the physics of the
CR$e^{-}$ propagation and diffusion, to better match the morphology of
the corresponding radio data. 
Later work by \citet{mh98} further tested this model using IRAS HiRes
images and found that this prescription worked on large scales across
galaxy disks.

In an attempt to better understand the FIR-radio correlation, we 
are using infrared data from {\it Spitzer} observations obtained as
part of the {\it Spitzer} Infrared Nearby Galaxies Survey (SINGS)
legacy science project \citep{rk03}.
These data allow us to probe galaxies with dramatically increased
angular resolution and sensitivity compared to past infrared missions,
especially at 24 and 70~$\micron$.
Using high resolution {\it Spitzer} imaging, we are also able to test
the smearing model of \citet{bh90} with greater accuracy, at higher
spatial resolution, and in more galaxies, with the aim of gaining
better insight into CR$e^{-}$ diffusion and confinement within galaxy
disks. 
In this paper we examine the FIR-radio correlation within four of
the nearest face-on galaxies in the SINGS sample for which we have
acquired both {\it Spitzer} MIPS and WSRT radio continuum data:
NGC~2403, NGC~3031 (M81), NGC~5194 (M51a), and NGC~6946. 
These galaxies are quite diverse in their Hubble type and star
formation activity, but their distances allow us to probe the
correlation on the scale of a few hundred parsecs within each of
their respective star-forming disks (see Table \ref{tblgaldat}).

The paper is organized as follows: 
In $\S2$ we describe the observations and data analysis procedures.
Then, in $\S3$, we present the empirical results of our work.
In $\S4$ we compare our results within disks to previous results on the
{\it global} FIR-radio correlation, 
and explore the role of cosmic ray propagation in the local FIR-radio
correlation.  
Finally, in $\S5$, we provide a brief summary of the paper.  

\section{Observations and Data Reduction}
\subsection{{\it Spitzer} Images}
{\it Spitzer} imaging was carried out for each galaxy using the
Multiband Imaging Photometer for {\it Spitzer} (MIPS; \citet{gr04}) as
part of the SINGS legacy science program.  
Accordingly, a detailed description of the basic observation strategy
can be found in \citet{rk03}, although a few modifications have been
made after receiving SINGS validation data on NGC~7331
(e.g. \citet{mr04}).  
Note that each target is mapped (or visited) twice so that
asteroids and other transient phenomena can be removed from the data if
necessary.   
The MIPS data were processed using the MIPS Data Analysis Tools (DAT)
versions 2.80-2.92 \citep{kdg05}.  
Due to residual artifacts such as latent images and background
curvature in the 24~$\micron$ data, as well as short term drifts in
the 70 and 160~$\micron$ signals, additional processing beyond that of
the standard MIPS DAT was necessary.
These exceptions in the standard data processing are listed below.

For the 24~$\mu$m data, a few additional steps were performed on the
data.  First, the flatfielding was performed in two steps.
Scan-mirror-position dependent flats, created from off-target data in
the scan maps from all SINGS MIPS campaign data, were first applied to
the data.  
Following this, scan-mirror-position independent flats, made from
off-target frames in each visit's scan map, were applied.  
Latent images from bright sources, erroneously high or low pixel
values, and unusually noisy frames were also masked out before the
data were mosaicked together.  
For the NGC~5194 and NGC~6946 data, mosaics of the data from each
visit were made, then linear backgrounds determined from sky regions
outside of the optical disks were subtracted.  
The two mosaics were then averaged together to produce the final maps.  
For the NGC~2403 and NGC~3031 data, backgrounds were measured as a
function of time in each scan leg and subtracted before mosaicking.  
After this background subtraction, the data from both visits were
mosaicked to form a single image.  
The final pixel scale and full-width at half-max (FWHM) of the point
spread function (PSF) are 0$\farcs$75 and 5$\farcs$7, respectively.  
The calibration factor applied to the final mosaic has an uncertainty
of $\sim$10\% and the RMS noise for the raw map  is listed in
Table \ref{galrms} for each galaxy.  

For the 70 and 160~$\micron$ data,  the major addition to the
processing beyond the standard MIPS DAT steps was the subtraction of
short term variations from a residual detector background drift.  
This step also removes the sky background emission.
The region that includes the galaxy is excluded from the drift
determination, so no extended emission is subtracted.  
The data from both visits were then used to make one mosaic, and a
residual offset measured in regions around the target was subtracted
from the maps.  
Some bright sources in the 70~$\micron$ data created negative latent
images that appeared as dark streaks in the data.  
As an artifact of the background subtraction, bright and dark streaks
appeared on opposite sides of these bright sources.  
These streaks, while visible in the images, are at a relatively low
signal level and should not significantly affect the analysis.  The
final pixel scales are 3$\farcs$0 and 6$\farcs$0, and the FWHM of the
PSFs are 17$\arcsec$ and 38$\arcsec$ at 70 and 160~$\micron$,
respectively.  The calibration factors applied to the final mosaics
have uncertainties of $\sim$20\% for each band and the RMS noise for
the raw 70~$\micron$ and 160~$\micron$ maps is listed in Table
\ref{galrms} for each galaxy.

While the calibration uncertainties will have a systematic effect
on the measured flux ratios, they will not cause artificial trends
as a function of signal strength.  
In contrast, the RMS noise will contribute to uncertainties in flux
ratios as a function of surface brightness, possibly causing low-level
artificial trends in the data.   
Accordingly, we only use pixels having a signal at least 3 times
above the RMS noise in our analysis to minimize these types of
effects.  

\subsection{Radio Continuum Images}
Radio continuum images at 22 and 18~cm were obtained using the Westerbork
Synthesis Radio Telescope (WSRT). 
Each target was observed for a twelve hour integration in 
the "maxi-short" array configuration, which has particularly good
sampling of short baselines (East-West baselines of 36, 54, 72 and
90 meters are all measured simultaneously) as well as a
longest baseline of about 2700 m.
The target observations were bracketed by observations of the primary
total intensity and polarization calibration sources 3C147 and 3C286,
yielding an absolute flux density calibration accuracy of better than
5\%. 
The observing frequency was switched every five minutes between two
settings (1366 and 1697~MHz).   
Each frequency setting was covered with eight sub-bands of 20~MHz
nominal width, but spaced by 16~MHz to provide contiguous,
non-attenuated coverage with a total bandwidth of 132~MHz. 
An effective integration time of 6 hours was realized at each
frequency setting.
All four polarization products and 64 spectral channels were obtained
in each sub-band. After careful editing of incidental radio frequency
interference, external total intensity and polarization calibration of
the data was performed in the AIPS package.  
Subsequently, each field was self-calibrated using an imaging pipeline
based on the Miriad package. 
Each of the eight sub-bands for a given frequency setting was first
processed and imaged independently; and these were subsequently
combined with an inverse variance weighting. 
Deconvolution of each sub-band image was performed iteratively within
a threshold mask based on a spatial smoothing of the previous
iteration. 
The individual frequency channels (of 312.5 kHz width) were
gridded during imaging, so that band-width smearing effects were
negligible. 
In this way, a moderately good reconstruction of the brightness
distribution was obtained for each target. 
The total detected flux density (scaled to a common reference
frequency of 1365 MHz) was 460, 610, 1410 and 1690~mJy for NGC 2403,
3031, 5194 and 6946 respectively. 
Although all of these values either agree with, or slightly exceed,
current estimates in the literature (387, 624, 1310 and 1432 mJy,
\citet{wb92}) 
they must still be regarded as lower limits, since the brightness
distribution declines so smoothly into the noise floor.
A more complete description of the processing steps will appear in
Braun et al. (2005, in preparation). 

Each final sub-band image was reprocessed to obtain a new output
point spread function, by first dividing the image FFT with the FFT of the
Gaussian CLEAN restoring beam and then convolving the result 
with a model of the MIPS 70~$\micron$ beam to permit an accurate joint
analysis with the MIPS data. 
The intrinsic FWHM of the radio beams is about 11\arcsec East-West by
11/sin($\delta$)\arcsec North-South at 1400 MHz and scales as
  1/frequency. This was in all cases smaller than the MIPS 70~$\micron$ beam. 
Accordingly, the MIPS 70~$\micron$ beam sets the spatial resolution for
the present study. 

As the frequency difference between the 22 and 18~cm emission is rather
small, with both wavelengths dominated by synchrotron emission, we
consider only the 22~cm data for the infrared-radio analysis
presented in this paper since the signal-to-noise ratio at 22~cm was
generally higher than at 18~cm.
The only exception is NGC~3031, for which a 20~cm map was created via
a variance weighted average of both the 22 and 18~cm data.
This was done in order to obtain good image quality for this very
challenging field, which has complications due to the low extended
surface brightness disk of NGC~3031, as well as calibration and
confusion problems due to the nearby starburst galaxy NGC~3034 (M82).
To allow for proper comparison with the 22~cm data, we scaled the
20~cm flux density to what is expected at 22~cm assuming a mean
spectral index of -0.7.
The RMS noise is given in Table \ref{galrms} for each galaxy.

The expected number density of background radio sources,
detectable at the 5~$\sigma$ level in our radio maps, is $\sim$0.17
arcminute$^{-2}$ \citep{hop03}. 
This number translates into $\sim$15 over the average area of a galaxy
disk studied in this paper.  
These background radio sources fall into two categories; galaxies
which are primarily star-forming, and those which are dominated by an
AGN. 
At flux densities $\ga$3~mJy, AGN dominate the radio source counts at
20~cm and are expected to be found at a frequency of $\sim$1 per the
average area of the sample galaxies \citep{rhb95}.  
Such sources can often be distinguished by their characteristic
(double or triple) morphologies and much higher surface brightnesses
compared to a galaxy's diffuse radio disk.  
At lower flux densities, star-forming galaxies will dominate the
counts. 
As such, they may introduce some additional scatter into our flux
ratios, but are unlikely to lead to a systematic bias since they will
affect $\sim$4\% of the total area analyzed within each galaxy.  
Accordingly, we do not expect significant contamination of our
analysis by background sources, but in general, very deep, high
resolution radio or FIR imaging would be required to determine if any
particular deviation from a constant FIR/radio ratio were due to a
confusing background source.

\subsection{Image Registration and Resolution Matching}
In the following analysis, we focus on the 24 and 70~$\micron$ {\it
  Spitzer} MIPS data since they have angular resolutions better
than, or similar to, our WSRT radio data.
The calibrated MIPS and radio continuum images for each galaxy
underwent a pre-analysis procedure to ensure that the different PSFs
and sampling at each wavelength did not introduce artifacts into our
results.  
Each image was first sky-subtracted using a variance weighted mean
calculated from regions surrounding the galaxy.
The images at different wavelengths were then cropped to a common
field-of-view, and regridded to a pixel scale of 3$\arcsec$.

The MIPS 24~$\micron$ images were then convolved to match the
70~$\micron$ beam using custom smoothing kernels.
The convolution kernels convert an input PSF into a lower resolution
output PSF using the ratio of Fourier transforms of the output to input
PSFs.  
As part of the creation of these kernels, the high frequency
noise in the input PSF is suppressed (for details see K. D. Gordon et
al. (2005, in preparation)).
The resulting 70 and 24~$\micron$ maps are displayed in the second and
third columns of Figure \ref{maps}, respectively, and the RMS noise of
the convolved 24~$\micron$ maps is listed in Table \ref{galrms}.
The final radio maps, having beams matched to the MIPS 70~$\micron$
PSF (see $\S$2.2), are displayed in the first column of Figure
\ref{maps}.
Finally, we cross-correlated the radio and MIPS images to measure and
remove any existing MIPS position offsets.

After the above image registration and PSF matching was carried out,
we constructed logarithmic infrared/radio ratio ($q$)-maps, where
 \begin{equation}
   q_{\lambda [\micron]} \equiv \log\left(\frac{f_{\nu}(\lambda)[{\rm
   Jy}]}{f_{\nu}(22~{\rm cm})[{\rm Jy}]}\right),
 \end{equation}
for $\lambda =$ 24 and 70~$\micron$.
The only exception, as mentioned in $\S$2.2, is the case of
NGC~3031 where a 20~cm radio continuum map was used.
The $q_{70}$ and $q_{24}$ maps for each galaxy are presented
in the fourth and fifth columns of Figure \ref{maps}, respectively, for
pixels having $>$3~$\sigma$ detections in each of the infrared and
radio images.  

We tested to ensure that using the monochromatic 70~$\micron$
emission does not significantly affect conclusions drawn about the
FIR-radio correlation by comparing $q_{\rm FIR}$ and $q_{\lambda}$
maps at matching resolutions.  
In order to do this we performed the same pre-analysis procedure
described above to properly match the radio continuum and the 24 and
70~$\micron$ images to the resolution and pixel scale at
160~$\micron$.
We then constructed a total infrared (3-1100~$\micron$, TIR) map
using Equation 4 from \citet{dd02} and estimated the FIR
fraction using the same \citet{dd02} spectral energy distribution
(SED) models for pixels having $>$3~$\sigma$ detections in each MIPS band.
Finally, we constructed the logarithmic FIR/radio ratio map
following the convention of \citet{gxh85}, such that
\begin{equation}
q_{\rm FIR} \equiv \log\left(\frac{\rm FIR}{3.75\times10^{12}
  ~{\rm W m^{-2}}}\right) - \log\left(\frac{S_{1.4~{\rm GHz}}}{\rm W
  m^{-2} Hz^{-1}}\right),
\end{equation}
for pixels which are also detected above the 3~$\sigma$ level in the
radio continuum image.  
The 160~$\micron$, FIR, and $q_{\rm FIR}$ maps for each galaxy
are presented in the left, middle, and right columns of Figure 
\ref{160maps}, respectively.
Results comparing the behavior of the monochromatic $q_{\lambda}$
ratios with $q_{\rm FIR}$ ratios at matching resolutions are presented
in $\S$3.1.  

\subsection{Aperture Photometry}
To probe the variations that exist within each galaxy, we compare
$q_{\lambda}$ values differentiating, to the extent possible, nuclear,
arm, inter-arm, disk, inner-disk, and outer-disk environments. 
This procedure is carried out over 'critical' apertures, defined by
diameters equal to the FWHM of the PSF.  Critical apertures of a given
angular extent naturally correspond to different projected physical
scales for galaxies at different distances.
Assuming distances to each galaxy given in Table \ref{tblgaldat}, 
the FWHM of the 70~$\micron$ beam corresponds to 'critical'
apertures of $\sim$0.3, 0.3, 0.5, and 0.75~kpc for NGC~2403, 3031,
6946, and 5194, respectively.  

Aperture masks were created using the 24~$\micron$ images in their
native resolution.
The different regions are defined as follows. 
Nuclear regions are the bright central point in the galaxy, co-spatial
in both the infrared and radio maps (except for NGC~2403 where the
nucleus was not identifiable).
Arm regions trace the spiral arms and are centered on individual
giant H~II regions where possible, but also contain the observed
emission in between discrete star-forming regions within each arm. 
Inter-arm regions probe the more quiescent areas in between spiral
arms.  
Inner-disk regions are circumnuclear regions ($r \la 5$~kpc)
that are both bright and not clearly associated with any large
coherent structures, such as an inner-ring of spiral arms.
We define the star-forming disk to contain all obvious star formation
sites visible in the 24~$\micron$ images.
Disk regions are areas within the star-forming disk, that are both
diffuse and not clearly associated with any type of coherent
structures.    
Outer-disk regions are identified as areas of diffuse emission
surrounding the the star-forming disk.  
Our aperture masks for each galaxy are illustrated in Figure
\ref{apmask}.
Results of the aperture photometry are discussed in $\S3$, and a
summary of statistical results,  including the mean
($<q_{\lambda}>$) and dispersion ($\sigma_{\lambda}$) of the
measured logarithmic infrared/radio ratios for each galaxy, is
presented in Table \ref{tblstats}.

\section{Results}
\subsection{Infrared/Radio Maps}
An inspection of the $q_{\lambda}$ maps, presented in the fourth
and fifths columns of Figure \ref{maps}, reveals structure in the
ratio images corresponding to the patterns of star-formation in each
galaxy disk.  
We also find that all galaxies have dynamic ranges in $q_{70}$ and
$q_{24}$ each spanning $\ga$1~dex.
For comparison, previous studies of the FIR/radio ratios within
NGC~3031 (M81) showed variations by a factor of 6, excluding its AGN
nucleus  \citep{kdg04}, and by an order of magnitude within M31
\citep{hoer98}. 

By comparing our $q_{\rm FIR}$ maps in Figure \ref{160maps} with the
$q_{70}$ and $q_{24}$ maps in Figure \ref{maps}, we find the
morphologies and associated trends are generally similar.
Quantitatively, we compare the dispersions in  $q_{\rm FIR}$
($\sigma_{\rm FIR}$), $q_{70}$ ($\sigma_{70}$), and $q_{24}$
($\sigma_{24}$) across each disk 
using projected 1.5~kpc diameter apertures. 
Looking at the computed dispersion for each disk in Table \ref{tbl160}
we find that the scatter generally decreases when using 70~$\micron$
data as opposed to the 24~$\micron$ data and is lowest in all cases
when using the estimated FIR emission.  
This suggests that the correlation between the radio and FIR
emission within each galaxy is tighter than the correlation between
the radio and either of the monochromatic 70 or 24~$\micron$
emission bands.
However, since the dispersion in $q_{70}$ is only $\sim$0.03~dex
larger than the dispersion in $q_{\rm FIR}$, we perform our
analysis at our best common resolution between the infrared and radio
data (i.e. at the 70~$\micron$ resolution) since the area of the beam
is a factor of $\sim$4 smaller than at the resolution of the
160~$\micron$ data.  

We find elevated infrared/radio ratios at 70 and 24~$\micron$
associated with bright structures appearing in the input infrared and
radio images of each galaxy.
The most obvious case is seen for the bright spiral arms of NGC~3031,
NGC~5194 and NGC~6946.
The spiral structure in all three of these galaxies is visible in their
infrared/radio ratio maps, which show enhanced values along the arms
with local peaks centered on H~II regions and depressed ratios located
in the quiescent inter-arm and outer-disk regions of each galaxy.
For NGC~2403, which does not have a grand-design spiral morphology, we
still see $q_{70}$ and $q_{24}$ peaks associated with H~II regions.

While the peaks in the $q_{24}$ and $q_{70}$ maps appear spatially
coincident, there is a slight difference in their morphologies.
Even after degrading the resolution of the 24~$\micron$ maps to match
the PSF at 70~$\micron$, the corresponding $q_{24}$ maps display a
more compact morphology around star-forming regions within each
galaxy.  
This observation is expected since 24~$\micron$ emission traces hotter
dust than emission at 70~$\micron$ and is therefore more localized
around active star-forming regions.
In comparing the $q_{24}$ and $q_{70}$ maps for NGC~3031, NGC~5194 and
NGC~6946, we find the spiral arms in each galaxy appear less broad
and have more strongly peaked H~II regions in the $q_{24}$ maps
compared to the $q_{70}$ maps. 
We find a similar result for the bright H~II regions in NGC~2403.
These observations are consistent with recent {\it Spitzer}
results by \citet{gxh04} who report 24~$\micron$ emission to be
strongly peaked in star-forming regions within NGC~300 and
consequently suggest that emission at 24~$\micron$ is an intimate
tracer of ongoing star formation.

\subsection{Infrared/Radio Ratios vs. Infrared Surface Brightness}
In order to see how $q_{\lambda}$ ratios vary with the strength of
infrared surface brightness within each galaxy, we produced
scatter plots using the infrared and radio flux densities extracted
from our aperture photometry scheme described in $\S$2.4.  
Since the measuring apertures are equal in diameter for each galaxy,
the measured flux densities are directly proportional to surface
brightnesses.
These results are illustrated for both $q_{70}$ and $q_{24}$ in
Figures \ref{q70f70} and \ref{q24f24}, respectively. 
As these plots have naturally correlated axes, we over-plotted
the relation expected if the radio disk were completely flat in
brightness across the entire galaxy.

In Figures \ref{q70f70} and \ref{q24f24}, we see a general trend of
increasing infrared/radio ratios with increasing infrared surface
brightness. 
The slopes of the regression lines within the scatter plots are
significantly lower (by a factor $\ga$2) than what would be expected
for a radio disk characterized by a constant surface brightness.  
This non-linearity of increasing infrared/radio ratio with increasing
infrared surface brightness within galaxies has been observed by other
authors \citep{mh95, hoer98, hip03}, and is opposite to the
non-linearity observed in the {\it global} FIR-radio correlation in
which the radio power of galaxies increases faster than infrared
luminosity \citep{fitt88, cox88, con91}.
The concern that this non-linearity may be a color effect is
unwarranted since the gradient in the color correction
FIR/$f_{\nu}(70\micron)$ would have to be $\sim$5 times steeper
than what is observed to eliminate this trend.
The measured dispersion in $q_{70}$ and $q_{24}$ is $\la$0.25~dex for
each galaxy (see Table \ref{tblstats}), which is only slightly larger
than the nominal dispersion of 0.2~dex measured in the {\it global}
FIR-radio correlation for late-type star-forming galaxies which do not
host powerful AGN \citep{gxh85}. 
Our measured dispersion, however, is in agreement with what was
found by \citet{yun01} using a much larger sample of galaxies
spanning a wider range of parameters than prior IRAS-based efforts.

In comparing Figures \ref{q70f70} and \ref{q24f24} for each galaxy, we
find that the dispersion in $q_{24}$ is a bit larger than 
what is found for $q_{70}$, except in the case of NGC~3031.  
However, if we look at the dispersion about each regression line, we
find the dispersion in $q_{24}$ to be generally smaller than what is
found for $q_{70}$.

\subsection{Environmental Trends}
In all galaxies, there are clear differences in $q_{\lambda}$ values
among the different galaxy disk environments.
The different environments are well separated in Figures \ref{q70f70}
and \ref{q24f24}, and tend to clump along the regression lines in
these Figures due to their relative surface brightnesses.
The measured dispersion for each environment appears to scale with the
range of star formation activity within it.
We also find that in the galaxies with a well defined infrared
nucleus, the infrared/radio ratios of the nuclei do not fall along the
regression line as seen in Figures \ref{q70f70} and \ref{q24f24}.
In NGC~3031 we find the nuclear $q_{70}$ and $q_{24}$ ratios lie
$\sim$1.7 and $\sim$1.2~dex below what is expected from the fitted
regression line, respectively.  
We also find that the circumnuclear regions of NGC~3031 display a
trend of decreasing infrared/radios ratios with increasing infrared
surface brightness. 
As the nucleus of NGC~3031 is known to host an AGN, this result is
expected.
The nucleus of NGC~5194 is categorized as an H~II/Sy2, and accordingly
we find the associated $q_{70}$ and $q_{24}$ ratios lie below
the expectation of the regression line by 0.3~dex.
The nuclear $q_{70}$ and $q_{24}$ ratios in NGC~6946 lie below the
regression line expectation by 0.4~dex, even though NGC~6946 is not
known to host an AGN which would provide extra radio emission.

\subsection{Radial Trends}
We identify any radial trends which might exist for $q_{70}$ and
$q_{24}$ in Figures \ref{q70rad} and \ref{q24rad}, respectively.  
For each galaxy in our sample, there is an obvious trend of decreasing
$q_{70}$ and $q_{24}$ ratios with increasing galactocentric radius.
A similar trend was also found by \citet{bh90} who, using IRAS scan
data, observed a decrease in 60~$\micron$ to 20~cm ratios with
increasing radius.
This result can be characterized by smaller scale-lengths for the
infrared disks than the radio disks.

What we find in Figures \ref{q70rad} and \ref{q24rad} is a slight
trend of increasing dispersion in the infrared/radio ratios with
radius.
The only exceptions are for NGC~2403 at 24~$\micron$, and
NGC~3031, in which the dispersion is large in the circumnuclear
region due a combination of the central AGN and the non-Gaussian
MIPS PSF. 
We also note that NGC~3031 displays anomalously low infrared/radio
ratios for a few apertures at a radius of $\sim$4~kpc because of
SN~1993J.
This trend of increasing dispersion in $q_{70}$ and $q_{24}$ with
radius does not seem to be an artifact of lower signal-to-noise
at larger radii as the general appearance of Figures \ref{q70rad}
and \ref{q24rad} persists even when we increased the detection
threshold from 3~$\sigma$ to 6~$\sigma$. 
We also find that the dispersion in $q_{70}$ and $q_{24}$ at
constant radius is much larger than at constant surface brightness.
To quantify this, we computed the dispersion in 1~kpc and 1.5~Jy
bins about the median radius and flux density, respectively, and 
find that the dispersion in $q_{70}$ and $q_{24}$ is larger at
constant radius than at constant surface brightness by an average of
$\sim$0.1~dex.
By moving farther out radially into the disks of galaxies two effects
occur.
There is a general drop in the disk surface brightness coupled with a
drop in the H~II region density.
These two effects likely drive the increase in scatter for $q_{70}$
and $q_{24}$ ratios with radius and is the reason that there is a more
firm correlation between infrared/radio ratio with surface brightness
than with radius.
This suggests that the distribution of star formation sites
within the disk is more important in determining the overall
appearance of the infrared/radio disk maps than the underlying
exponential disk elements, such as the ISM mass distribution and the
older stellar population of galaxies.

\section{Discussion}
\subsection{Infrared/Radio Relations Inside and Among Galaxies}
The goal of this study is to improve our understanding of the physical
processes governing the FIR-radio correlation.  
Accordingly, we compare the results of {\it global} FIR-radio
studies with local ($\sim$kpc scale) FIR-radio studies. 
Identifying similarities and differences in the observed trends
between local and global studies can help to constrain the physical
scales and associated processes responsible for the FIR-radio
correlation.  

\subsubsection{Relating IRAS $q_{60}$ to {\it Spitzer} $q_{70}$} 
Since most of the previous work on the FIR-radio correlation has been
done using IRAS data, we had to convert IRAS 60~$\micron$ flux
densities to the nearby {\it Spitzer} 70~$\micron$ flux densities for
comparison.  
In order to convert IRAS 60$~\micron$ to {\it Spitzer} 70~$\micron$
flux densities used IRAS 60/100~$\micron$ flux density ratios
along with the SED models of \citet{dd02}.
The models allow for IRAS 60/100~$\micron$ flux density ratios
in the range of 0.2847 to 1.635 corresponding to a range in {\it
  Spitzer} 70/IRAS 60~$\micron$ flux density ratios between
0.9585 and 1.568.
This relation between {\it Spitzer} $f_{\nu}(70~\micron)$ to IRAS
$f_{\nu}(60~\mu\rm{m})$ and $f_{\nu}(100~\micron)$ flux densities is
approximated by 
\begin{equation}
f_{\nu}(70~\micron) = \left \{
\begin{array}{cc}
f_{\nu}(60~\micron) \displaystyle\sum_{i=0}^{3}
\xi_{i}\left(\frac{f_{\nu}(60~\micron)}
   {f_{\nu}(100~\micron)}\right)^{i} 
& ~~~~~0.3044 \leq \frac{f_{\nu}(60~\micron)}{f_{\nu}(100~\micron)} \leq
1.635 \\ 
1.532 f_{\nu}(60~\micron) & ~~~~~~~0.2847 \leq
\frac{f_{\nu}(60~\micron)}{f_{\nu}(100~\micron)} < 0.3044\\
\end{array}
\right .,
\end{equation}
where \([\xi_{0},\xi_{1},\xi_{2},\xi_{3}] = [1.976, -1.582,
  0.9632, -0.2308]\).  
We derived these coefficients using a singular value
decomposition solution to an overdetermined set of linear equations as
described in $\S$15.4 of \citet{press02}.

\subsubsection{Comparison with {\it Global} Infrared/Radio Ratios}
For a comparison of our results to previous {\it global} FIR-radio
correlation data, we made use of IRAS and NRAO VLA Sky Survey
(NVSS; \citet{con98}) data collected for a sample of 1809 galaxies by
\citet{yun01}.  
Of these 1809 galaxies, 1752 had IRAS 60/100~$\micron$ flux
density ratios compatible with the range of \citet{dd02} SED
models, and for this sub-sample we converted the observed IRAS
60~$\micron$ flux densities into estimated {\it Spitzer} 70~$\micron$
flux densities (see $\S$4.1.1).  
Using the 1.4~GHz NVSS data and estimated distances to the sources
\citep{yun01}, we plot global infrared/radio ratios along with our  
local infrared/radio ratios for 1.5~kpc diameter apertures versus
luminosity in the top portion of Figure \ref{globloc}.  
Although the NVSS is a snapshot survey, and is therefore prone
to miss extended emission from galaxies having large angular
extents,  \citet{yun01} pay proper attention to these effects and
derive unbiased 1.4~GHz fluxes. 
It should be noted that this sample, however, has been found to
contain both confused IRAS measurements and AGN missed by automated
procedures, which are not expected to obey the FIR-radio correlation.
We also compare our $q_{24}$ results within galaxies to global
 $q_{24}$ ratios from data obtained as part of the {\it Spitzer}
First Look Survey (FLS) in the bottom portion of Figure
\ref{globloc}.  
A total of 179 sources are plotted using 24~$\micron$ and VLA 1.4~GHz
measurements which have been $k$-corrected using an SED-fitting method
as described in \citet{pa04}.
Objects having a $q_{24}$ value well below 0 are likely to be galaxies
hosting an AGN and are not expected to follow any correlation found in
our aperture work within galaxies. 

In Table \ref{tblglobstats} we list the mean and standard deviation of
$q_{70}$ and $q_{24}$ found within and among galaxies, as well as the
number of measurements used to calculate each.
In this comparison we present all data points including the few
outliers in each sample.
For our four sample galaxies, the local dispersions in $q_{24}$ and
$q_{70}$ are nearly identical.
The dispersion in the $\sim$kpc scale $q_{70}$ ratios is comparable
to what is measured globally, but the dispersion in the global
$q_{24}$ ratios is 0.13~dex higher than what is measured within
galaxies.  
This increase in dispersion may be due to sample selection as the FLS
contains galaxies at $z\la$1, while the \citet{yun01} sample contains
objects only up to $z\la$0.15.
Samples at higher redshifts will likely include a larger number of AGN
and perhaps less evolved galaxy disks compared to samples limited to
lower redshifts, and both effects may increase the dispersion.

In both panels of Figure \ref{globloc} the global infrared/radio
ratios of our sample galaxies appear to be slightly 
higher than the median of the corresponding local kpc-scale values.
This offset is not statistically significant, as the global value
is never greater than the median by more than 1~$\sigma$, and likely
due to the brightest regions in galaxies contributing a large fraction
of the global flux.  
What clearly appears as a significant difference between the local
and global infrared/radio ratios is the behavior of $q_{\lambda}$
versus increasing luminosity.  
In both the FLS and \citet{yun01} samples the infrared/radio ratios
are roughly constant with increasing galaxy luminosity while, within
each disk, the infrared/radio ratios clearly increase with luminosity.   
We will see in $\S$4.2.2 that the difference in 
$q_{\lambda}$ versus luminosity within and among galaxies is likely
due to the diffusion of CR$e^{-}$ within the galaxy disks.  
Although we do not see a strong non-linearity in either sample of
global infrared/radio ratios, we note that a non-linearity in the
global FIR-radio correlation is known to exist.
However, this trend in global FIR/radio ratios is the opposite of what
we find on kpc scales within galaxies, which is that galaxies with low
FIR luminosities have radio luminosities lower than expected from a
linear fit to the correlation \citep{fitt88, cox88, con91}.

\subsection{Cosmic Ray Diffusion}
\subsubsection{Image-Smearing Model Technique}
The phenomenological image-smearing model of \citet{bh90} predicts
that the radio morphology of a galaxy can be reproduced by convolving
the FIR image with a specific smearing kernel, $\kappa({\bf r})$,
containing the diffusion information of the galaxy's cosmic ray
electrons (CR$e^{-}$s).
As previous work relied on IRAS maps made using the maximum
correlation method (MCM), described by \citet{afm90}, to achieve
'super'-resolution of $\la1\arcmin$ (i.e. \citet{mh98}), 
it is worth repeating this analysis using the current {\it Spitzer}
maps obtained at the natural resolution of the instruments.
We performed a simple image-smearing analysis for both the {\it Spitzer}
24 and 70~$\micron$ data and look for a preference among Gaussian and
exponential smearing kernels projected either in the plane of the sky
or in the plane of the galactic disk.  
The choice of Gaussian and exponential kernels are due to their
differences in describing the diffusion and confinement
characteristics of CR$e^{-}$s.  
Gaussian kernels suggest a simple random walk diffusion scenario for
CR$e^{-}$s in each disk.
Exponential kernels, having broader tails than Gaussian kernels of the
same scale-length, are suggestive of CR$e^{-}$ escape on time scales
less than or comparable to the diffusion time scales, and correspond to
empirical ``leaky box'' models \citep{bh90}.  

The kernel, $\kappa({\bf r})$, is a function of a two-dimensional
angular position vector ${\bf r}$ with magnitude \(r = (x^{2} +
y^{2})^{1/2}\), where $x$ and $y$ are the right ascension and
declination offsets on the sky. 
Let $R({\bf r})$ and $I({\bf r})$ denote the {\it observed} radio and
infrared images respectively.
Let us denote the type of function and projection of the parameterized
kernel by the subscripts $t,p$ such that Gaussian (G) and exponential 
(e) kernels projected in the plane of the galaxy (g), $\kappa({\bf
  r})_{t,g}$,  take the form of
\(\kappa({\bf r})_{G,g} = e^{-r^{2}/(r_{\rm o}^{2})}\) and
\(\kappa({\bf r})_{e,g} = e^{-r/(r_{\rm o})}\),
respectively where
\begin{equation}
r_{\rm o} = \frac{l\cos i}{[1 - (x\sin \theta +
  y\cos\theta)^{2}\sin^{2}i/r^{2}]^{1/2}}.
\end{equation}
The quantities $i$ and $\theta$ are the inclination, where $i = 0$
defines a face-on projection, and position angle of the tilt axis of
the galactic disk measured East of North, respectively, and $l$ is the
$e$-folding length of the smearing kernel.
When the kernels are oriented in the plane of the sky (s), denoted as
$\kappa({\bf r})_{t,s}$, both $i$ and $\theta$ are equal to 0 which
sets $r_{\rm o} = l$. 
We define the quantity,
\begin{equation}
\phi(Q,t,p,l) = \frac{\sum[Q^{-1}\tilde{I}_{j}(t,p,l) -
    R_{j}]^{2}}{\sum R_{j}^2},
\label{phi}
\end{equation} 
where \(Q=\frac{\sum \tilde{I}_{j}({\bf r})}{\sum R_{j}({\bf r})}\) is
used as a normalization factor (i.e. \(\log(Q) = q_{\lambda}(global)\)),
$\tilde{I}({t,p,l})$ represents the infrared image after smearing with
kernel of type ($t,p$) having scale length $l$, and the subscript $j$
indexes each pixel.
This quantity $\phi$ was minimized to determine the best fit smearing
kernel for each galaxy in our sample.  
We normalize by the squared sum of radio flux density to allow for
proper comparison of our galaxies which vary in intrinsic surface
brightness. 
Estimation of the residuals was carried out after first removing
pixels not detected at the 3~$\sigma$ level, and then editing out
identifiable contaminating background radio sources and SNe.
Because the AGN nucleus of NGC~3031 is also identifiable in the
infrared images, it was removed before smearing the infrared images
and in the calculation of the residuals.
In the case of NGC~5194, its companion galaxy (NGC~5195) was removed
before calculating the residuals.
We find the best fit kernel by determining the minimum in \(\log(\phi)\)
as a function of smearing scale-length $l$, as shown in Figures
\ref{res70} and \ref{res24}.  
The quantity,
\begin{equation}
\Phi = \log\left(\frac{\phi(Q,t,p,0)}{{\rm min}(\phi(Q,t,p,l))}\right),
\end{equation}
which is the maximum depth of each residual trough, is a measure of
how much the correlation is improved by smoothing the infrared image.

Because $\phi$ and $\Phi$ characterize the residual behavior for the
entire galaxy as single quantities, we also constructed residual
maps for the best fit smearing kernels to inspect the spatial
variations of the residuals.  
The residual image is defined as,
\begin{equation}
{\rm Residual~image} = \log(Q^{-1}\tilde{I}({\bf r})) - \log(R({\bf r})).
\end{equation}
In Figures \ref{smear70} and \ref{smear24} we plot residual maps for
the best fit exponential kernels oriented in the plane of the sky
for the 70 and 24~$\micron$ data, respectively.  

Making the assumption that propagation of CR$e^{-}$s is symmetric in
the plane of the sky, we can crudely attempt to measure the CR$e^{-}$
diffusion length within each galaxy disk. 
The infrared images, $I({\bf r})$, might be considered a 'smeared'
version of the distribution of the original sources of infrared
luminosity.  
They are 'smeared' due to both the heating of dust by UV photons,
having a scale-length $l_{\rm UV}$ some hundreds of parsecs away from
their originating sources, and the angular resolution of the
telescope, having a beam width $l_{\rm beam}$.  
Accordingly, the infrared image, $\tilde{I}({\bf r})$, artificially
smeared by a kernel having a scale-length $l$, has an effective
scale-length of $l_{\tilde{I}}$, such that
\begin{equation}
l_{\tilde{I}}^{2} = l_{\rm beam}^{2} + l_{\rm UV}^{2} + l^{2}.
\end{equation}
The radio images are initially 'smeared' by both the angular
resolution of the telescope and propagation of CR$e^{-}$s, having a
scale-length of $l_{{\rm CR}e^{-}}$, such that the total scale-length
of the radio continuum image, $l_{\rm R}$ is approximated by, 
\begin{equation}
l_{\rm R}^{2} = l_{\rm beam}^{2} + l_{{\rm CR}e^{-}}^{2}.
\end{equation}
Assuming the smearing model holds, we set \(l_{\rm R}^{2} =
l_{\tilde{I}}^{2}\) and find a general relation between the
scale-length of the smearing kernel and the scale-lengths of the UV
heating photons and CR$e^{-}$s such that,
\begin{equation}
  l^{2} = l_{{\rm CR}e^{-}}^{2} - l_{\rm UV}^{2}.
\end{equation}
The scale-length of the best fit smearing kernel is a combination of
the CR$e^{-}$ and UV photon scale-lengths, and we cannot separate
their effects with the current data.
However, the scale-length of the CR$e^{-}$s is probably significantly
larger than that for the UV photons (e.g. \citet{hb93}), and therefore
is the dominant term in the scale-length of the best fit smearing kernel.
Once we have determined the best fit smearing kernels, we use the
corresponding smeared infrared maps to perform the same aperture
photometry as described in $\S$2.4 to calculate the mean and dispersion
in $q_{\lambda}$ on sub-kpc scales within each galaxy.
The smearing kernel scale-lengths ($l$), $\Phi$, and dispersion in
$q_{\lambda}$ are given in Table \ref{tblsmear} for each galaxy and
kernel type.

\subsubsection{Image-Smearing Model Results}
We look to see whether the image-smearing model works to significantly
improve the correlation between the infrared and radio morphologies of
galaxies.  
Determining the functional form and scale-length of the best fit
smearing kernel provides insight into the propagation and diffusion
characteristics of cosmic rays within galaxy disks.

An examination of Figures \ref{res70} and \ref{res24} shows that the
the image-smearing technique improves the overall correlation between
the radio maps and the 70 and 24~$\micron$ images by an average of 0.2
and 0.6~dex in \(\log(\phi)\), respectively.
Even though the infrared wavelengths being studied trace two different
temperature regimes and grain species, we find similar preferences in
kernels for both the smeared 70~$\micron$ and 24~$\micron$ images.
We find exponential kernels are preferred independent of
projection, since they have $\Phi$ values at least 0.04~dex larger
than those from Gaussian kernels.  
This result differs from \citet{mh98}, who found Gaussian and
exponential kernels to work equally well, but is consistent with
\citet{bh90}.  
Since Gaussian kernels describe a simple random-walk diffusion
for CR$e^{-}$s, additional processes such as escape and decay appear
necessary to describe the evolution of CR$e^{-}$s through the galaxy
disks, as was suggested by \citet{bh90}.  
Each galaxy displays a marginal difference between exponential kernels
oriented in the plane of the sky and those oriented in the plane of
the galaxy. 
This poor discrimination for kernel projections is not surprising due
to the low inclination of these face-on targets.  

We choose to only present residual maps for exponential kernels oriented 
in the plane of the sky (Figures \ref{smear70} and \ref{smear24})
since this kernel typically had the largest value of $\Phi$ for each
galaxy.   
Residual images for the other kernels appear very similar.  
It is obvious from the residual maps that a simple symmetric function
cannot properly fit each part of the galaxy as arm, inter-arm, and
individual giant H~II regions strongly deviate from having zero
residuals.
However, the residuals in Figures \ref{smear70} and \ref{smear24},
show two distinct and opposite trends.  
In the more active star-forming galaxies (NGC~5194 and NGC~6946: SFR
$>$4 M$_{\sun}$/yr) we see that star-forming regions (i.e. arms) have
residuals with infrared excesses while inter-arm and outer-disk
regions display residuals having radio excesses. 
For the two less active star-forming galaxies (NGC~2403 and NGC~3031:
SFR $<$1 M$_{\sun}$/yr), star-forming regions in the disk display
radio excesses while the inter-arm and outer-disk regions generally
have infrared excesses. 
Because the less active star-forming galaxies have larger
smearing scale-lengths, the bright star-forming regions that appear
in the infrared images of these galaxies are over-smoothed. 
This over-smoothing of the star-forming regions redistributes a
larger amount of flux into the more quiescent parts of the galaxy
disks than is needed to match what is observed in the radio image.  
The larger scale-lengths and over-smoothing of discrete star-forming
sites in these less active star-forming galaxies may be due to the
diffuse emission dominating the appearance of the disk.
We propose that the relative paucity of H~II regions translates into a
deficit of recent CR$e^{-}$ injection into the ISM, and a longer
effective timescale for CR$e^{-}$ diffusion.
While this picture is self-consistent and accounts for the data
on the four sample galaxies, further tests are required to
ascertain its applicability outside of this small sample of
galaxies.

The CR$e^{-}$ scale-lengths for the exponential kernels oriented in
the plane of the sky range from a few hundred parsecs to a couple
kpc.
Three out of the four galaxies have smearing scale-lengths for the
70~$\micron$ maps that are smaller than what is found for the
24~$\micron$ maps by a couple hundred parsecs.   
This result is not surprising since the 24~$\micron$ emission
is associated with hotter dust, and is more centrally peaked around
bright star-forming regions than the 70~$\micron$ emission.
The exception is NGC~3031 which does not have well determined
scale-lengths, perhaps 
because the galaxy is mainly quiescent and lacks a large number of
luminous star-forming regions for the size of its disk, so the
majority of the  24~$\micron$ emission is as diffuse as the
70~$\micron$ emission.  
While the smearing scale-lengths at 70 and 24~$\micron$ for NGC~3031
are not well determined, those for NGC~2403 do seem relatively well
determined even though both galaxies have similar star formation rate
surface densities.  
A comparison of HII~region luminosity functions shows that NGC~3031
has fewer high luminosity H~II regions than NGC~2403 \citep{pet88,
  siv90}, which may account for this difference.
  
The dispersion across each disk is found to be lower by at least
$\sim$0.04~dex using the smeared infrared images to construct the 
infrared/radio ratio maps, except for NGC~3031, which is probably due
to the inclusion of its AGN in calculating the dispersion.
We also find that the slopes in the $q_{70}$ and $q_{24}$ versus
infrared surface brightness relations are, on average, factors of
$\sim$2 and $\sim$1.5 times smaller, respectively, when using the
smeared infrared images.  
Accordingly, this reduction in slope improves the linearity of
$q_{\lambda}$. 
Since this non-linearity within disks is suppressed by smearing
the infrared images, it may well be a result of  the diffusion of
CR$e^{-}$ away from star-forming sites.  
In this case, such a non-linearity should not be found in the {\it
  global} correlation when integrating the flux over entire galaxies,
which is indeed what is observed in Figure \ref{globloc}. 

Assuming that the differences between the radio and infrared
distributions are due to the diffusion to CR$e^{-}$s, we can relate
the best fit smearing scale-lengths to the mean ages of the CR$e^{-}$
populations of each galaxy.    
Comparing the shapes of the residual curves in Figures \ref{res70} and
\ref{res24} for each galaxy, we find quite distinct behaviors between
our two more active star-forming galaxies (NGC~5194 and NGC~6946) and
our two less active star-forming galaxies (NGC~2403 and NGC~3031). 
The curves for NGC~5194 and NGC~6946 have clearly defined minima for
each kernel type. 
In contrast, the curves in NGC~2403 and NGC~3031 are less well
behaved, with the initial decrease in residuals being less smooth  
and, in the case of NGC~2403 at 70~$\micron$, displaying clearly
defined first and second minima having scale-lengths greater and
less than 1~kpc.  
We speculate that the double-minimum behavior, especially present for
NGC~2403 in Figure \ref{res70}, results from a superposition of two
populations of CR$e^{-}$s: 
those from an older episode of star formation, now associated with the
diffuse radio disk, and those from a more recent episode of star
formation, associated with the prominent H~II regions.   
We also speculate that NGC~2403 has gone through a period of relative
quiescence between the two episodes of star formation.
The relatively small scale-lengths ($\la$1~kpc) found in the first
minimum of NGC~2403 are consistent with the scale-lengths observed in
NGC~6946 and NGC~5194. 
They correspond to the spreading scale-length values expected in
galaxies with typical ISM density $\ga$5~cm$^{-3}$ and CR$e^{-}$ ages
$\la$$5\times10^{7}$yrs \citep{hb93}.
These relatively young CR$e^{-}$s are thought to have been recently
accelerated in star-forming regions.
Scale-lengths $\ga$1~kpc, however, are expected for older CR$e^{-}$
which have been diffusing through the ISM for $\ga$$5\times10^{7}$ yr.    
The double-minimum behavior is more apparent in NGC~2403 presumably
because of the relative luminosities of the diffuse disk emission
compared to the H~II regions, and because of the geometry of the star
formation sites with respect to the disk.

\section{Summary}
We present an initial look at the FIR-radio correlation within
galaxies using infrared and radio data taken as part of the {\it
  Spitzer} Infrared Nearby Galaxies Survey (SINGS) and the parallel
WSRT SINGS project.
We report on four of the most nearby objects in our sample, allowing
us to probe physical scales down to 0.3~kpc, for a 17$\arcsec$
beam, and analyze the variations in the logarithmic ratios of 70 and
24~$\micron$ dust emission to 22~cm radio continuum emission. 

We have also performed preliminary modeling of CR$e^{-}$ diffusion
using the image-smearing technique of \citet{bh90}.
We find that this phenomenological model of smoothing the infrared
maps to match the morphology of the radio maps does indeed improve the
correlation. 
This model relies on the fact that CR$e^{-}$s emit synchrotron
radiation as they diffuse away from the same star-forming regions
that heat dust, effectively creating a smoother version of the
infrared image at radio wavelengths.  
Characterizing the optimal smoothing kernel for the infrared map
provides insight into the evolution of the CR$e^{-}$s, including an
estimate of their diffusion scale-lengths.  
In our image-smearing analysis we have tested the differences between
Gaussian and exponential smoothing kernels, oriented in the planes of
the galaxy and sky, on the 70 and 24~$\micron$ maps of each galaxy.

As emission from 70 and 24~$\micron$ probe different grain populations,
the variations in the logarithmic 70~$\micron$/22~cm ($q_{70}$) and
24~$\micron$/22~cm ($q_{24}$) surface brightness ratios across each
galaxy disk are not identical. 
From comparisons of the $q_{70}$ and $q_{24}$ behavior within our
sample, along with our image-smearing analysis, we find the following:
\begin{enumerate}
\item 
  $q_{70}$ and $q_{24}$ generally decrease with declining surface
  brightness and increasing radius. 
  However, the dispersion measured in $q_{70}$ and $q_{24}$ at
  constant surface brightness is found to be smaller than at constant
  radius by $\sim$0.1~dex suggesting that the distribution of star
  formation sites is more important in determining the infrared/radio
  disk appearance than the underlying exponential disk elements,
  such as the ISM mass distribution and the older stellar population.
\item
  The $q_{24}$ ratio-maps are more strongly peaked on star-forming
  regions than the $q_{70}$ ratio maps at matching resolution and,
  consequently, the dispersion in $q_{70}$ for each disk is
  generally smaller ($\la$0.03~dex) than what is found for $q_{24}$,
  except in the case of NGC~3031.
  This is consistent with the 24~$\micron$ emission being more
  closely correlated spatially with sites of active star formation
  than the cooler 70~$\micron$ dust emission, as was found by
  \citet{cal05} and \citet{gxh04}. 
\item
  The ratio of FIR (42-122~$\micron$) to radio emission within
  galaxies displays less scatter than the monochromatic $q_{70}$ and
  $q_{24}$ ratios at matching resolution. 
  However, the dispersion in $q_{70}$ is never more than
  $\sim$0.03~dex larger than the dispersion in $q_{\rm FIR}$ for
  each galaxy. 
\item
  The dispersion in the {\it global} FIR-radio correlation is
  comparable to the dispersion measured in $q_{70}$ and $q_{24}$
  within the galaxy disks on 1.5~kpc scales.
  Also, the trend of increasing infrared/radio ratio with increasing
  infrared luminosity within each galaxy is not observed in the {\it
    global} correlation probably due to the diffusion of cosmic ray
  electrons.
\item
  The phenomenological modeling of cosmic ray electron (CR$e^{-}$)
  diffusion using an image-smearing technique is successful, as it
  both decreases the measured dispersion in $q_{70}$ and $q_{24}$ by
  an average of $\sim$0.05~dex, and reduces the slopes in the $q_{70}$
  and $q_{24}$ versus infrared surface brightness relations, on
  average, by a factor of $\sim$1.75.
  This reduction in slope suggests that the non-linearity in
  $q_{\lambda}$ within galaxies may be due to the diffusion of
  CR$e^{-}$s from star-forming regions. 
\item
  The image-smearing models with exponential kernels work marginally
  better to tighten the correlation between the radio and
  infrared maps than Gaussian kernels, independent of projection.  
  This result suggests that CR$e^{-}$ evolution is not well described
  by random-walk diffusion in three dimensions alone and requires
  additional processes such as escape and decay.
\item
  Using a simple symmetric smearing kernel to smooth the infrared
  image does not provide a perfect fit to the radio continuum image,
  and leaves organized structures such as arms and H~II regions still
  visible.  
  Our two less active star-forming galaxies display radio excesses
  around star-forming regions in their residual maps while our two
  more active galaxies have infrared excesses around star-forming
  arms. 
  This difference in the appearance of the residual maps may be
  due to time-scale effects in which there has been a deficit of
  recent CR$e^{-}$ injection into the ISM in the two less active
  star-forming galaxies, thus leaving the underlying diffuse disk as
  the dominant structure in the morphology.
\end{enumerate}

\acknowledgements
The authors would like to thank the anonymous referee for helpful
comments and suggestions that have improved the content and
presentation of this paper.    
E. J. M. would like to acknowledge support for this work provided by
the {\it Spitzer} Science Center Visiting Graduate Student program.
As part of the {\it Spitzer} Space Telescope Legacy Science Program,
support was provided by NASA through Contract Number 1224769 issued
by the Jet Propulsion Laboratory, California Institute of Technology
under NASA contract 1407.

\clearpage
\begin{deluxetable}{cccccccccccc}
  \tablecaption{Galaxies\label{tblgaldat}}
  \tablewidth{0pt}
\tabletypesize{\scriptsize}
  \tablehead{
    \colhead{} & \colhead{R.A.} & \colhead{Decl.} &
    \colhead{$D_{25}$} & \colhead{} & \colhead{} & \colhead{$V_{r}$} &
    \colhead{Dist.} & \colhead{$i$} & \colhead{PA} & \colhead{SFR} &
    \colhead{} \\
    \colhead{Galaxy} & \colhead{(J2000)} & \colhead{(J2000)} &
    \colhead{(arcmin)} & \colhead{Type} & \colhead{Nuc.} & 
    \colhead{(km s$^{-1}$)} & \colhead{(Mpc)} & \colhead{($\degr$)} &
    \colhead{($\degr$)} & \colhead{(M$_{\sun}$/yr)} &
    \colhead{$q_{\rm FIR}$}\\ 
    \colhead{(1)} & \colhead{(2)} & \colhead{(3)} & \colhead{(4)} & 
    \colhead{(5)} & \colhead{(6)} & \colhead{(7)} & \colhead{(8)} &
    \colhead{(9)} & \colhead{(10)} & \colhead{(11)} & \colhead{(12)}
  }
  \startdata
  NGC~2403 & 7 36 51.4 & +65 36 09 & 21.9$\times$12.3 & SABcd & H~II & 
  131 & 3.5 & 57 & 127 & 0.77 & 2.50\\
  NGC~3031 & 9 55 33.2 & +69 03 55 & 26.9$\times$14.1 & SAab  & Lin &
  -34 & 3.5 & 60 & 157 & 0.86 & 2.45\\
  NGC~5194 & 13 29 52.7 & +47 11 43 & 11.2$\times$6.9 & SABbc & H~II/Sy2 &
  463 & 8.2 & 53 & 163 & 6.1 & 2.09\\
  NGC~6946 & 20 34 52.3 & +60 09 14 & 11.5$\times$9.8 & SABcd & H~II & 
  48 & 5.5 & 32 & 69 & 4.1 & 2.28\\
  \enddata
  \tablecomments{Col. (1): ID. Col. (2): The right ascension in the
    J2000.0 epoch. Col. (3): The declination in the J2000.0
    epoch. Col. (4): Major- and minor-axis diameters. Col. (5): RC3
    type. Col. (6): Nuclear type: H~II: H~II region; Lin: LINER; Sy:
    Seyfert (1, 2). Col. (7): Heliocentric velocity. Col. (8):
    Flow-corrected distance in Mpc, for $H_{o}$ = 70 km s$^{-1}$
    Mpc$^{-1}$ \citep{rk03}. Col. (9): Inclination in
    degrees. Col. (10): Position Angle in degrees. Col. (11): IRAS based
    star formation rate (SFR) \citep{efb03}. Col (12): 
    \(q_{\rm FIR} \equiv \log\left(\frac{\rm FIR}{3.75\times10^{12}
      ~{\rm W m^{-2}}}\right) - \log\left(\frac{S_{1.4~{\rm GHz}}}{\rm
      W m^{-2} Hz^{-1}}\right)\) \citep{gxh85}.} 
\end{deluxetable}

\clearpage
\begin{deluxetable}{cccccc}
  \tablecaption{RMS Noise of {\it Spitzer} Infrared and WSRT Radio
  Maps \label{galrms}}
  \tablewidth{0pt}
  \tablehead{
    \colhead{Galaxy} & \colhead{160~$\micron$} & \colhead{70~$\micron$} 
    & \colhead{24~$\micron$} & \colhead{24~$\micron$$^{a}$} &
    \colhead{22~cm}\\
    \colhead{} & \colhead{MJy/sr} & \colhead{MJy/sr} & \colhead{MJy/sr} &
    \colhead{MJy/sr} & \colhead{$\mu$Jy/beam}
  }
  \startdata
  NGC~2403 & 0.61 & 0.35 & 0.044 & 0.0085 & 26\\
  NGC~3031 & 0.96 & 0.37 & 0.042 & 0.011  & 24$^{b}$\\
  NGC~5194 & 0.94 & 0.43 & 0.044 & 0.019  & 29\\
  NGC~6946 & 1.4  & 0.58 & 0.068 & 0.029  & 37\\
  \enddata
  \tablecomments{($^{a}$) 24~$\micron$ map convolved to match the
  70~$\micron$  PSF.  ($^{b})$ RMS measurement based on a 20~cm radio
  map.} 
\end{deluxetable}

\clearpage
\begin{deluxetable}{cccccccc}
  \tablecaption{Aperture Photometry Statistics For Galaxy
    Regions\label{tblstats}} 
  \tablewidth{0pt}
  \tabletypesize{\footnotesize}
  \tablehead{
    \colhead{} & \colhead{Nucleus} & \colhead{Inner-Disk}
    & \colhead{Disk} & \colhead{Outer-Disk} & \colhead{Arm} &
    \colhead{Inter-Arm} & \colhead{Total}
  }
  \startdata
  \cutinhead{NGC~2403}
  $<q_{70}>$    &\nodata &2.56 &2.43 &2.08 &2.40 &2.24 &2.24\\
  $\sigma_{70}$ &\nodata &0.05 &0.07 &0.18 &0.08 &0.13 &0.22\\
  $<q_{24}>$    &\nodata &1.42 &1.28 &0.90 &1.27 &1.06 &1.08\\
  $\sigma_{24}$ &\nodata &0.08 &0.10 &0.18 &0.16 &0.10 &0.24\\
  \cutinhead{NGC~3031}
  $<q_{70}>$    &1.31    &2.44 &2.34 &2.07 &2.29 &1.91 &2.20\\
  $\sigma_{70}$ &\nodata &0.26 &0.18 &0.14 &0.19 &0.14 &0.25\\
  $<q_{24}>$    &0.58    &1.20 &1.12 &0.90 &1.11 &0.78 &1.02\\
  $\sigma_{24}$ &\nodata &0.18 &0.16 &0.16 &0.19 &0.15 &0.22\\
   \cutinhead{NGC~5194}
  $<q_{70}>$    &1.94    &1.99 &\nodata &1.74 &2.02 &1.95 &1.90\\
  $\sigma_{70}$ &\nodata &0.09 &\nodata &0.17 &0.17 &0.17 &0.20\\
  $<q_{24}>$    &0.97    &1.00 &\nodata &0.63 &1.01 &0.91 &0.86\\
  $\sigma_{24}$ &\nodata &0.10 &\nodata &0.17 &0.20 &0.16 &0.23\\
   \cutinhead{NGC~6946}
  $<q_{70}>$    &2.13    &2.19 &2.02 &1.81 &2.11 &1.86 &1.94\\
  $\sigma_{70}$ &\nodata &0.05 &0.09 &0.15 &0.16 &0.16 &0.20\\
  $<q_{24}>$    &1.62    &1.39 &1.11 &0.77 &1.13 &0.87 &0.95\\
  $\sigma_{24}$ &\nodata &0.09 &0.08 &0.16 &0.15 &0.17 &0.23\\
  \enddata
\end{deluxetable}

\clearpage
\begin{deluxetable}{cccc}
  \tablecaption{Aperture Photometry Statistics for All Galaxy Regions
    (at 160~$\micron$ Resolution) \label{tbl160}}
  \tablewidth{0pt}
  \tablehead{
    \colhead{Galaxy}&
    \colhead{$\sigma_{24}$} & \colhead{$\sigma_{70}$} &
    \colhead{$\sigma_{\rm FIR}$}
  }
  \startdata
  NGC~2403& 0.21& 0.17& 0.15\\
  NGC~3031& 0.18& 0.19& 0.16\\
  NGC~5194& 0.16& 0.12& 0.11\\
  NGC~6946& 0.18& 0.14& 0.12
  \enddata
  \tablecomments{Dispersions ($\sigma_{\lambda}$) were calculated for
    $q_{24}$, $q_{70}$, and $q_{\rm FIR}$ within each galaxy disk
    using apertures having projected diameters of 1.5~kpc.}
\end{deluxetable}

\clearpage
\begin{deluxetable}{c|ccc|ccc}
  \tablecaption{Statistics {\it within} and {\it among} Galaxies
    \label{tblglobstats}}
  \tablewidth{0pt}
  \tablehead{
    \colhead{}&
    \colhead{$<q_{70}>$} & \colhead{$\sigma_{70}$} & \colhead{N} &
    \colhead{$<q_{24}>$} & \colhead{$\sigma_{24}$} & \colhead{N}
  }
  \startdata
      {\it global} & 2.30& 0.27& 1752& 0.92& 0.35& 179\\
      {\it within} & 2.02& 0.24&  282& 0.95& 0.23& 282
  \enddata
  \tablecomments{Statistics {\it within} galaxies using apertures
    of 1.5~kpc in diameter.  The {\it global} 70~$\micron$ data was
  taken from the \citet{yun01} sample and the {\it global} 24~$\micron$
  data was taken from the {\it Spitzer} FLS sample \citep{pa04}.  
  All data points, including outliers, were used in calculating these
  statistics for each set of data.}  
\end{deluxetable}

\clearpage
\begin{deluxetable}{cccc|ccc|ccc|ccc}
  \tablecaption{Smearing Kernel Results \label{tblsmear}}
  \tablewidth{-1pt}
  \tablehead{
    \multicolumn{1}{c}{}&
    \multicolumn{3}{c}{\(\kappa({\bf r})_{e,g}\)}&
    \multicolumn{3}{c}{\(\kappa({\bf r})_{e,s}\)}&
    \multicolumn{3}{c}{\(\kappa({\bf r})_{G,g}\)}&
    \multicolumn{3}{c}{\(\kappa({\bf r})_{G,s}\)}\\    
    \cline{2-13}
    \colhead{Galaxy} &
    \colhead{{\it l}} & \colhead{$\Phi$} &
    \colhead{$\sigma_{\lambda}$} &
    \colhead{{\it l}} & \colhead{$\Phi$} &
    \colhead{$\sigma_{\lambda}$} &
    \colhead{{\it l}} & \colhead{$\Phi$} &
    \colhead{$\sigma_{\lambda}$} &
    \colhead{{\it l}} & \colhead{$\Phi$} &
    \colhead{$\sigma_{\lambda}$} 
  }
 \startdata 
 \cutinhead{Smearing 70~$\micron$ maps}
 NGC~2403 &1300&0.07&0.15& 900&0.08&0.15&3000&0.03&0.16&2100&0.04&0.16\\
 NGC~3031 &4100&0.26&0.24&2500&0.31&0.23&8300&0.24&0.24&4900&0.29&0.23\\  
 NGC~5194 & 700&0.27&0.16& 500&0.29&0.16&1400&0.24&0.16&1100&0.26&0.17\\
 NGC~6946 & 300&0.17&0.17& 300&0.17&0.16& 500&0.11&0.18& 500&0.11&0.18\\  
 \cutinhead{Smearing 24~$\micron$ maps (matched to 70~$\micron$ resolution)}
 NGC~2403 &1400&0.83&0.15&1000&0.81&0.16&3200&0.77&0.16&2300&0.75&0.17\\
 NGC~3031 &1000&0.34&0.25&1500&0.33&0.24&1800&0.30&0.27&4100&0.30&0.24\\  
 NGC~5194 & 900&0.36&0.17& 600&0.36&0.18&1700&0.31&0.19&1200&0.32&0.20\\
 NGC~6946 & 800&1.03&0.16& 700&1.03&0.16&1600&0.89&0.18&1400&0.89&0.18\\  
 \enddata
 \tablecomments{Scale-length {\it l} given in units of parsecs; The
  newly measured dispersion in $q_{\lambda}$ ($\sigma_{\lambda}$) 
  calculated over the same apertures as done for $\sigma_{70}$ and
  $\sigma_{24}$ in Table \ref{tblstats}, including the regions which
  were removed before calculating residuals.}
\end{deluxetable}

\clearpage
\thispagestyle{empty}
\begin{figure}[!ht]
  \begin{center}
    \vspace{-1cm}
    \resizebox{3.75cm}{!}{
      \rotatebox{90}{
      \plotone{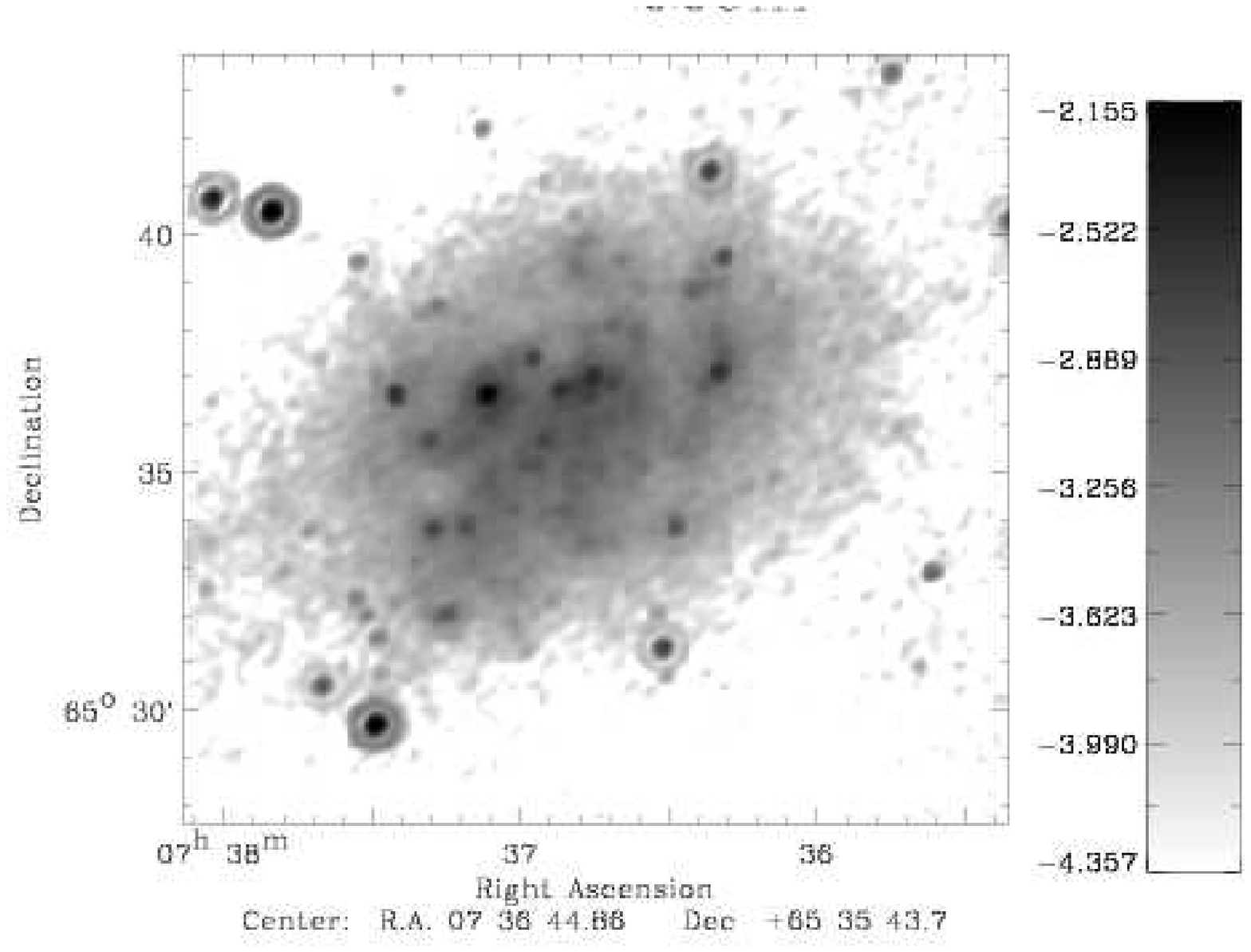}\hspace{-1.5cm}
      \plotone{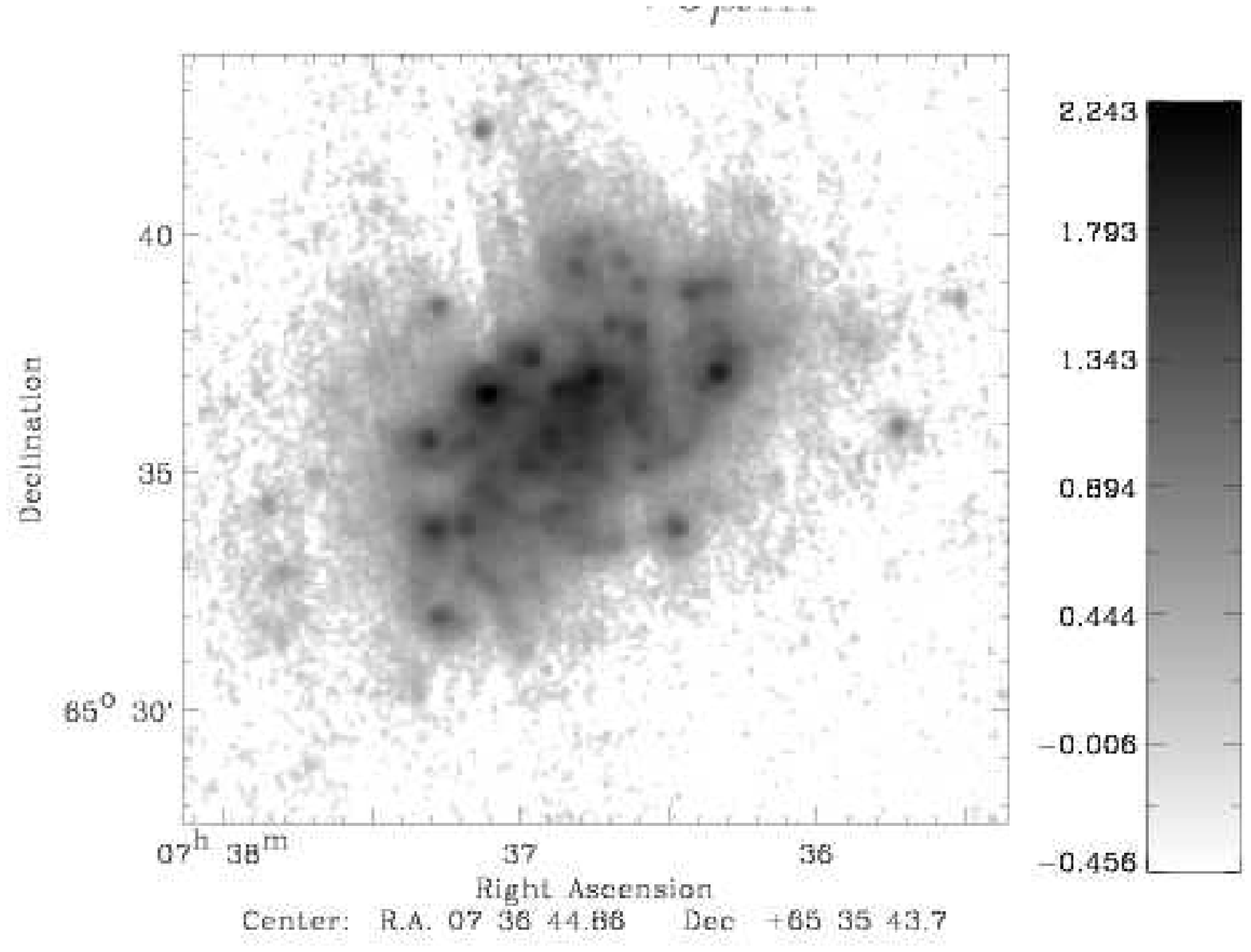}\hspace{-1.5cm}
      \plotone{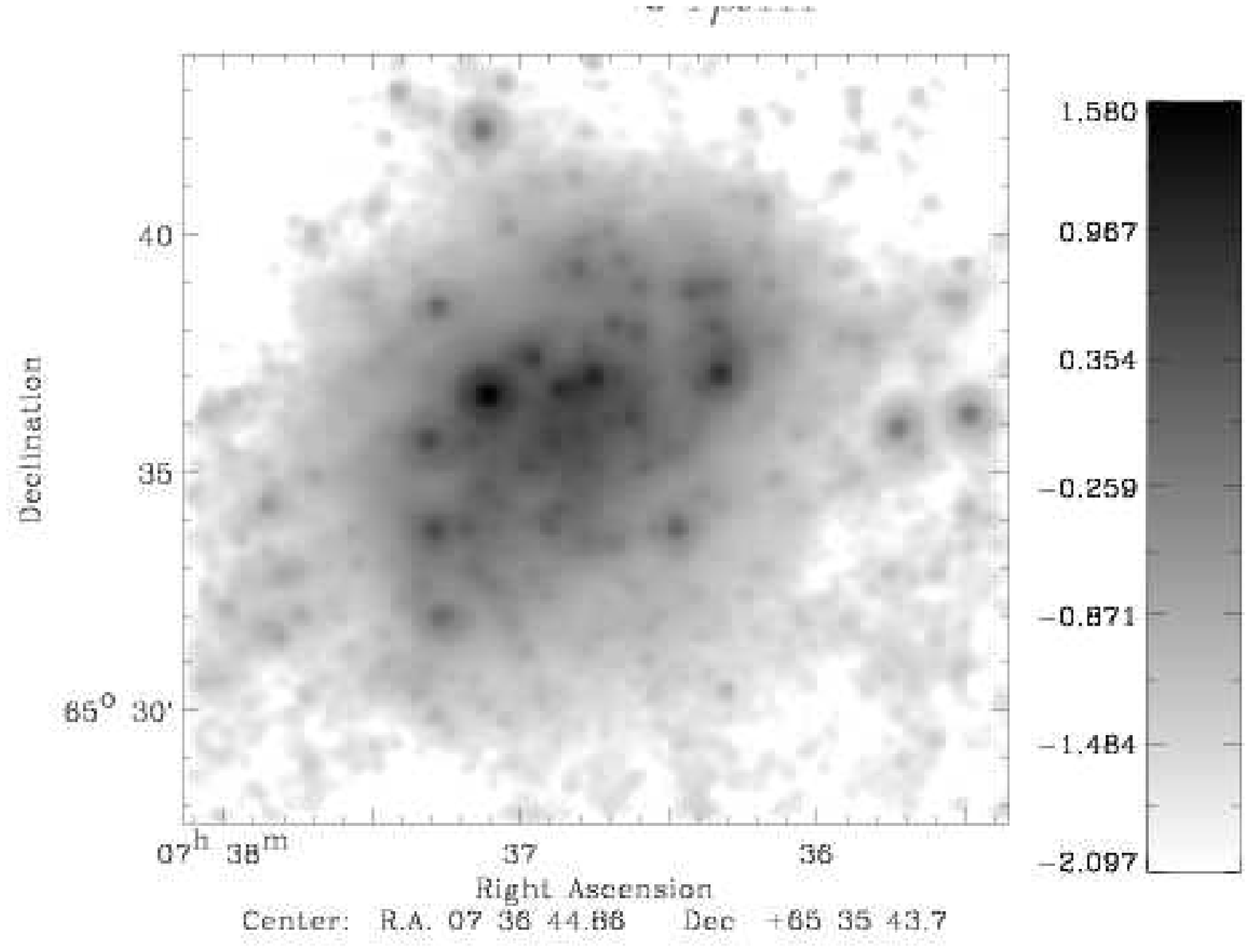}\hspace{-1.5cm}
      \plotone{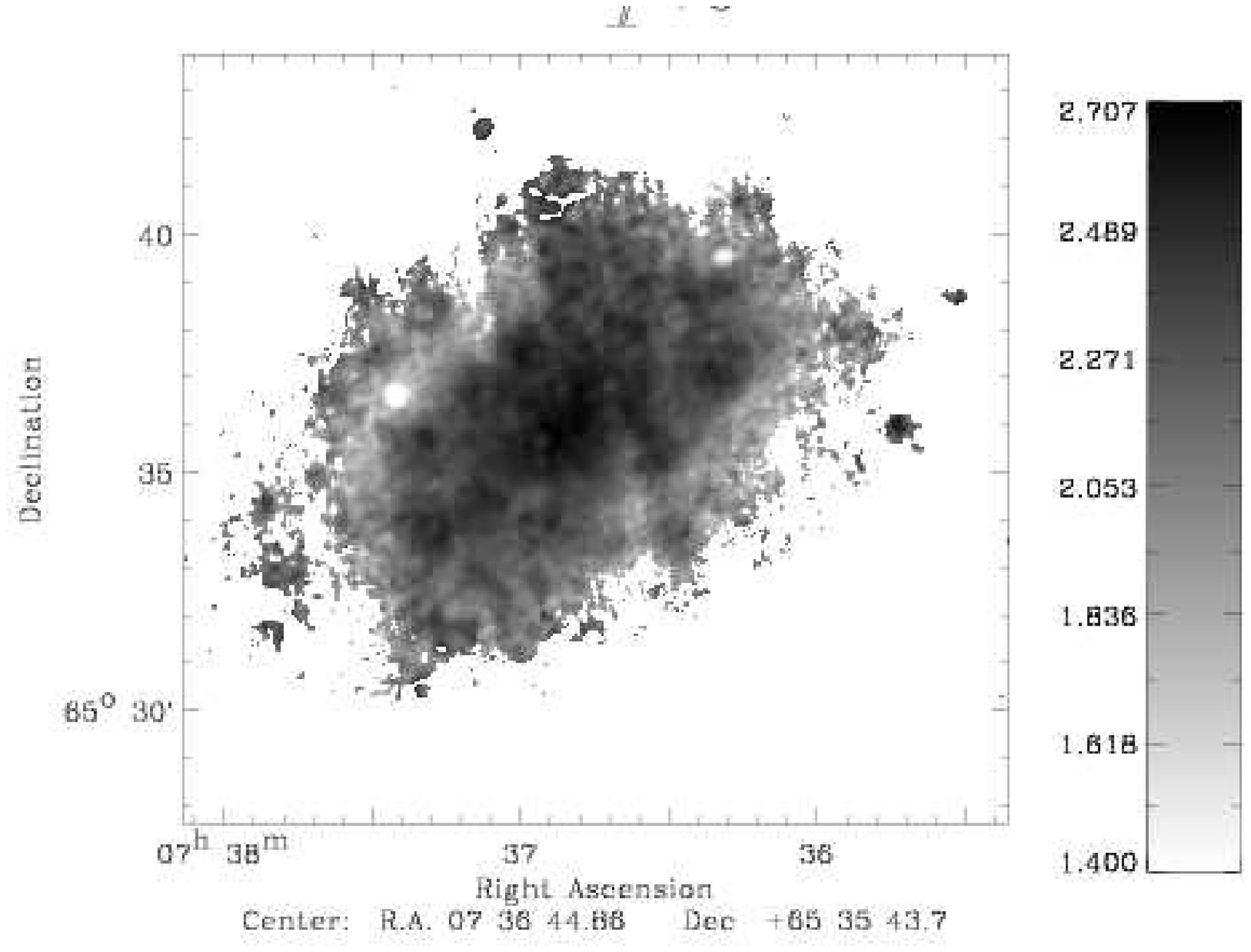}\hspace{-1.5cm}
      \plotone{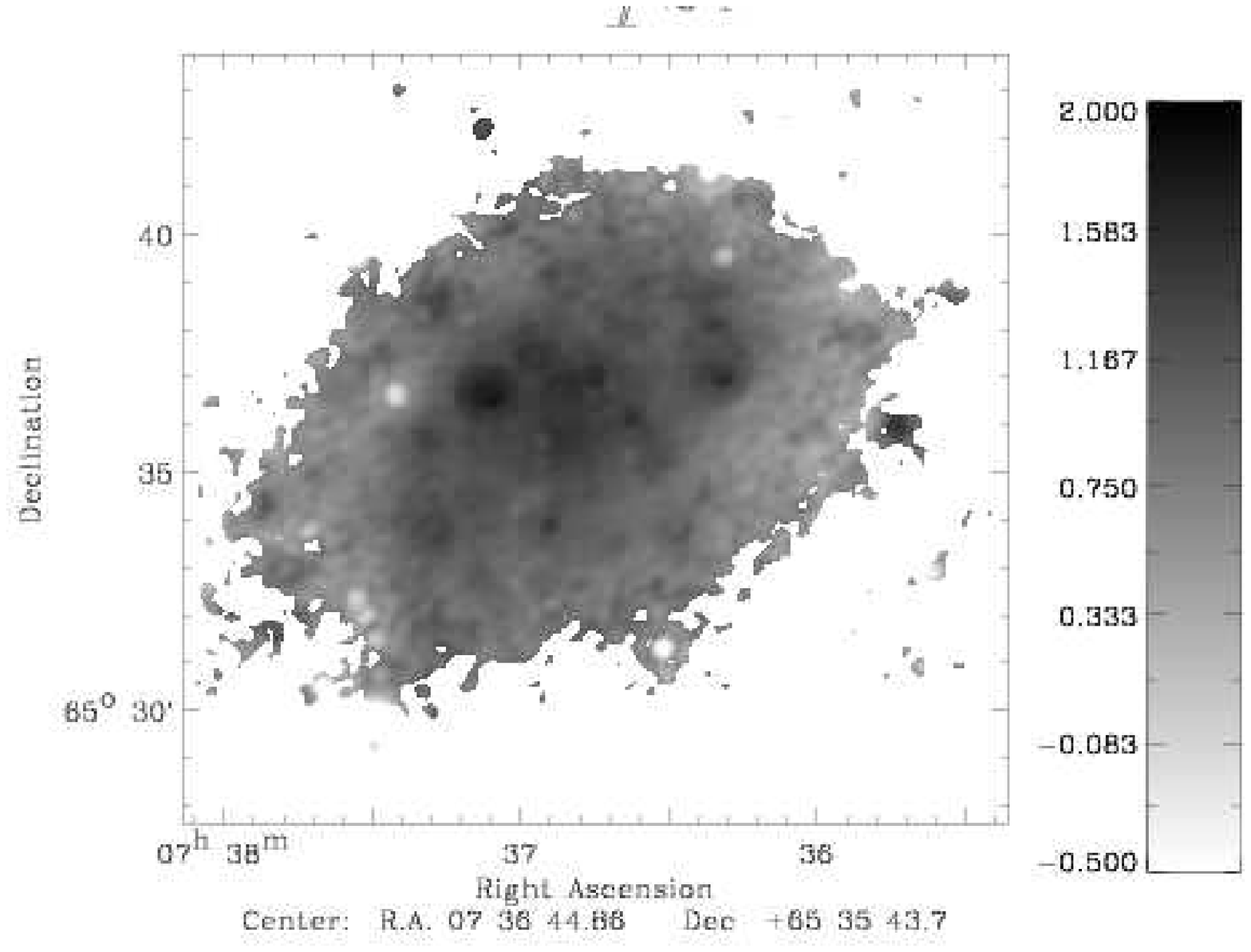}}}
  \hspace*{.1cm}
  \vspace*{-.8cm}
  \resizebox{3.75cm}{!}{
    \rotatebox{90}{
     \plotone{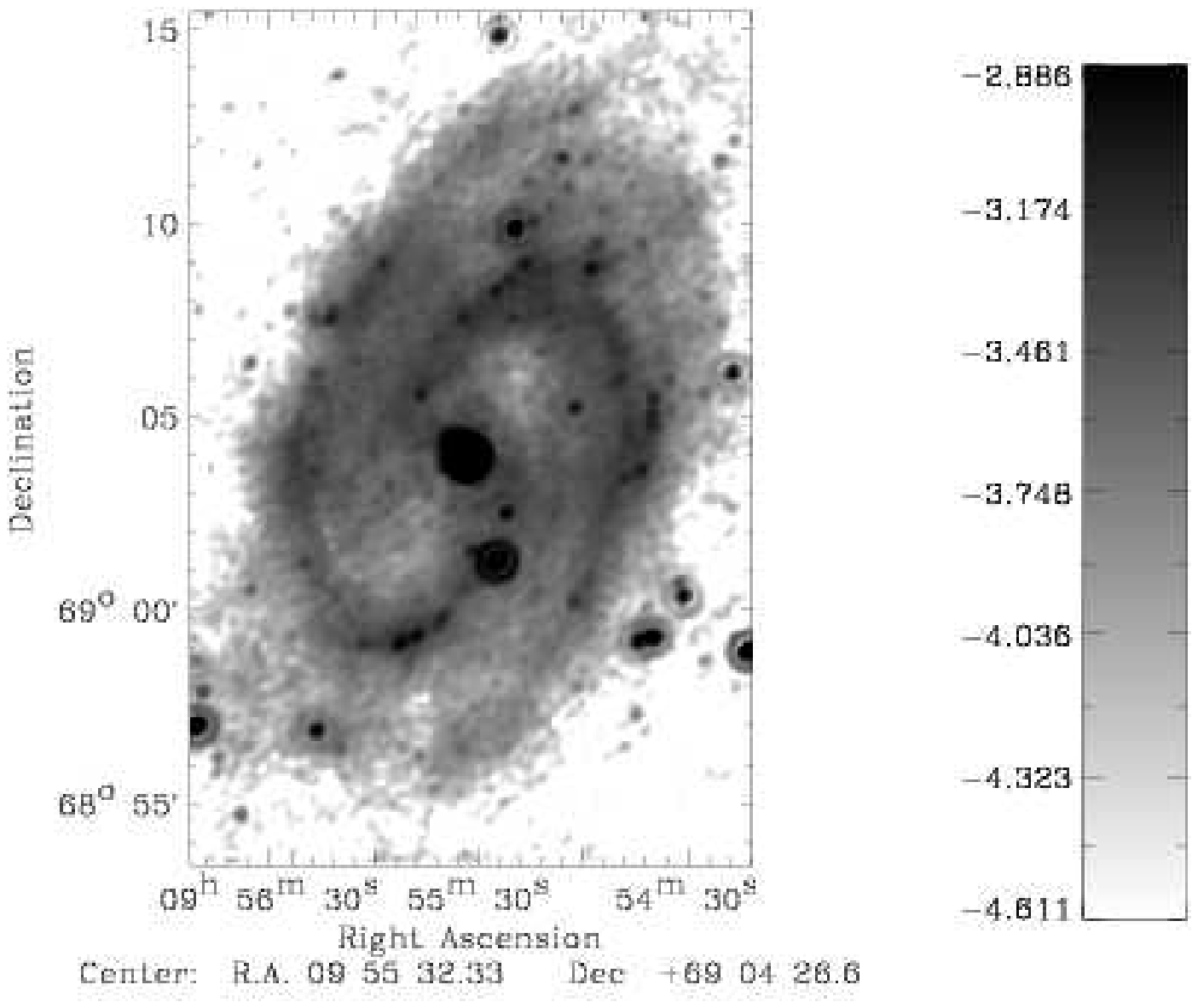}\hspace{-1.5cm}
     \plotone{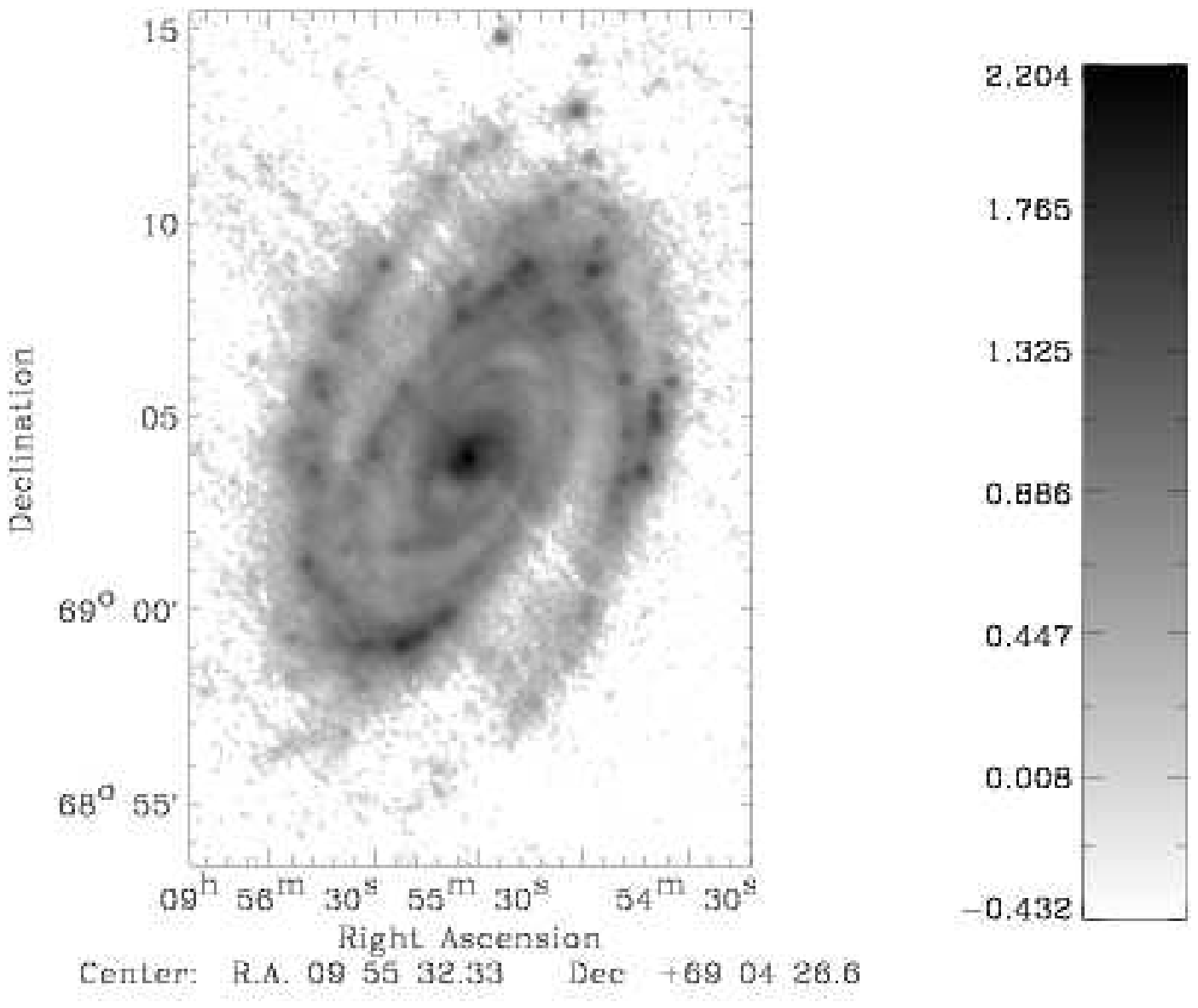}\hspace{-1.5cm}
     \plotone{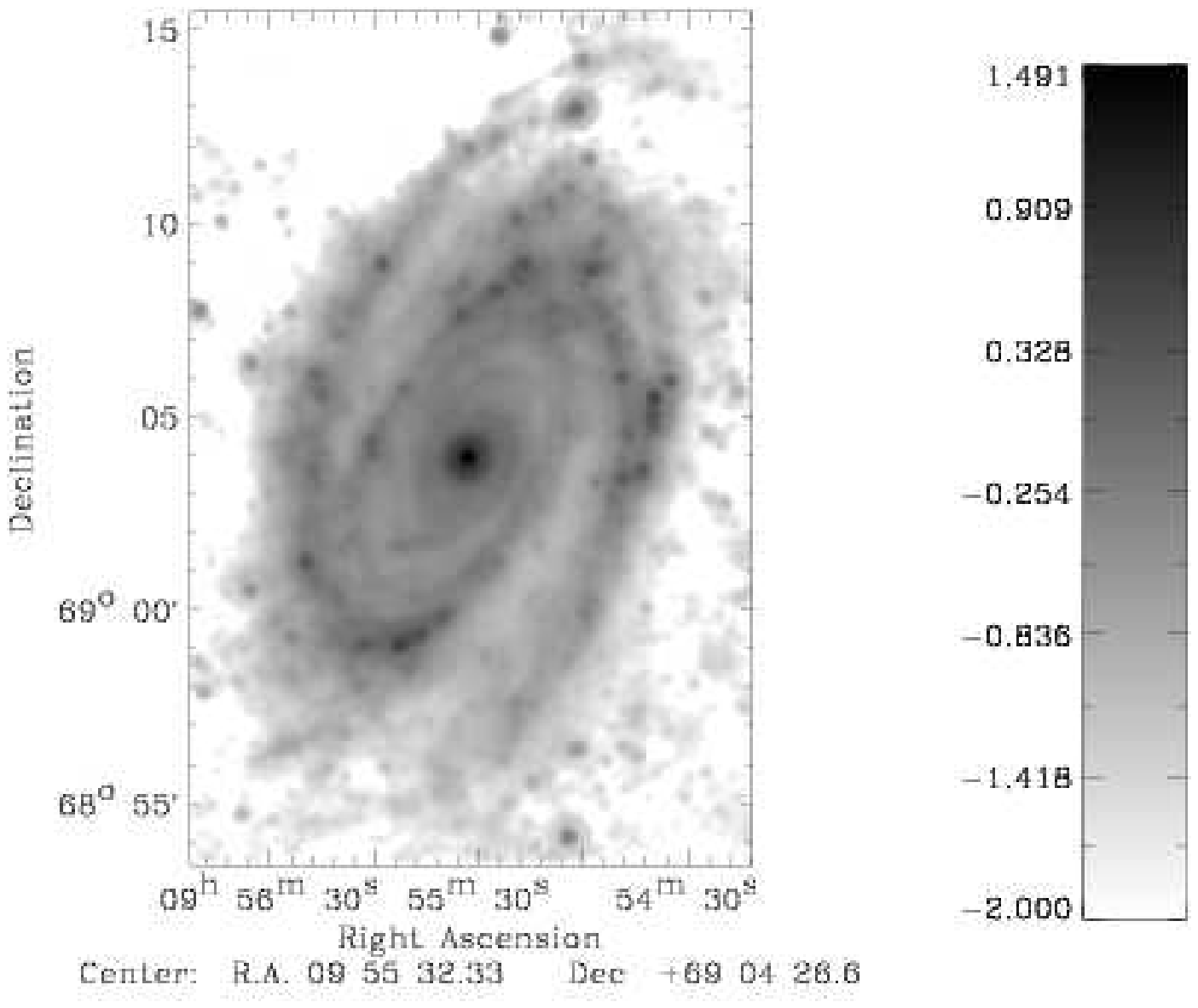}\hspace{-1.5cm}
     \plotone{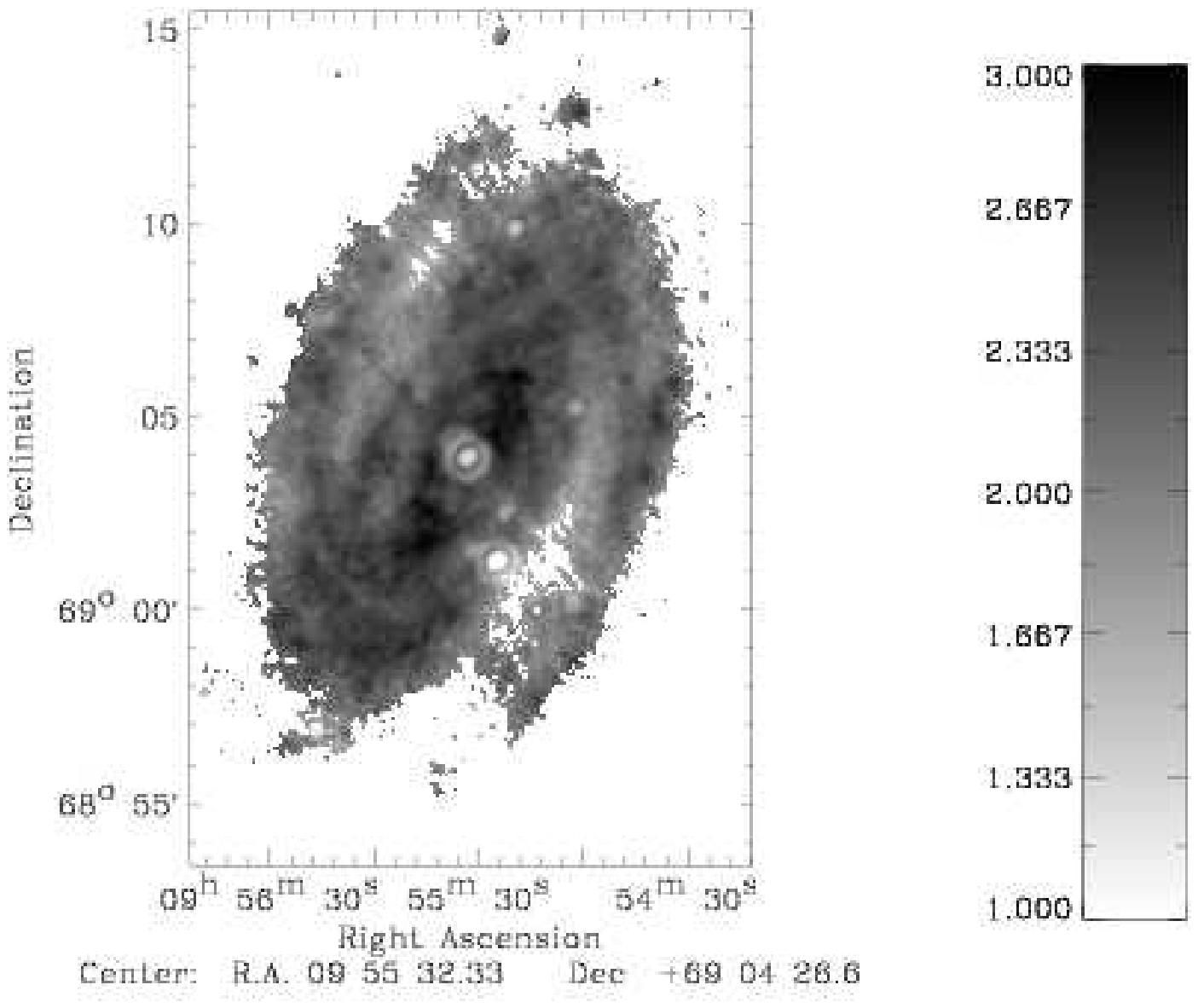}\hspace{-1.5cm}
     \plotone{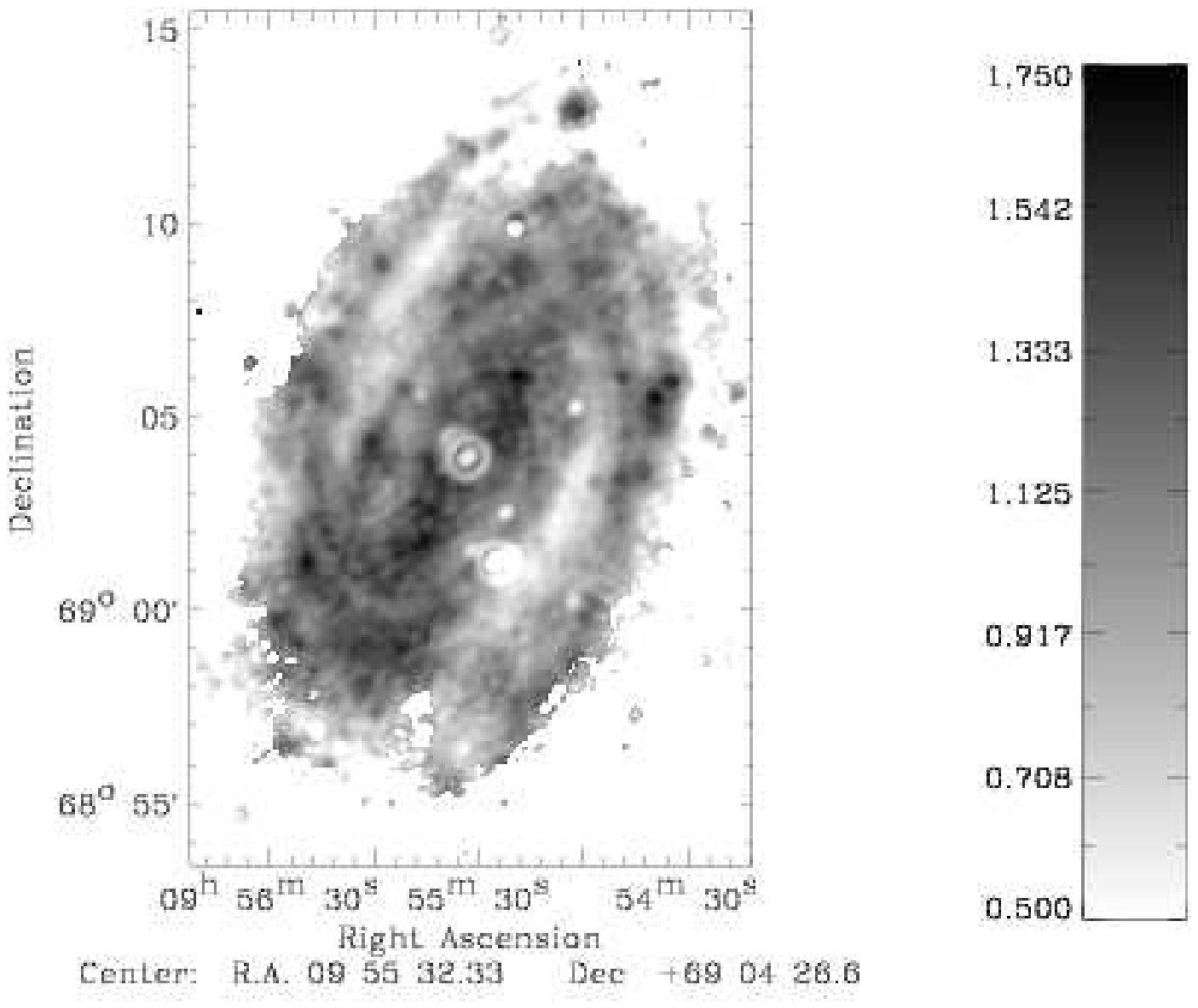}}}
 \hspace*{.1cm}
 \vspace*{-.8cm}
 \resizebox{3.75cm}{!}{
   \rotatebox{90}{
     \plotone{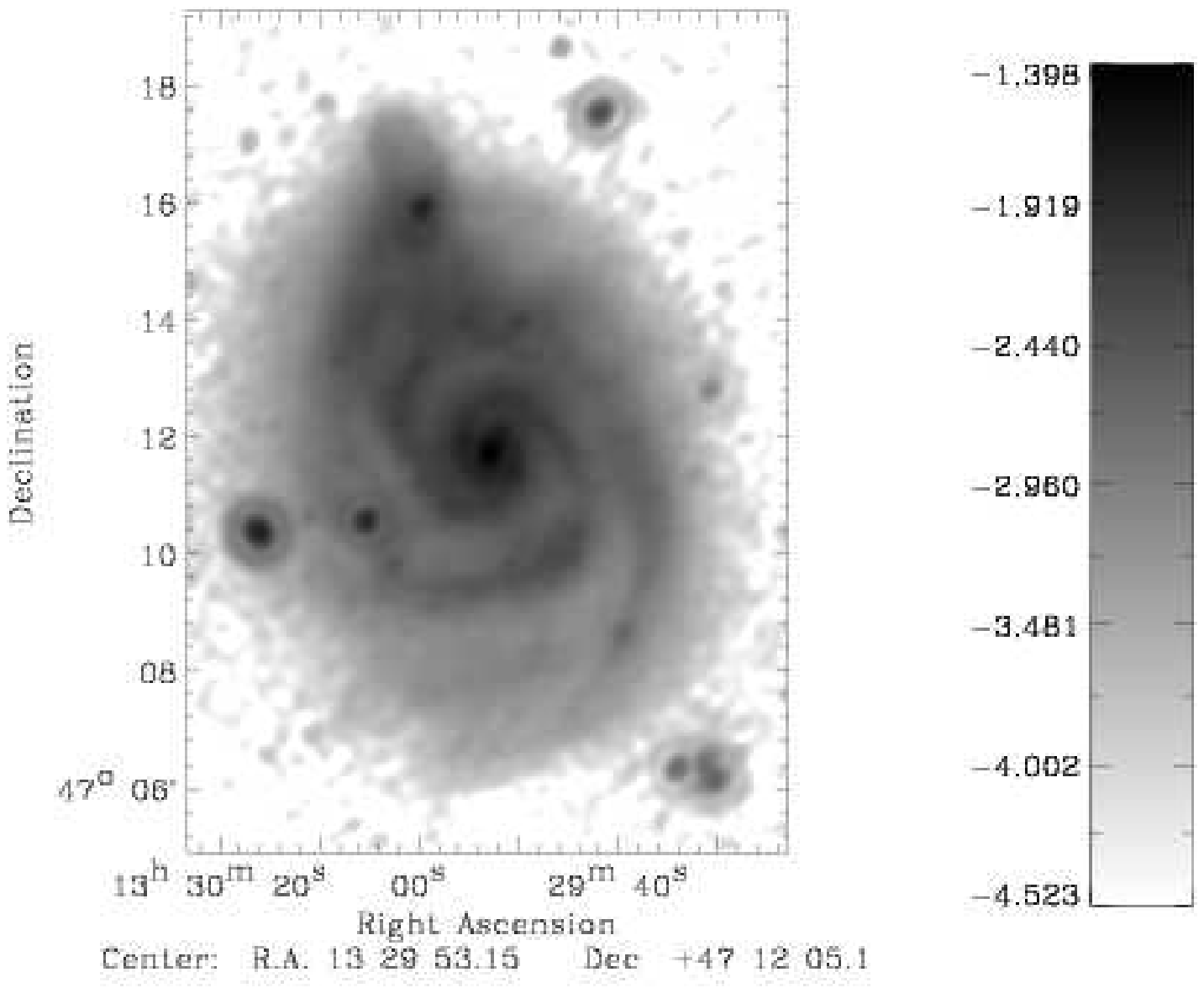} \hspace{-1.5cm}
     \plotone{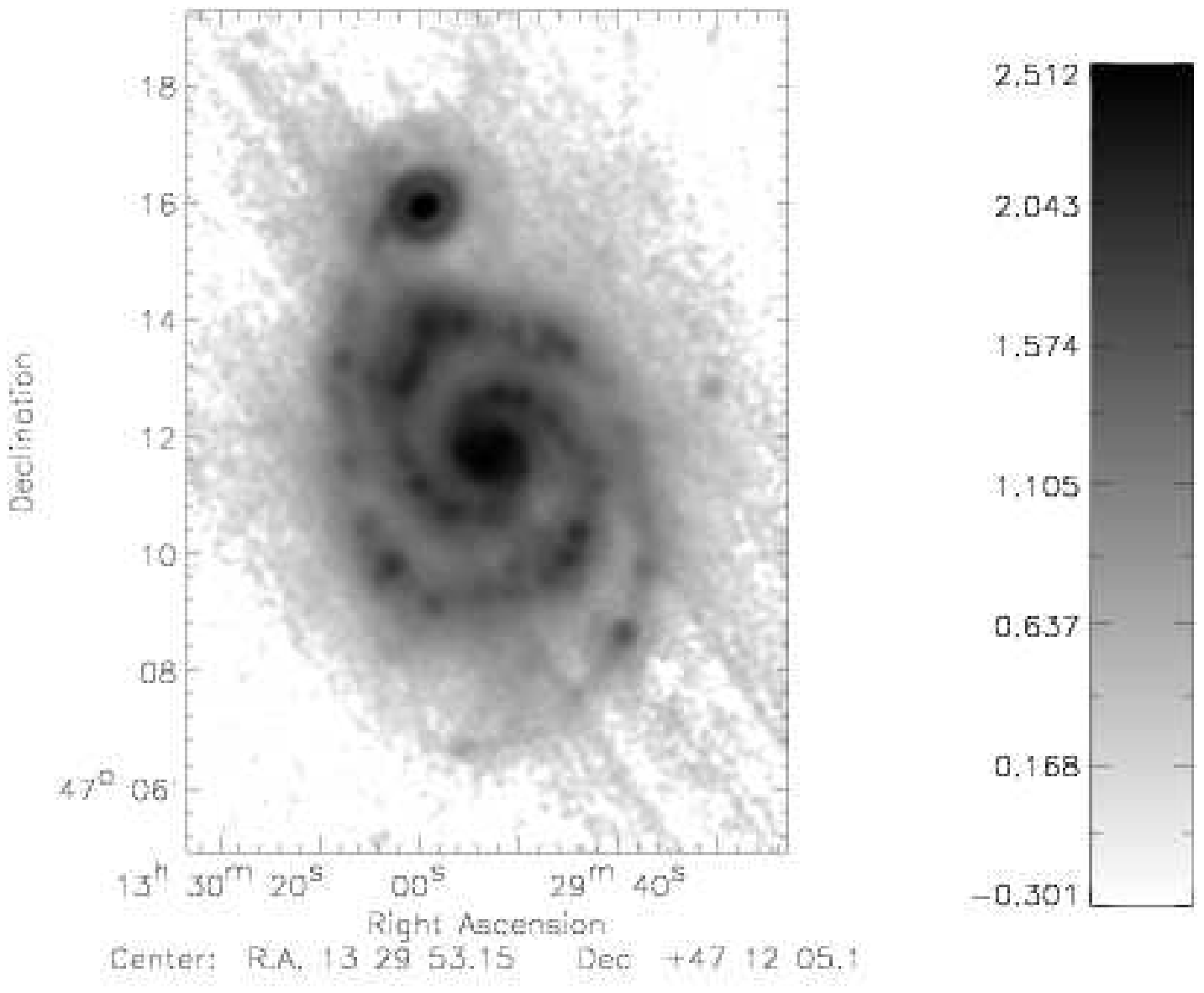} \hspace{-1.5cm}
     \plotone{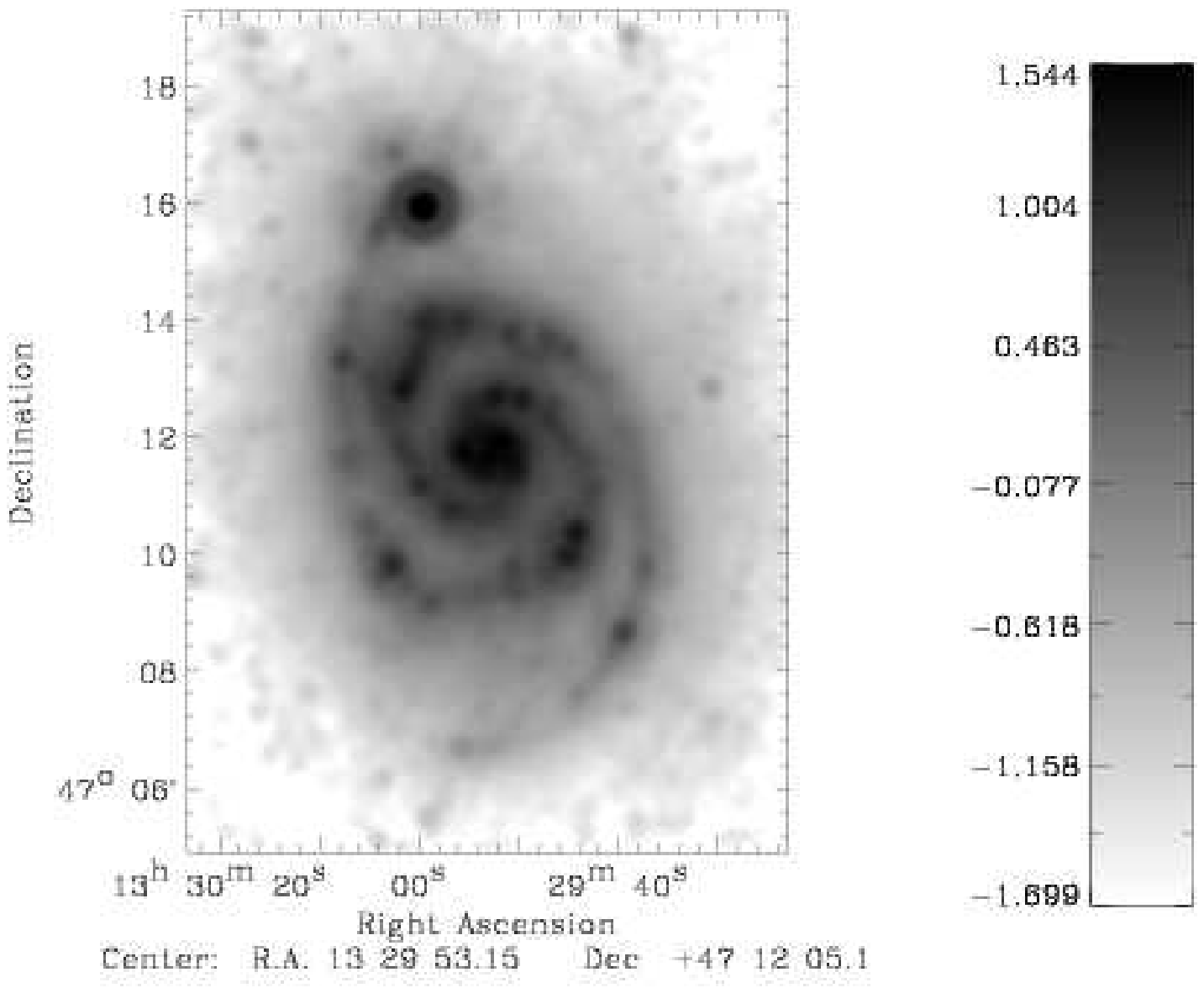} \hspace{-1.5cm}
     \plotone{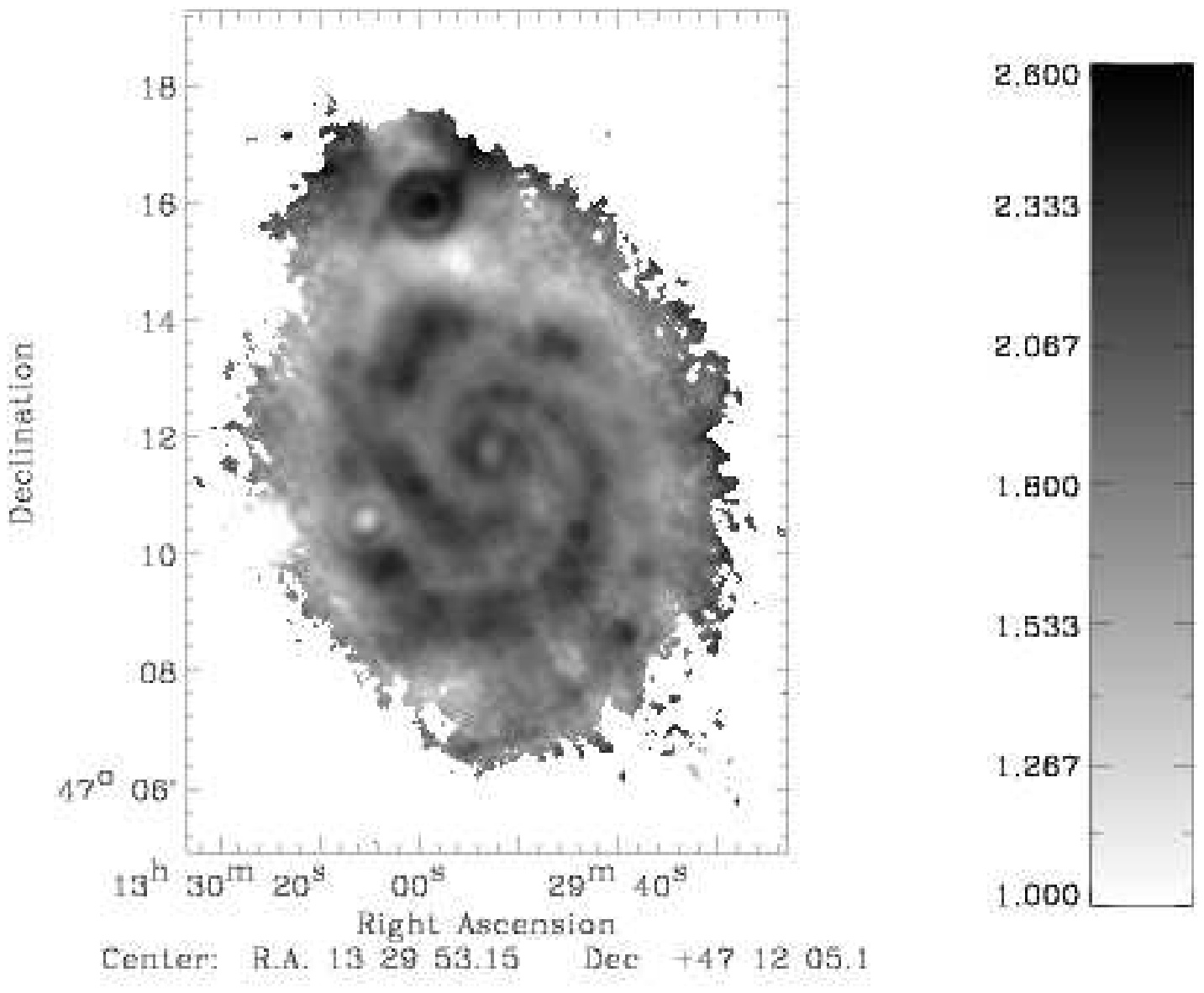}\hspace{-1.5cm}
     \plotone{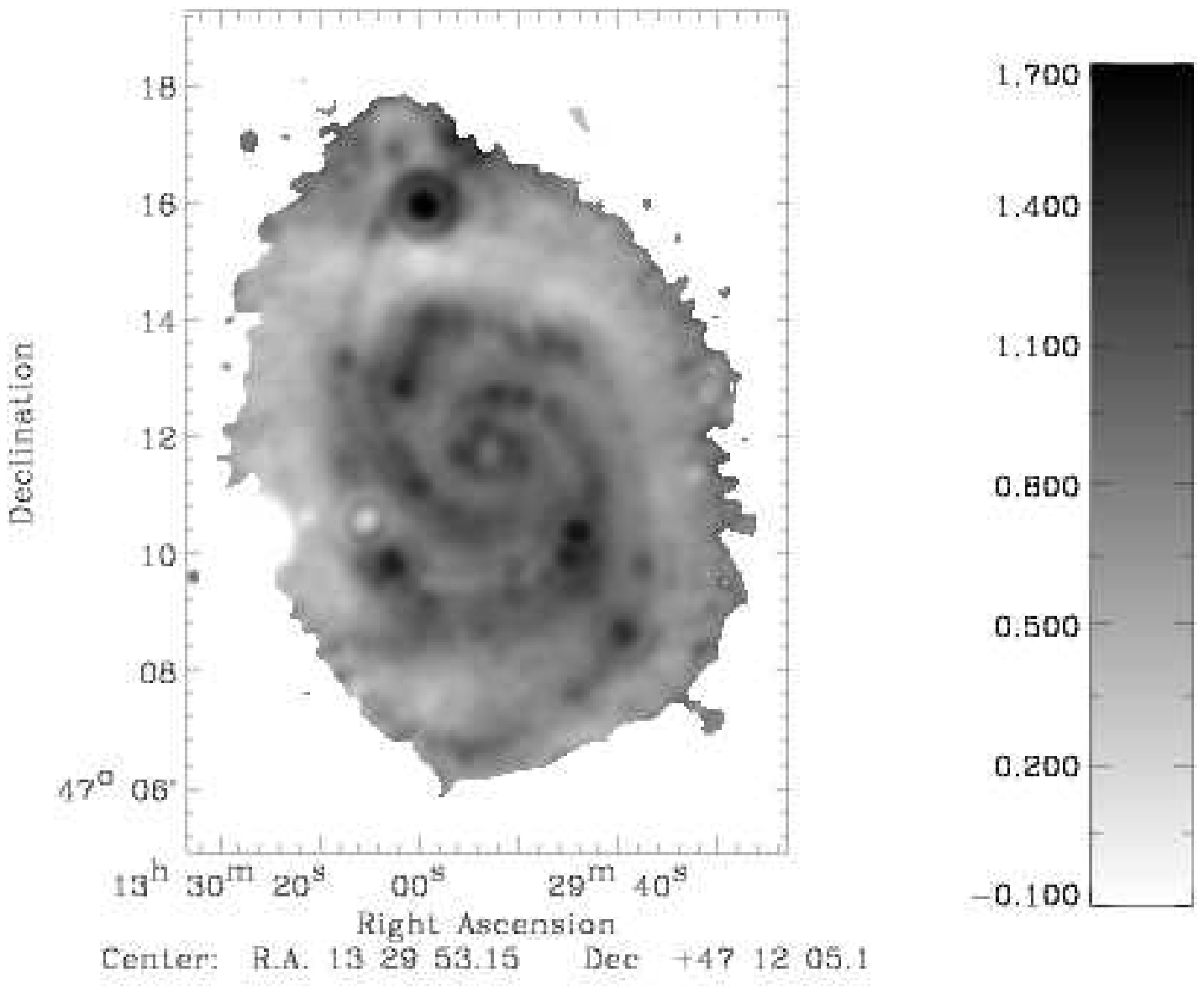}}}
 \hspace*{.1cm}
 \vspace*{-.8cm}
  \resizebox{3.75cm}{!}{
      \rotatebox{90}{
      \plotone{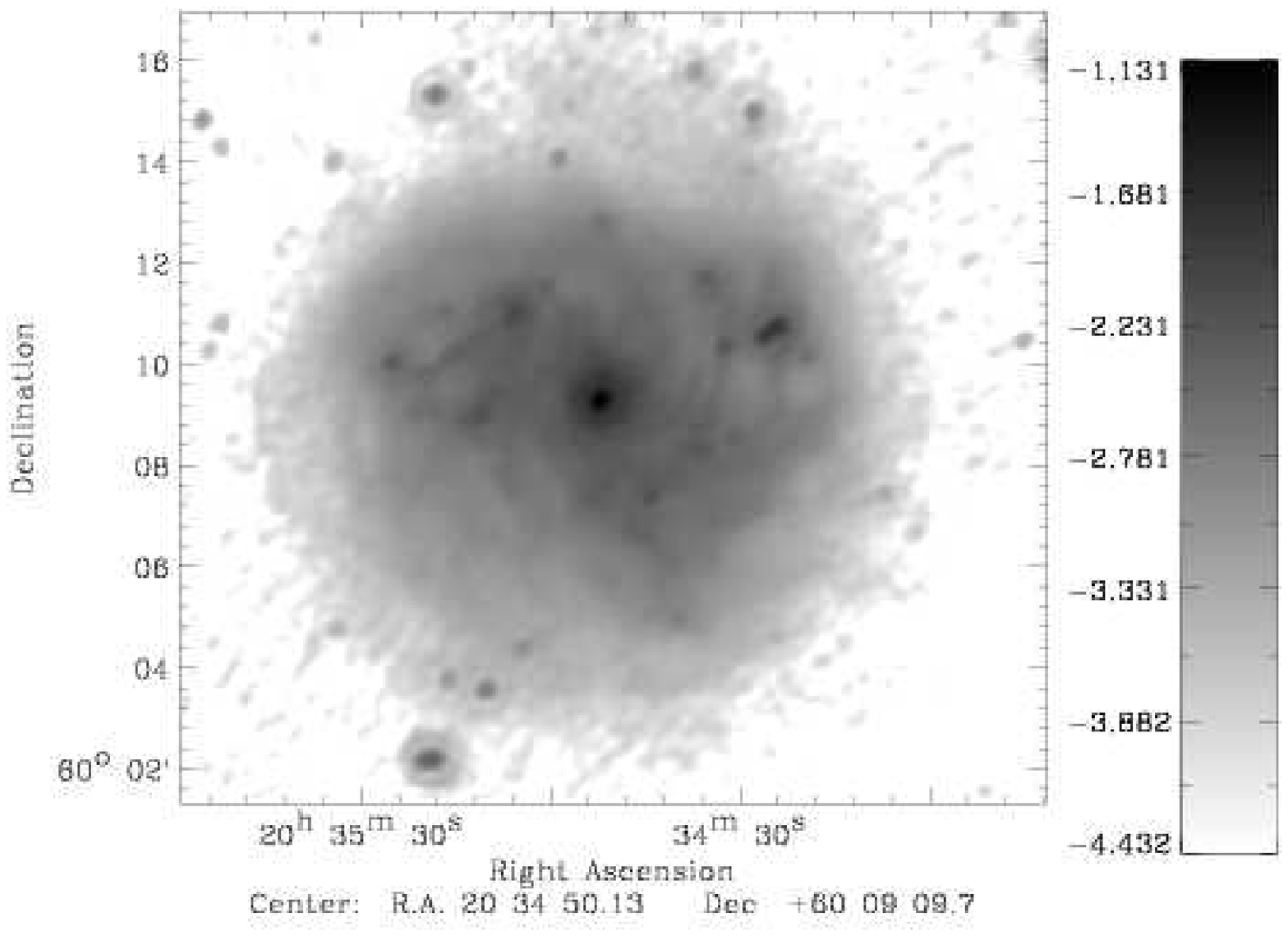}\hspace{-1.5cm}
      \plotone{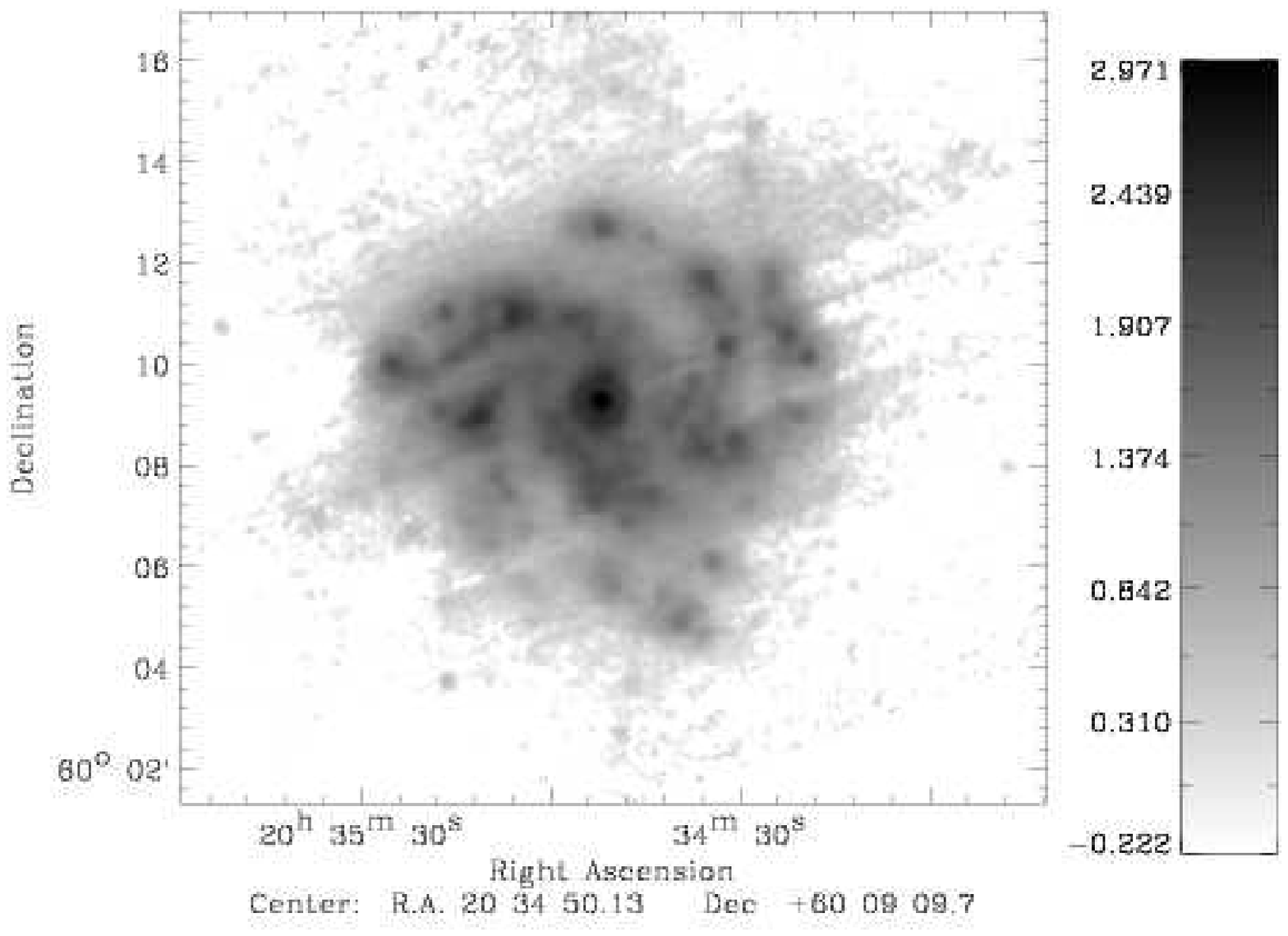}\hspace{-1.5cm}
      \plotone{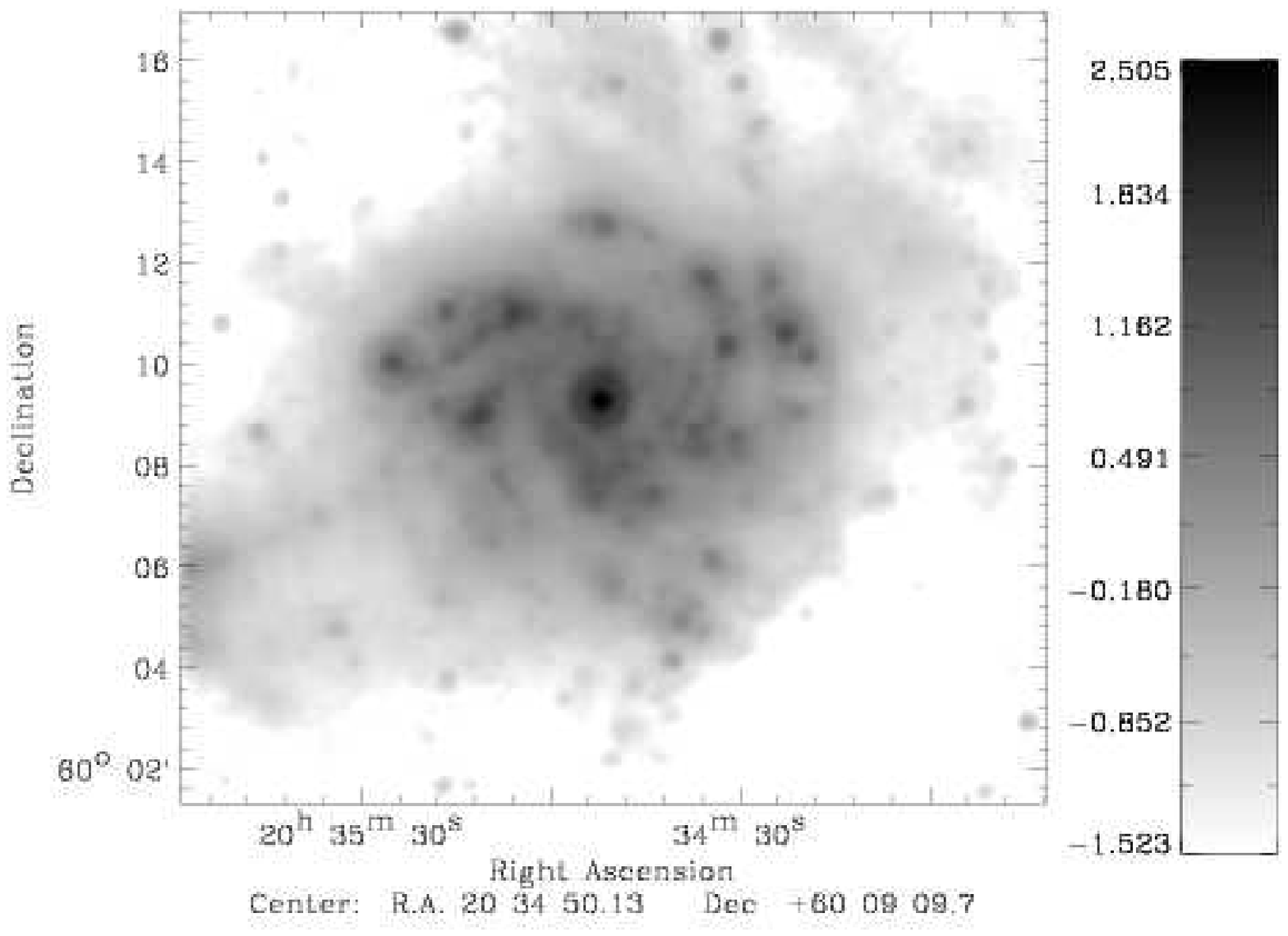}\hspace{-1.5cm}
      \plotone{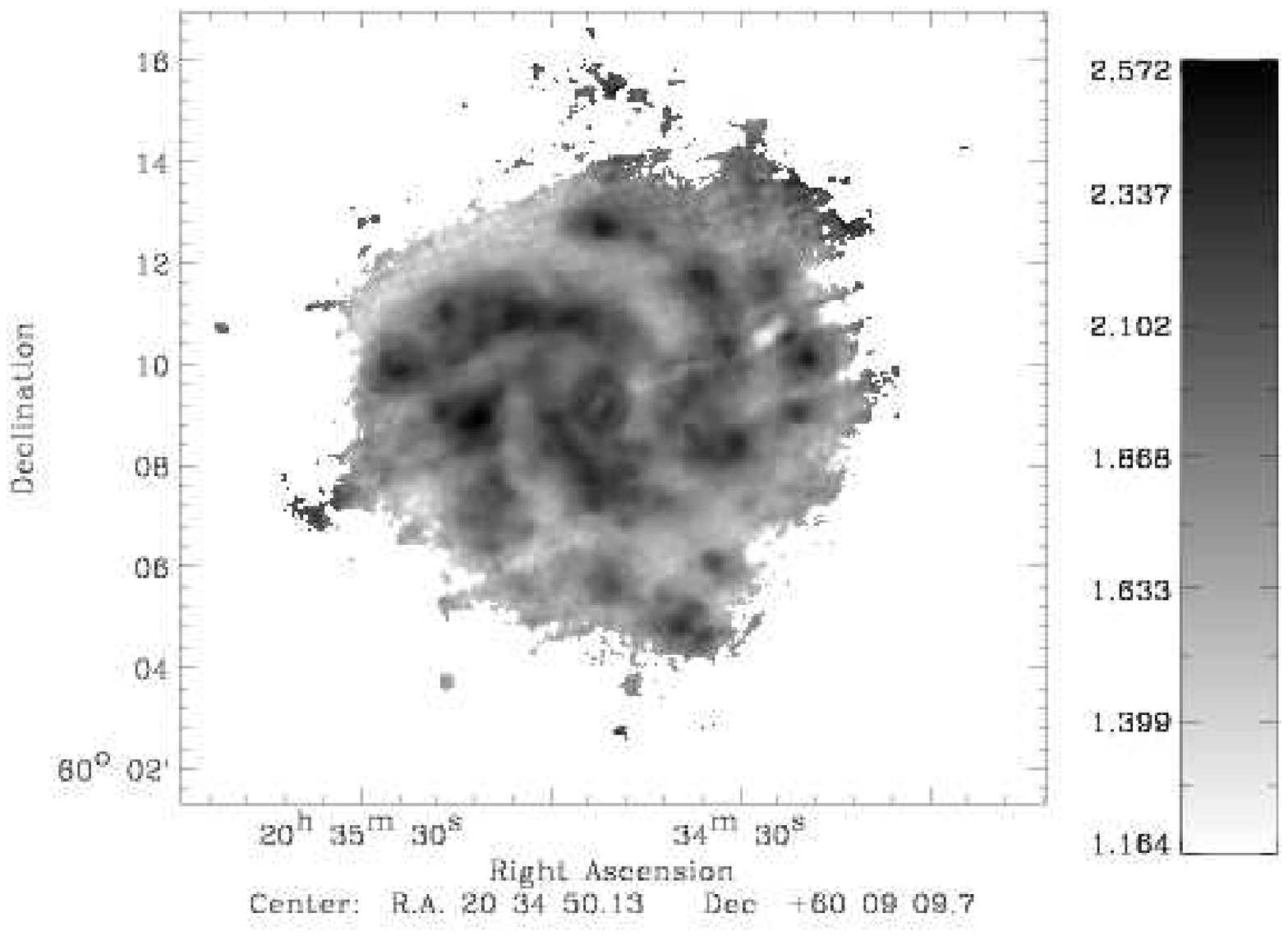}\hspace{-1.5cm}
      \plotone{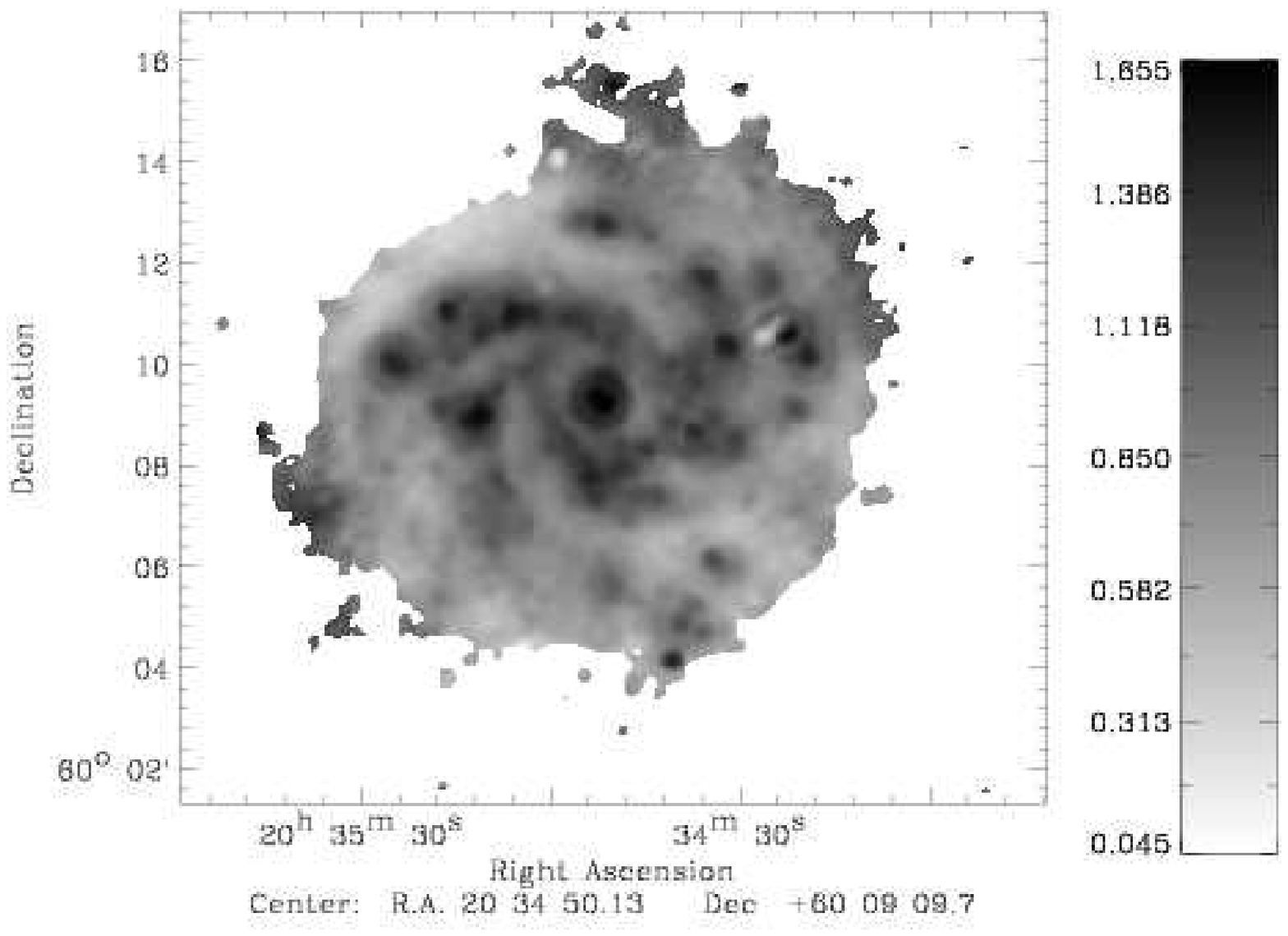}}}
  \vspace*{1.65cm}
  \caption{\label{maps}}
  \vspace*{-1cm}
  \end{center}
\end{figure}

\clearpage
\setcounter{figure}{0}
Caption for Figure \ref{maps} to be placed horizontally below the rows
of maps (along the long edge) of the page.

\begin{figure}[!ht]
\caption{From left to right for each galaxy: 22~cm radio map (except
  for case of NGC~3031 in which a 20~cm radio map is plotted);
  70~$\micron$ map; 24~$\micron$ map (matched to the 70~$\micron$
  resolution); $q_{70}$ map for pixels having a 3~$\sigma$ detections
  in both the input radio and 70~$\micron$ maps; $q_{24}$ map for
  pixels having a 3~$\sigma$ detections in both the input radio and
  24~$\micron$ maps.  The units of the radio maps are in $\log({\rm
  Jy/beam})$ and the infrared maps are in units of $\log({\rm
  MJy/sr})$.  All maps are displayed with a stretch ranging from the
  RMS background level to the maximum surface brightness in
  the galaxy disk. In the radio map of NGC~3031, estimation of the
  maximum surface brightness excluded its AGN nucleus and SN~1993J,
  located in the arm South of the nucleus.  
  Note that H~II regions and spiral arms are visible in the
  $q_{\lambda}$ maps, indicating that they have an excess of infrared
  emission relative to radio continuum emission} 
\end{figure}

\clearpage
\begin{figure}[!ht]
  \resizebox{18cm}{!}{
    {\plotone{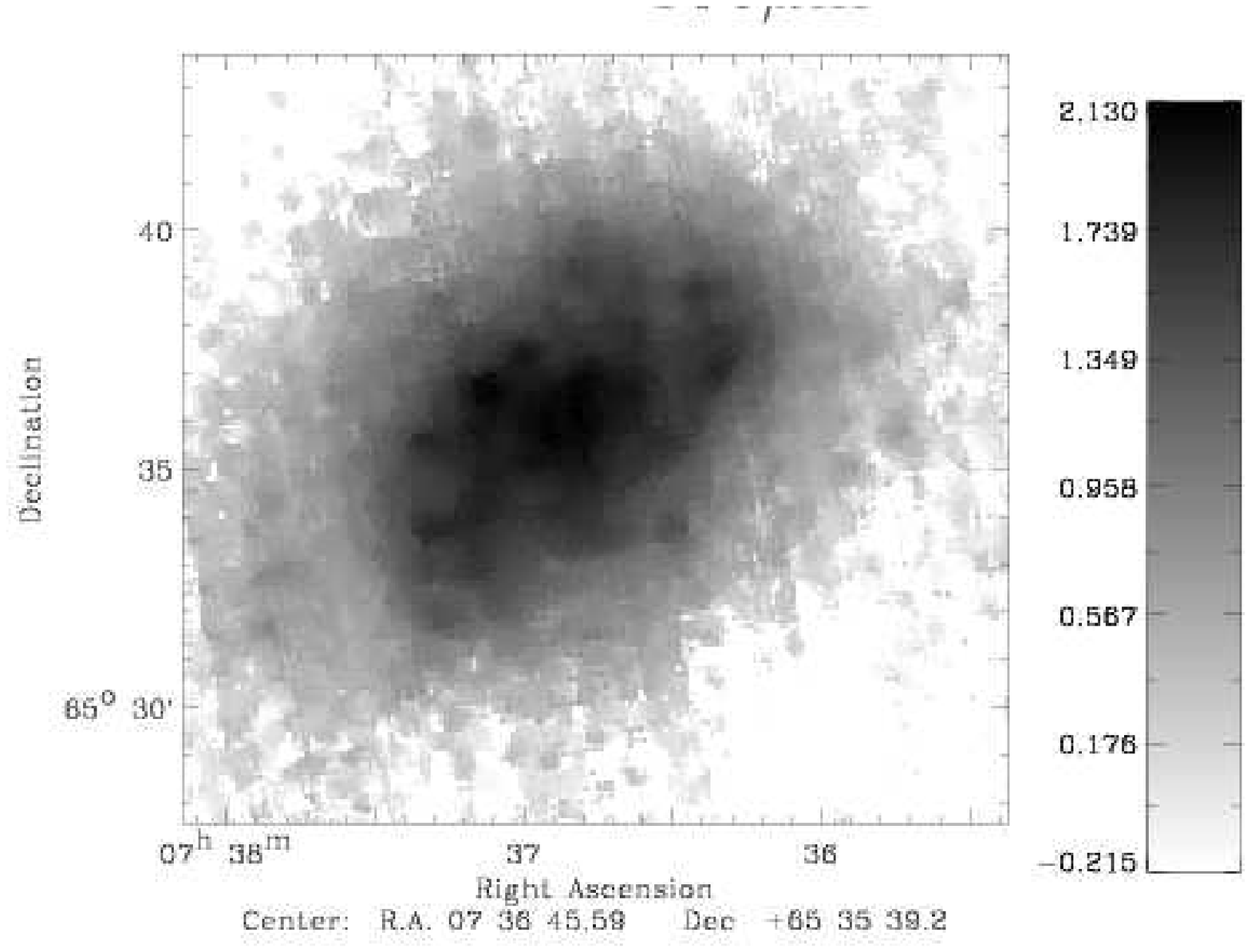}
      \plotone{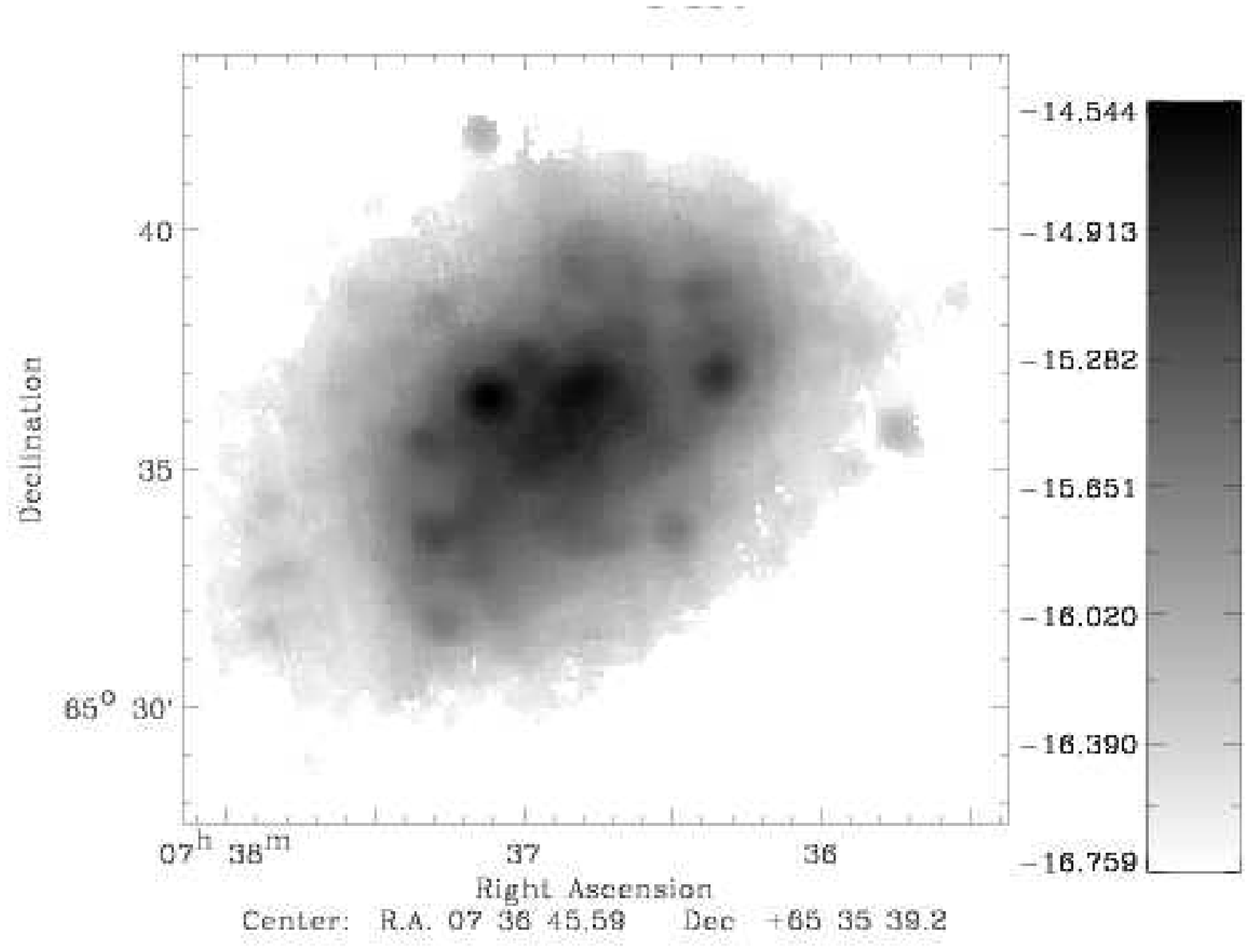}
      \plotone{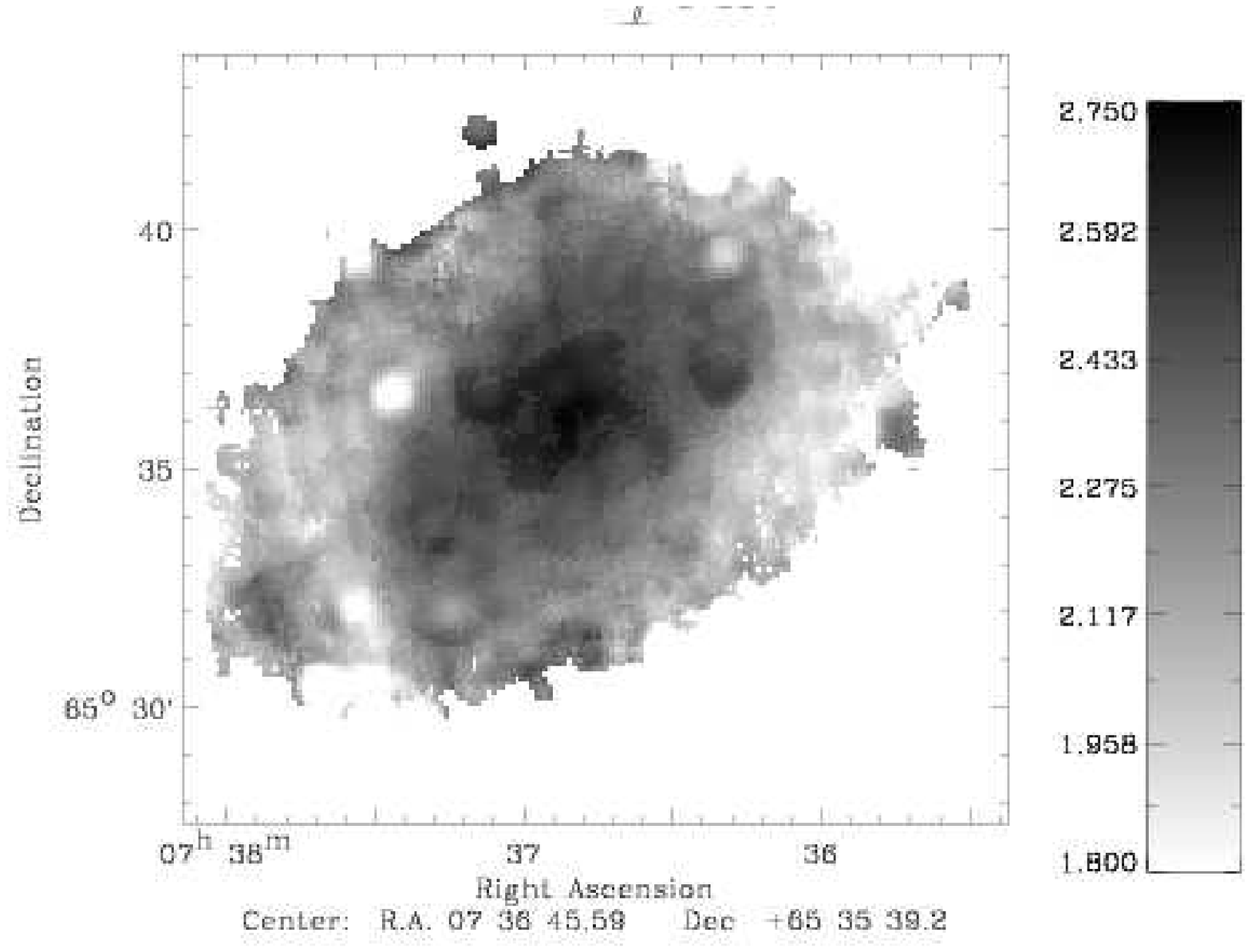}}}
  \vskip 0.75cm 
  \resizebox{18cm}{!}{
    {\plotone{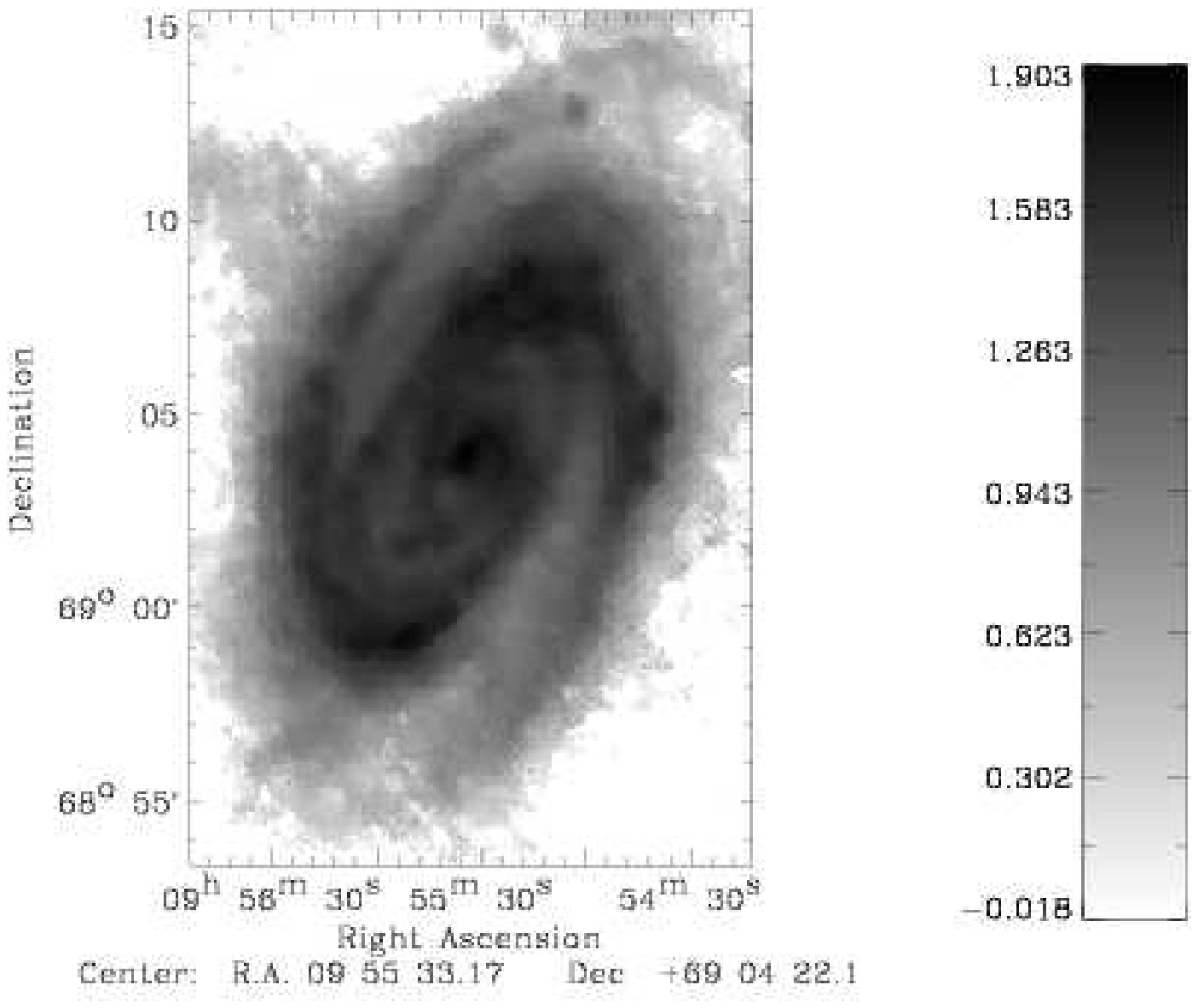}
      \plotone{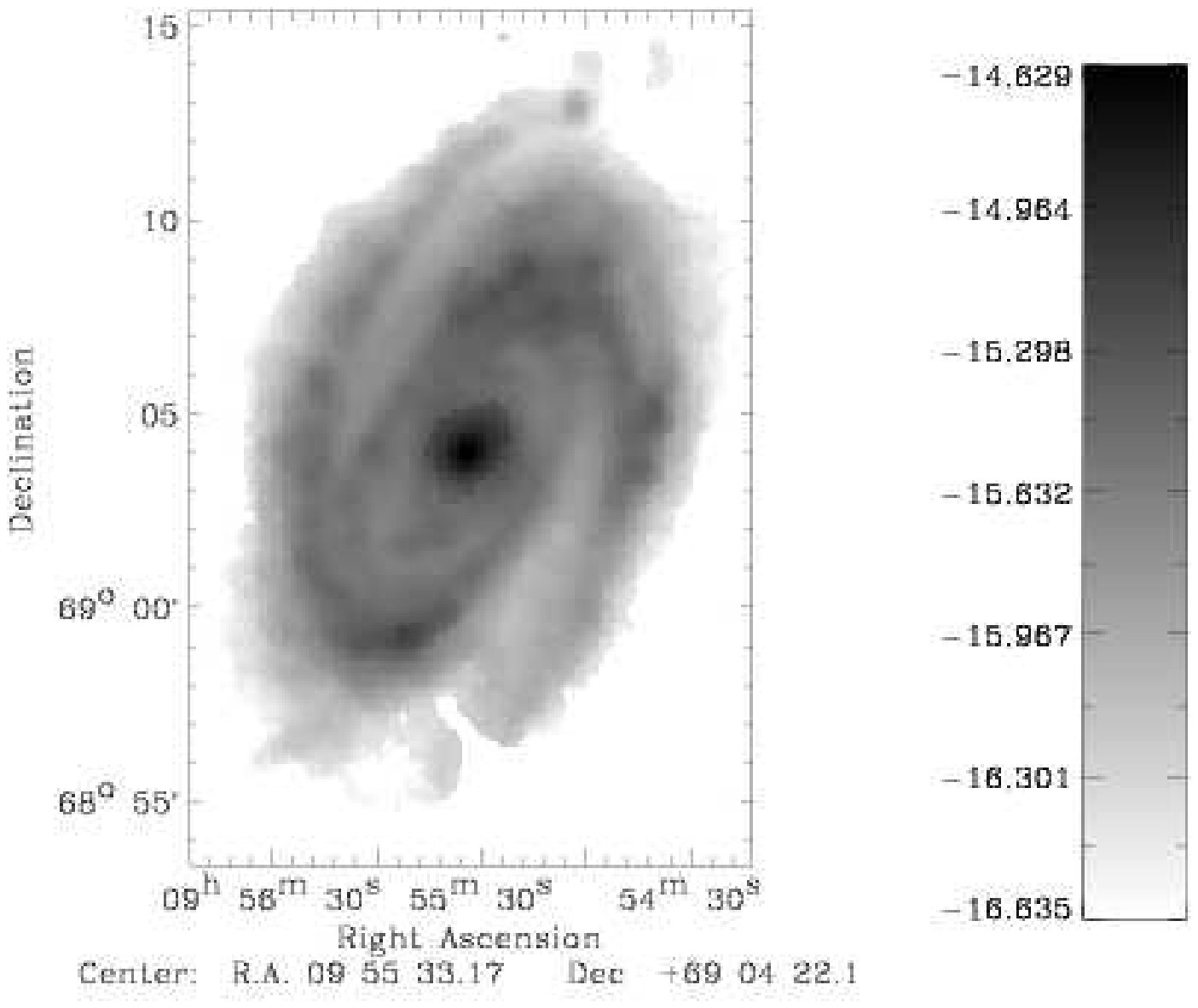}
      \plotone{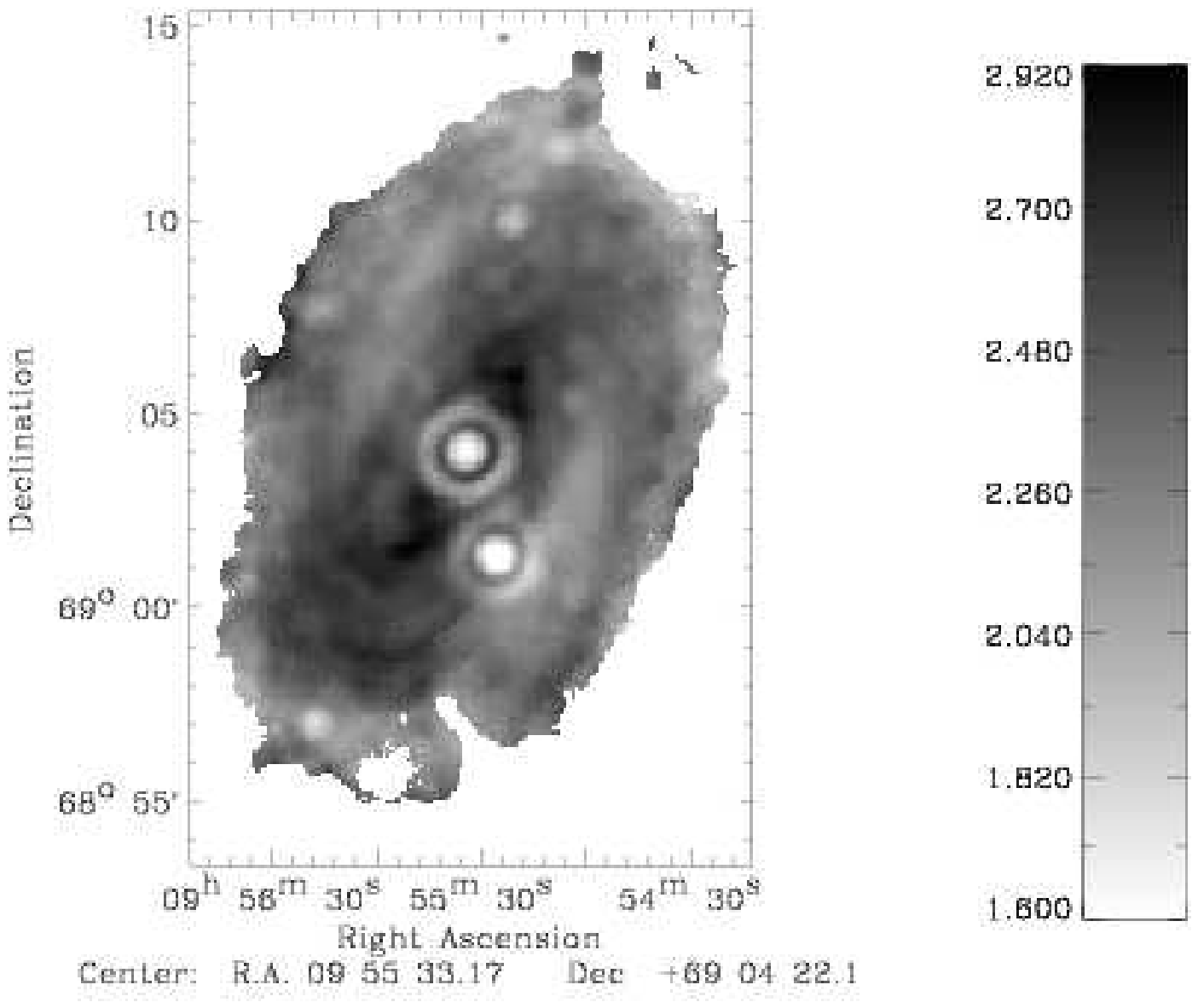}}}
  \vskip 0.75cm 
  \resizebox{18cm}{!}{
    {\plotone{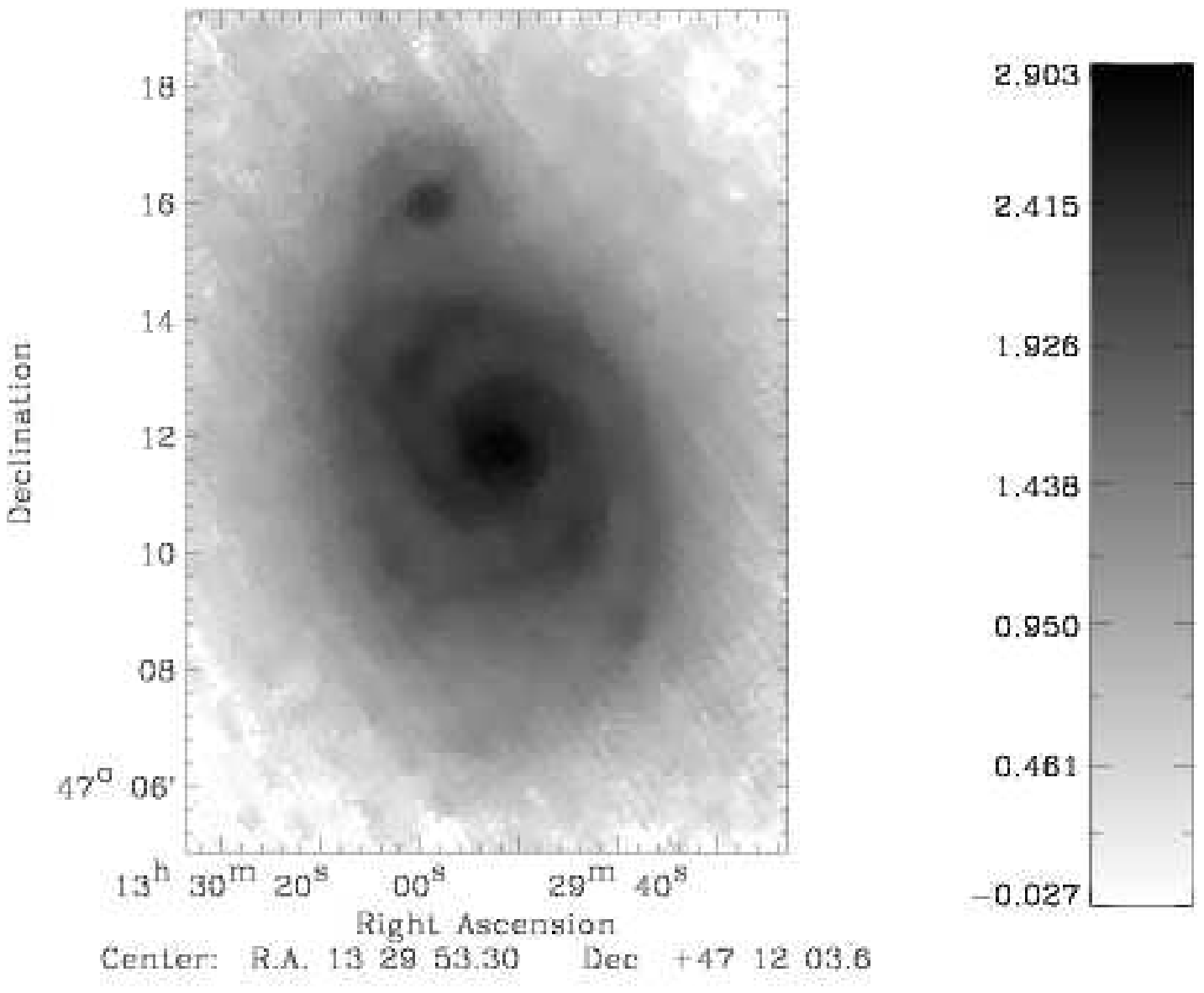}
      \plotone{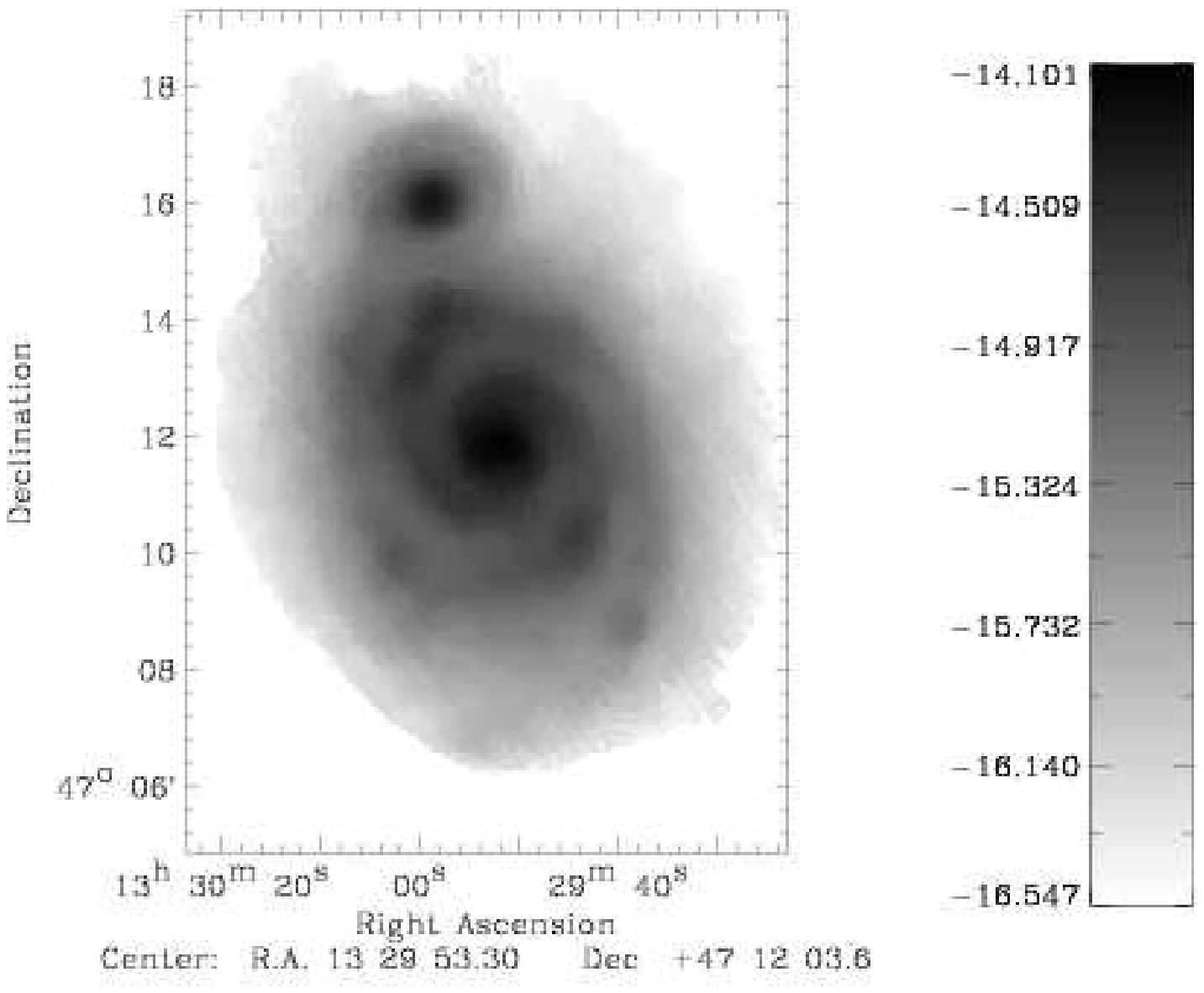}
      \plotone{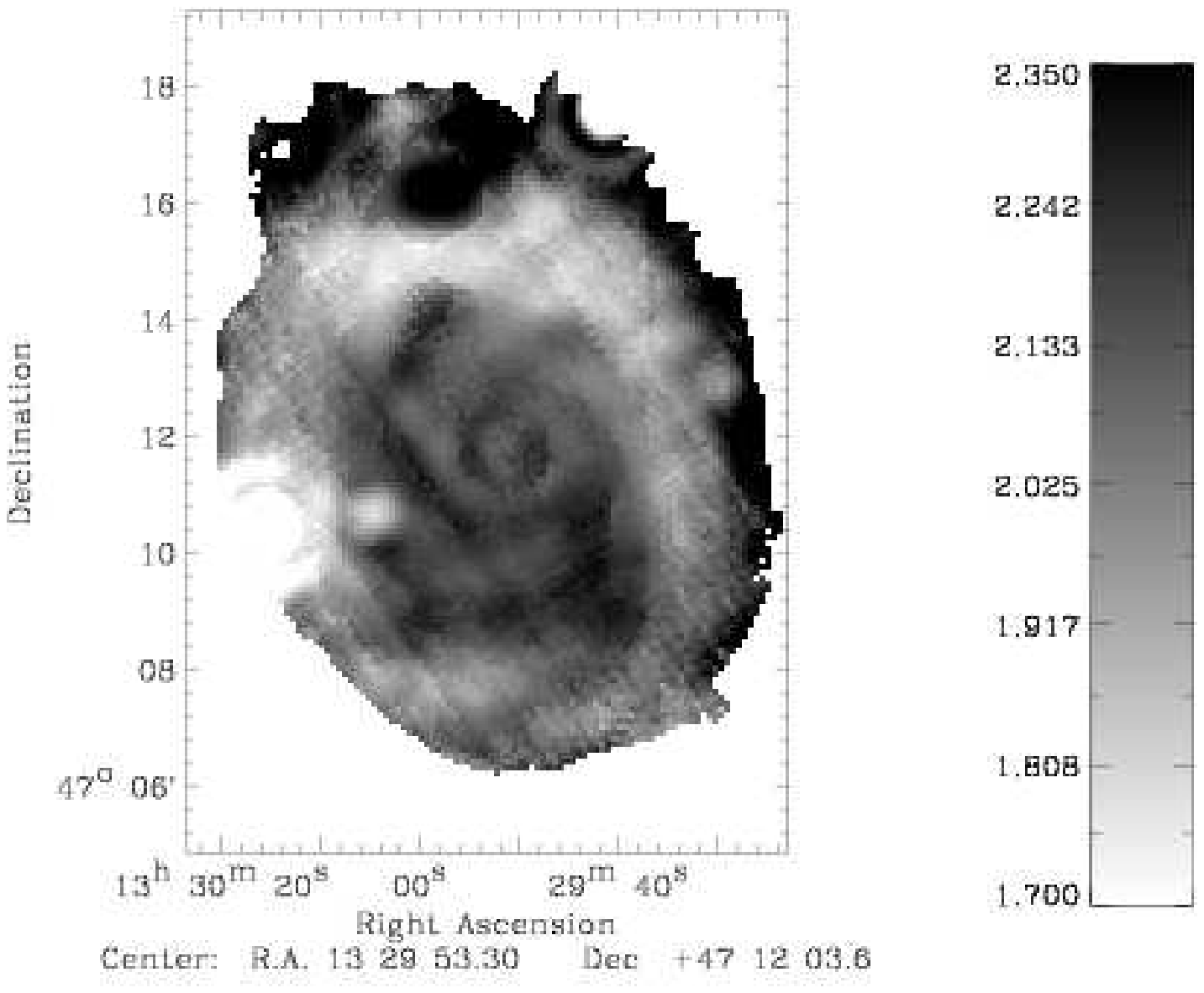}}}
  \vskip 0.75cm 
  \resizebox{18cm}{!}{
    {\plotone{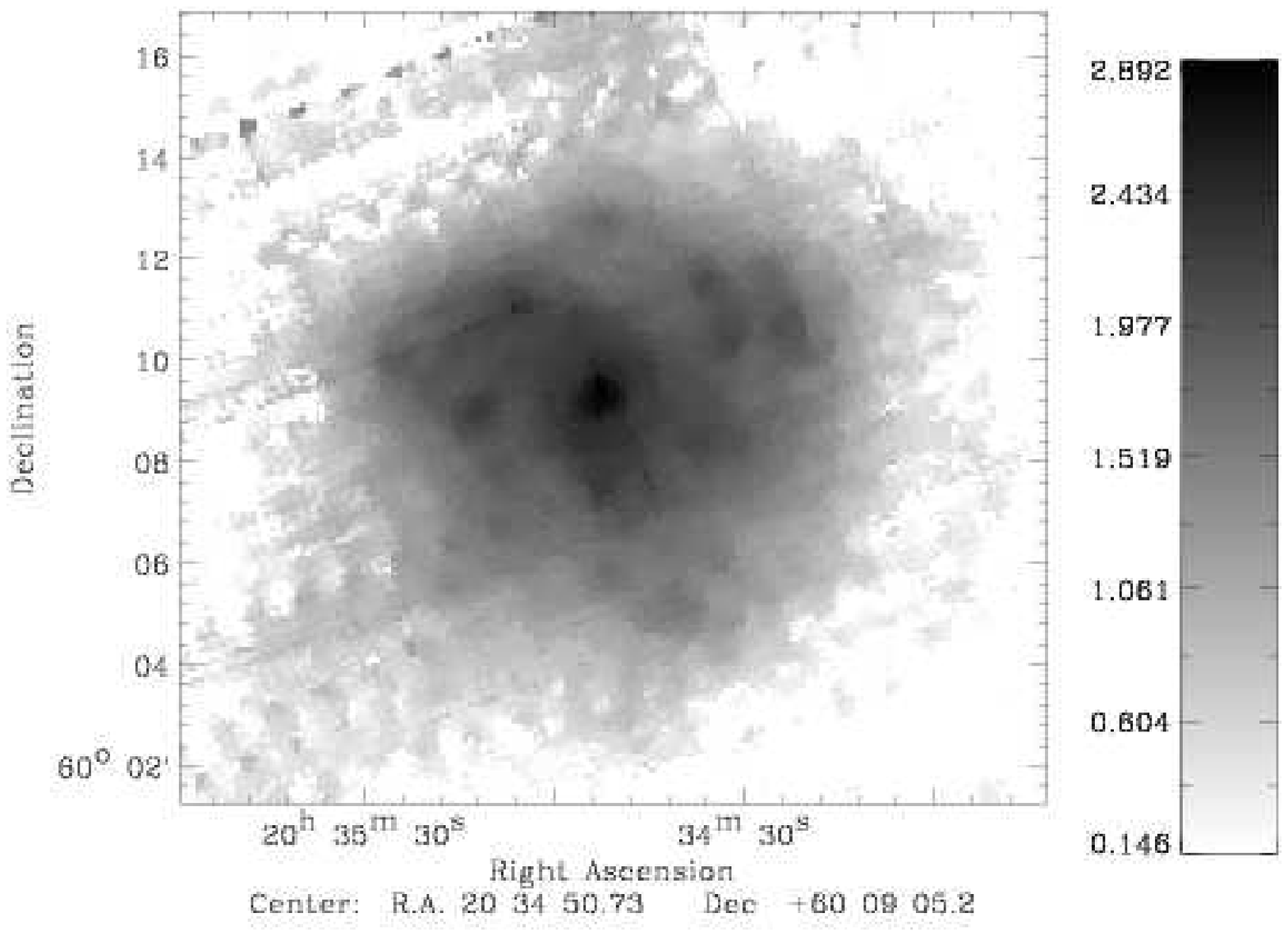}
      \plotone{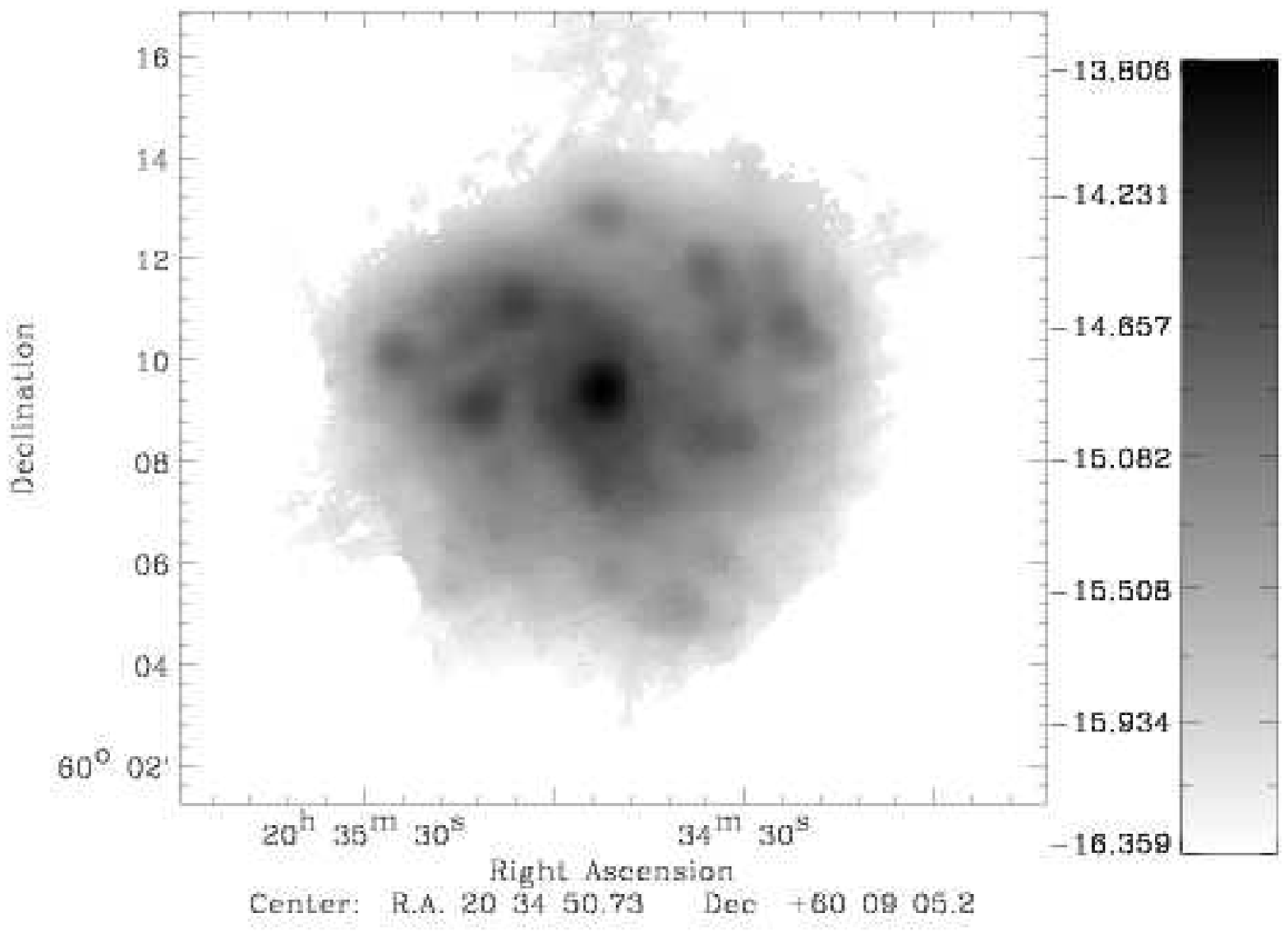}
      \plotone{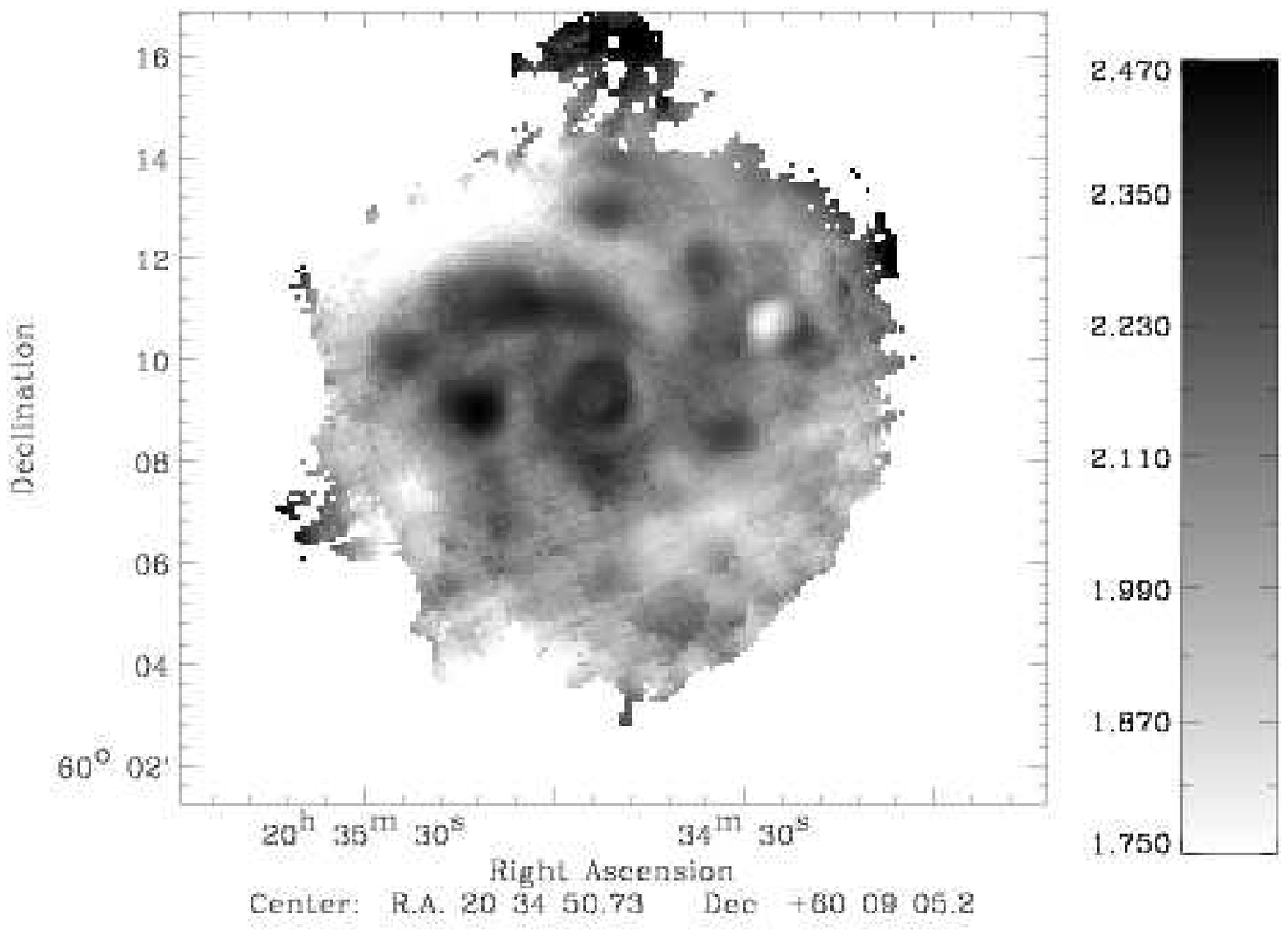}}}
  \vskip .5cm
  \caption{From left to right for each galaxy: 160~$\micron$ map; 
  far infrared (FIR) map calculated using a linear combination of the
  3 MIPS bands according to the SED models of \citet{dd02} (see
  $\S$2.3); $q_{\rm FIR}$ map for pixels having a 3~$\sigma$
  detections in all 3 MIPS bands and the input radio map.
  The 160~$\micron$ maps are in units of $\log({\rm
  MJy/sr})$ while the FIR maps are given in $\log({\rm W/m^{2}})$.  
  The 160~$\micron$ and FIR maps are displayed with a stretch ranging from
  the 1~$\sigma$ and 3~$\sigma$ background level to the maximum
  surface brightness in the galaxy disk, respectively. \label{160maps}}
\end{figure}

\clearpage
\begin{figure}[!ht]
  \begin{center}
    \resizebox{10cm}{!}{
      {\resizebox{10cm}{!}{
	  \plotone{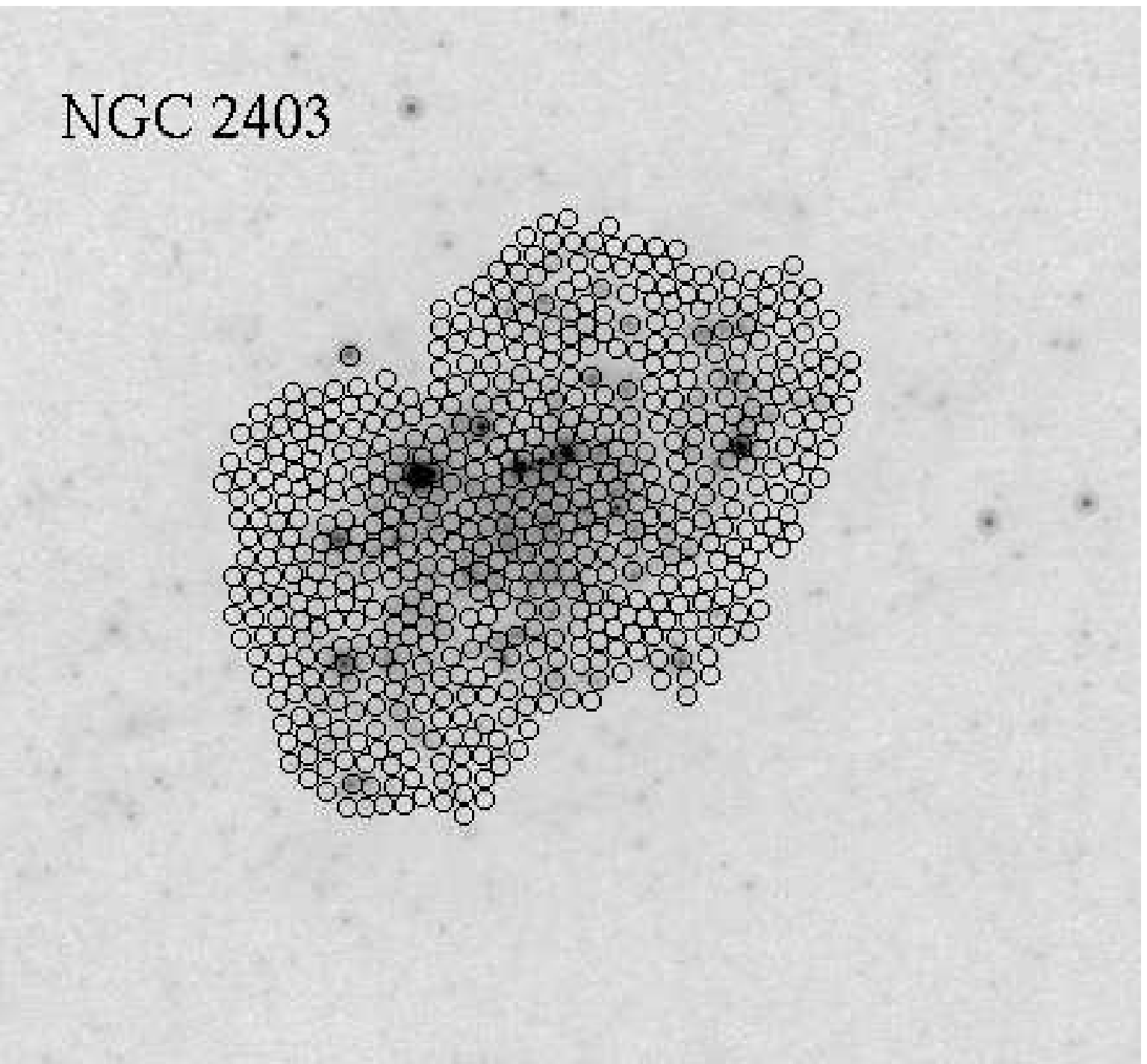}}}
      \hspace*{3cm}
	      {\resizebox{8cm}{!}{
		  \plotone{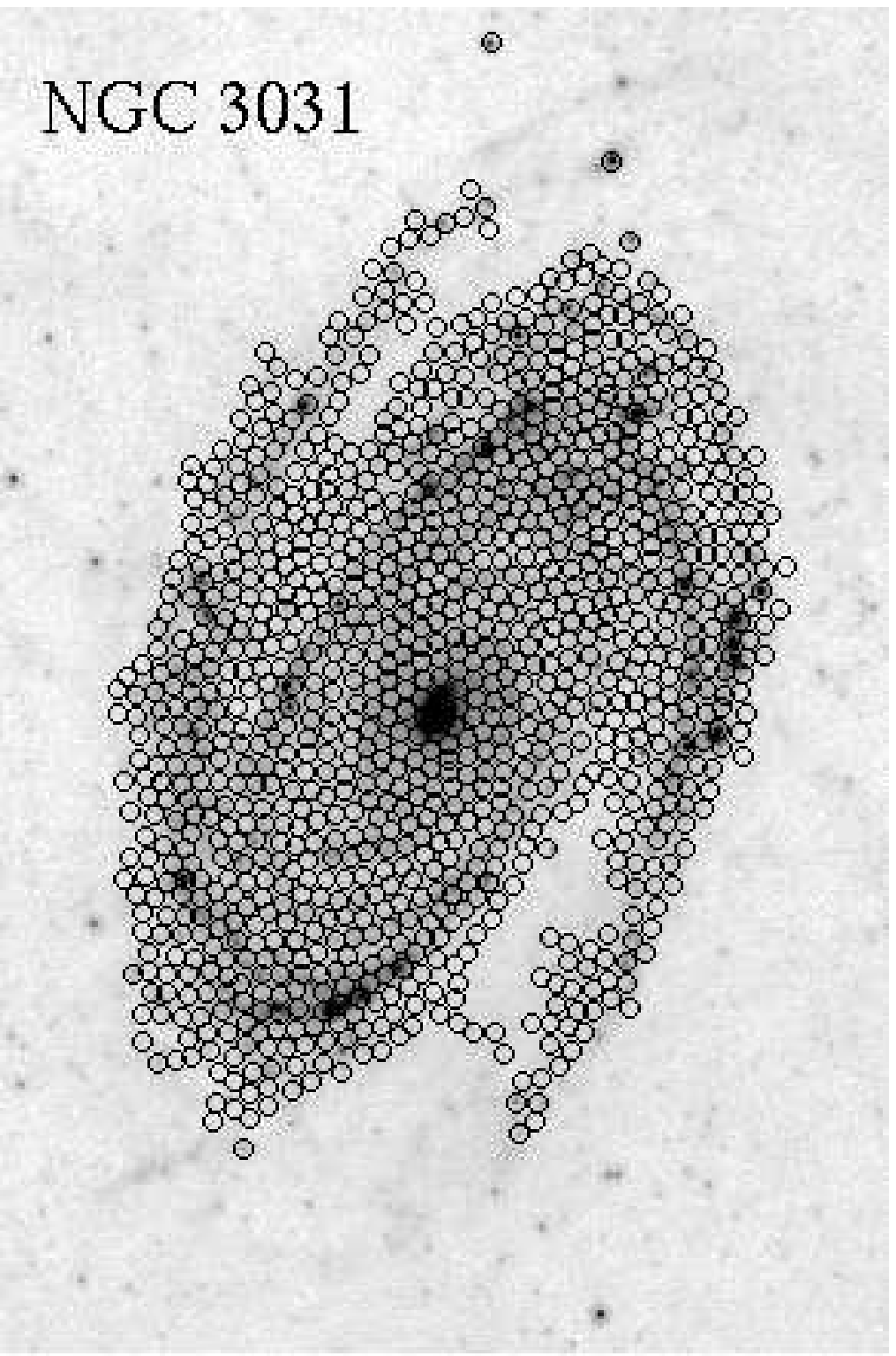}}}}
    \resizebox{10cm}{!}{
      {\resizebox{7cm}{!}{	
	  \plotone{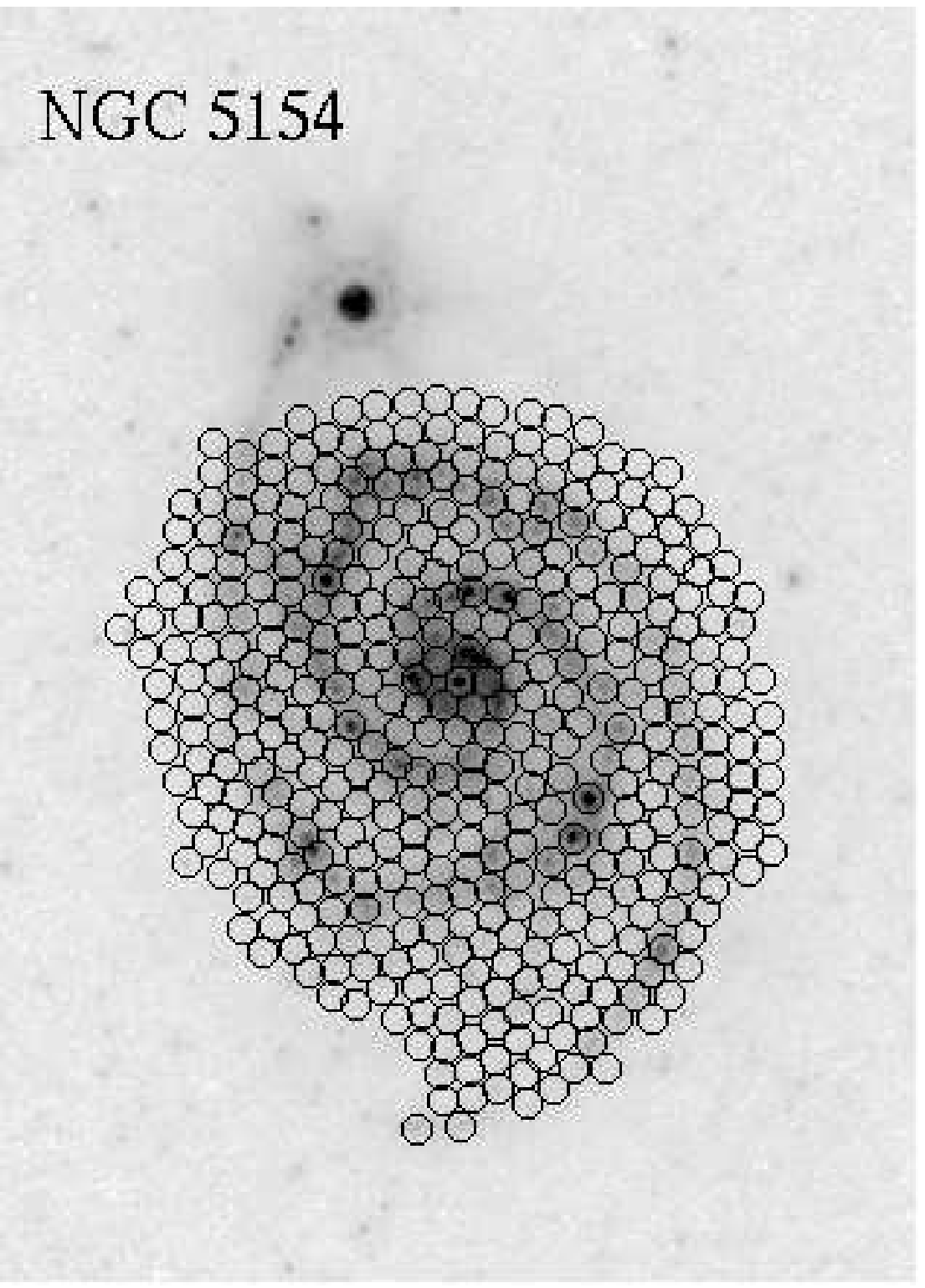}}}
      \hspace*{3cm}
	      {\resizebox{10cm}{!}{
		  \plotone{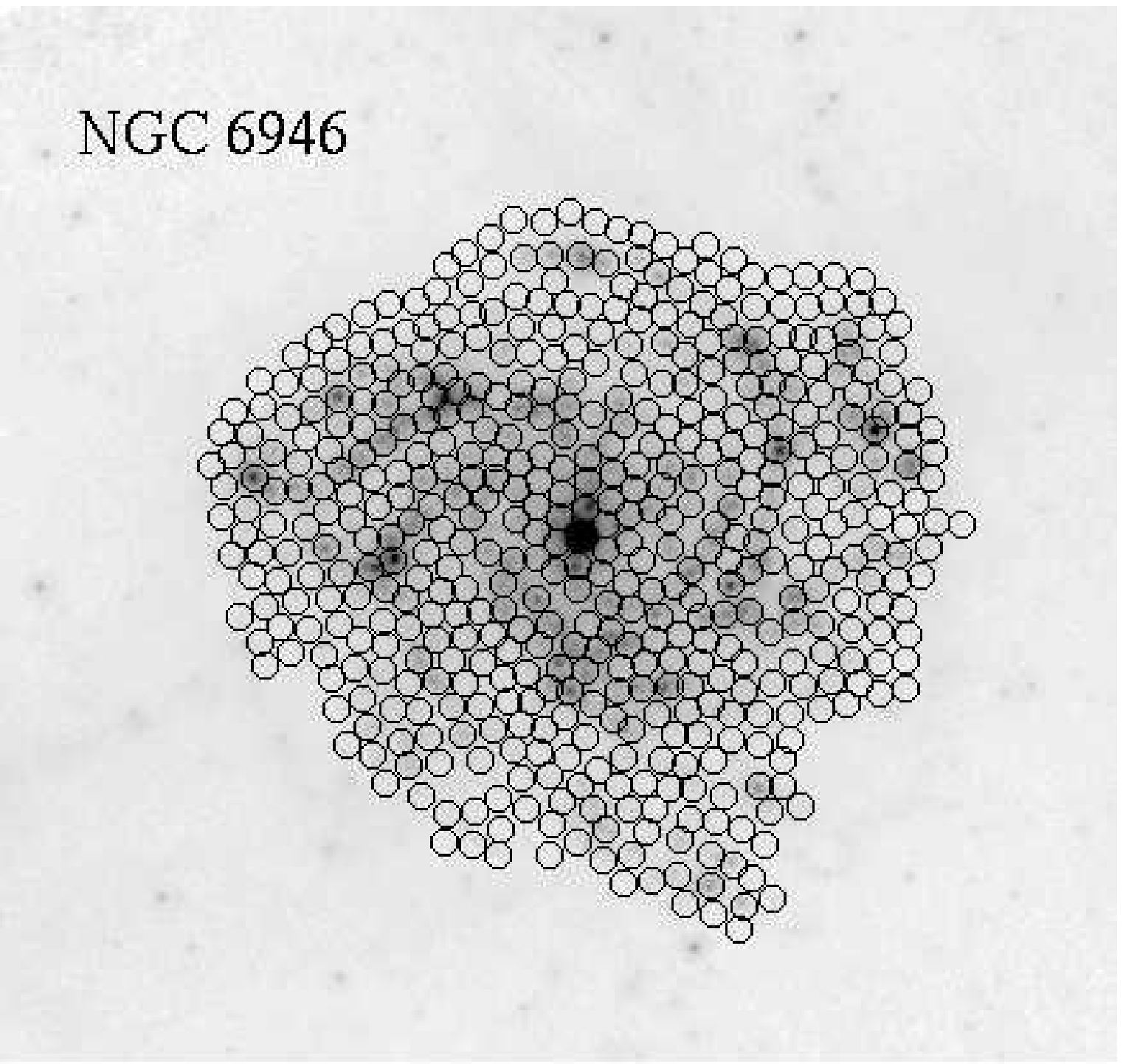}}}}
    \caption{Aperture masks plotted on 24~$\micron$ images of each
      galaxy using critical apertures defined with diameters equal to
      the FWHM of the 70~$\micron$ PSF ($\sim$17$\arcsec$).
      (The color scheme used in the electronic edition is as follows:
      nucleus (cyan), inner-disk (red), disk (magenta), outer-disk
      (yellow), arm (blue), and inter-arm (green).)  See $\S$2.4 for
      more details about the aperture definitions. \label{apmask}}  
  \end{center}
\end{figure}

\clearpage
\begin{figure}[!ht]
  \resizebox{16.5cm}{!}{
    {\plotone{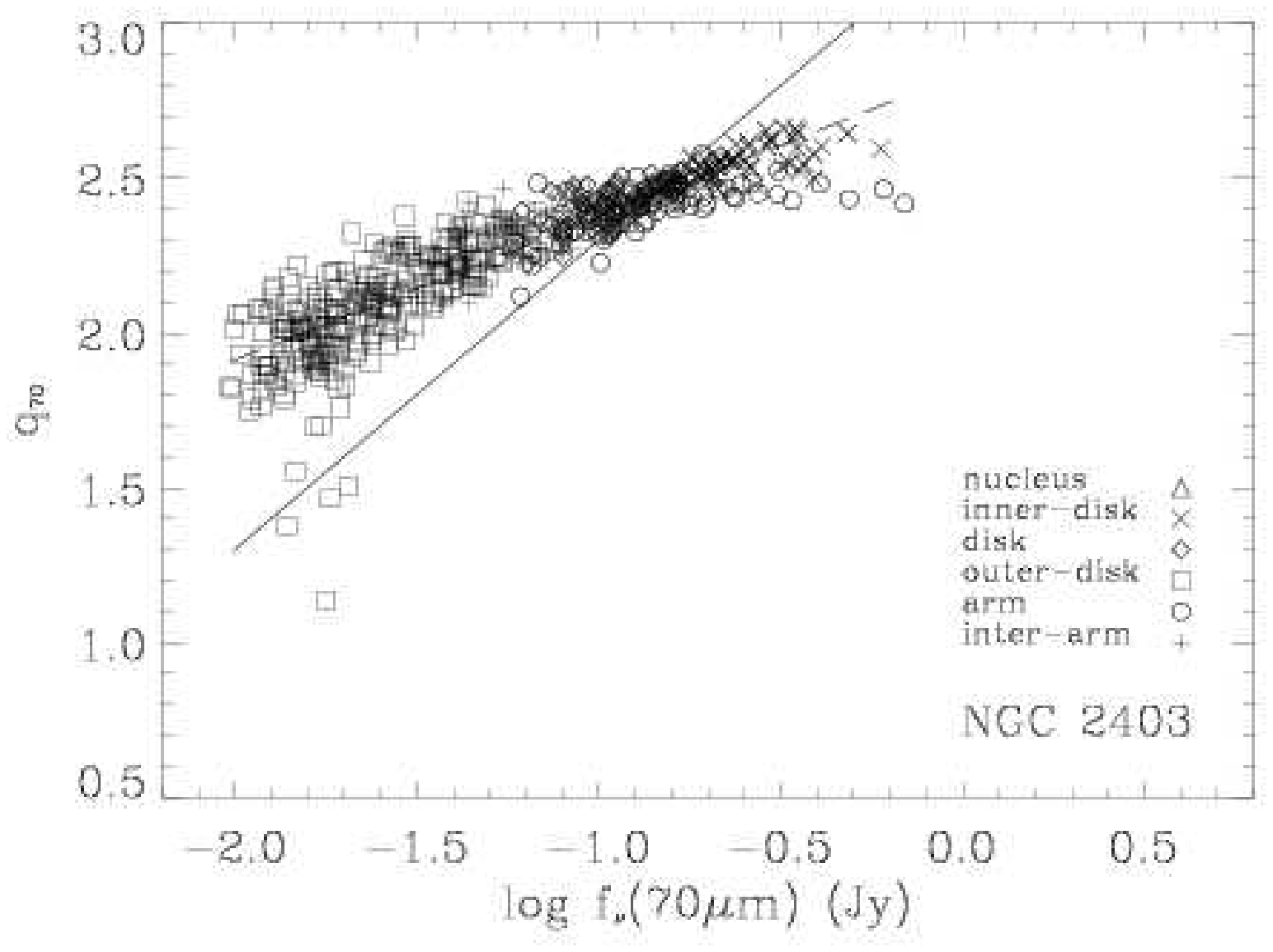}
      \plotone{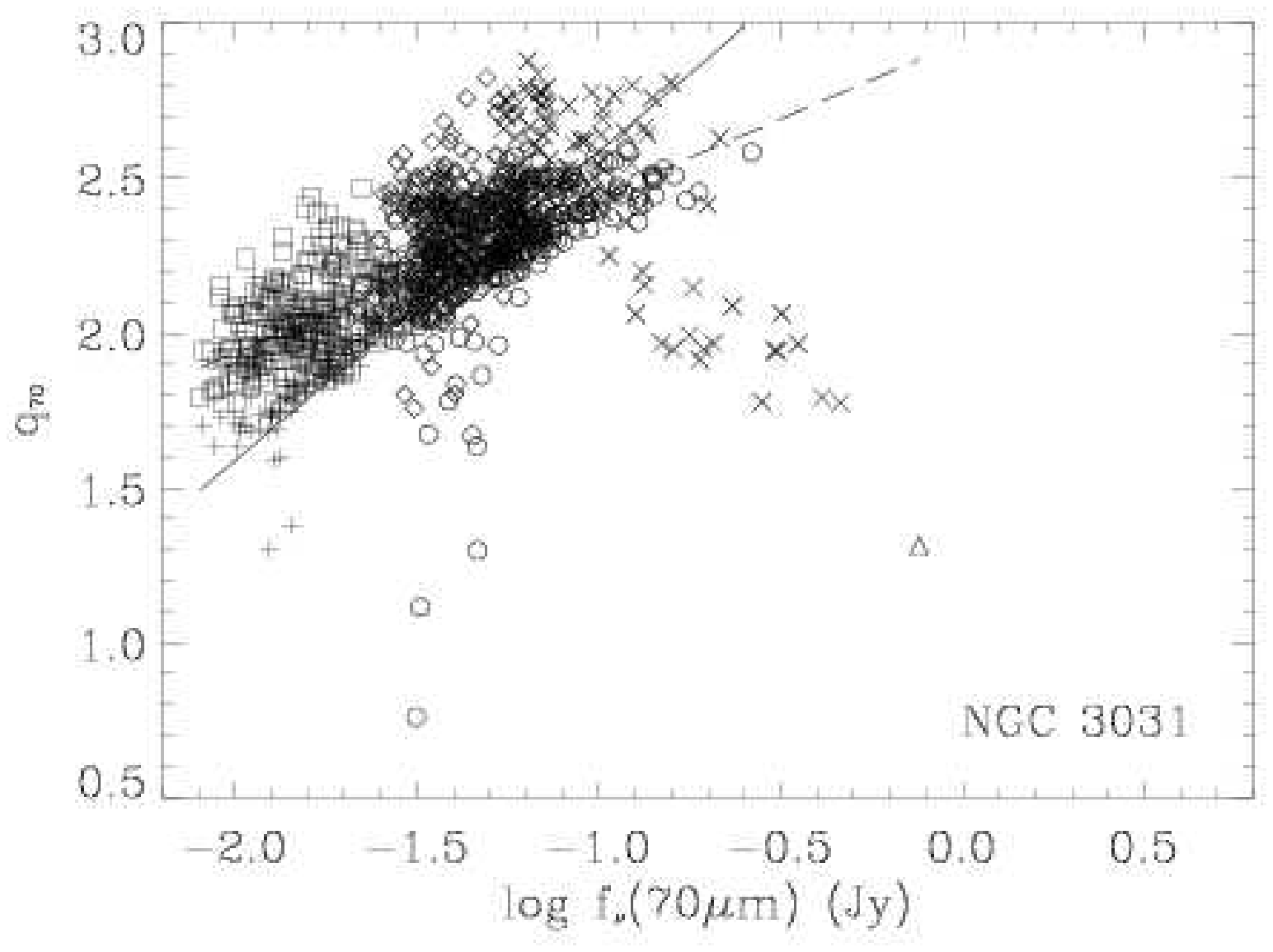}}}
  \vspace*{-.1cm}
\resizebox{16.5cm}{!}{
    {\plotone{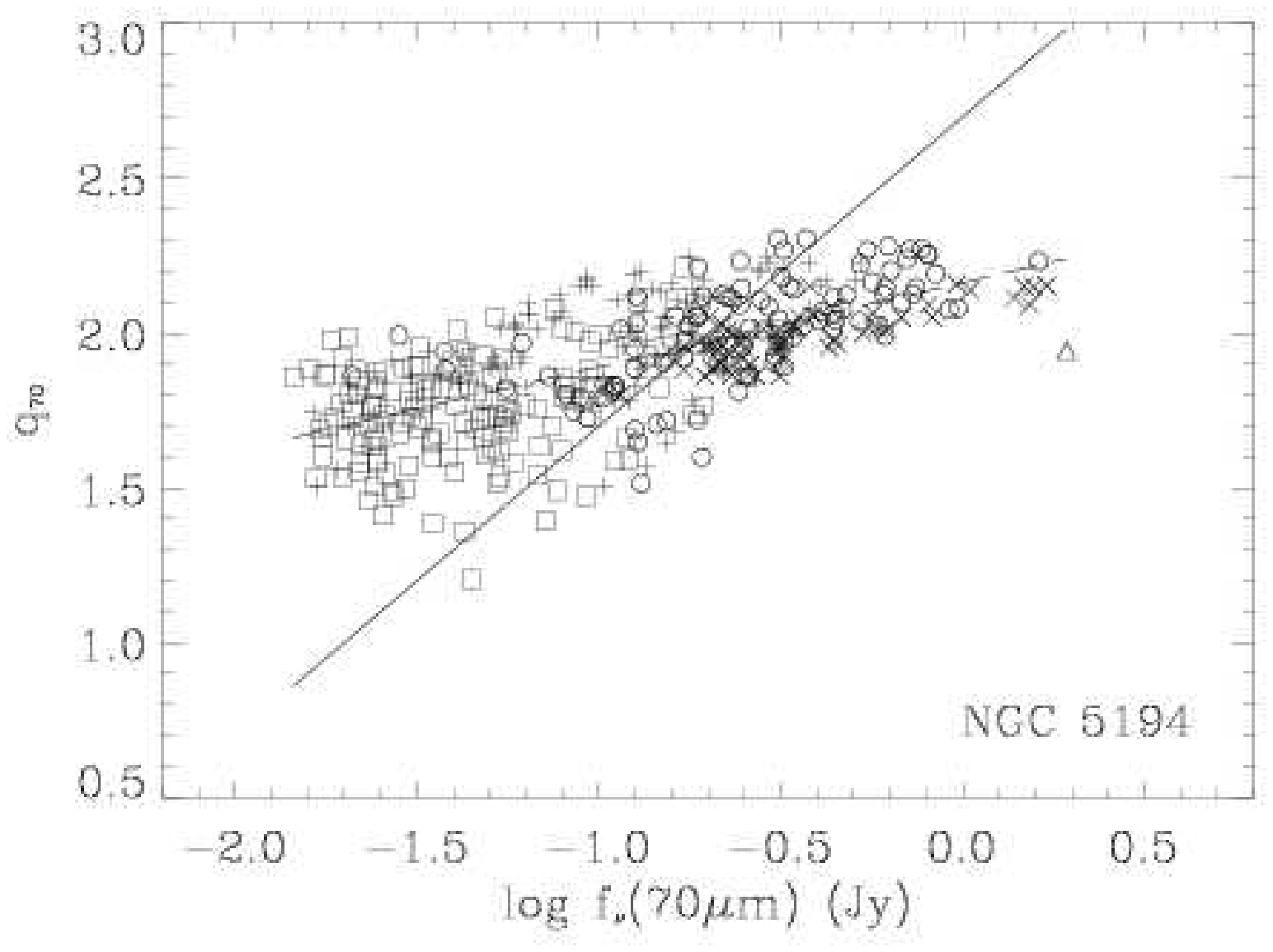}
      \plotone{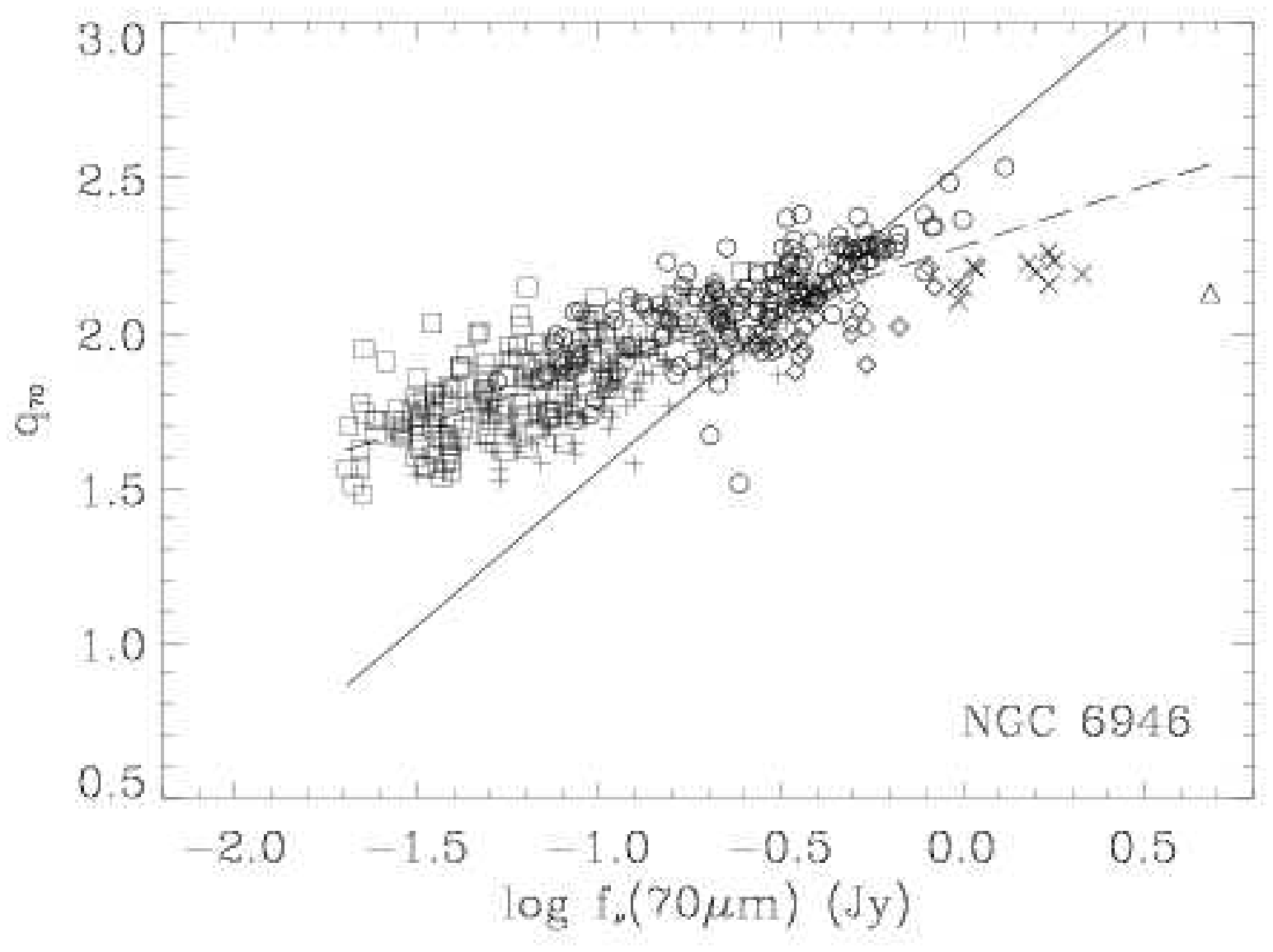}}}
\caption{$q_{70}$ plotted as a function of the 70~$\micron$ flux
  density grouped by different environments within each galaxy disk.
  Since the apertures are equal in diameter for each galaxy, the
  measured flux densities are directly proportional to surface
  brightnesses.  
  The solid line is the expected trend if the galaxy disk was
  characterized by a constant radio surface brightness.  The dashed
  line is the least-squares fit to the data.\label{q70f70}} 
\end{figure}

\clearpage
\begin{figure}[!ht]
  \resizebox{16.5cm}{!}{
    {\plotone{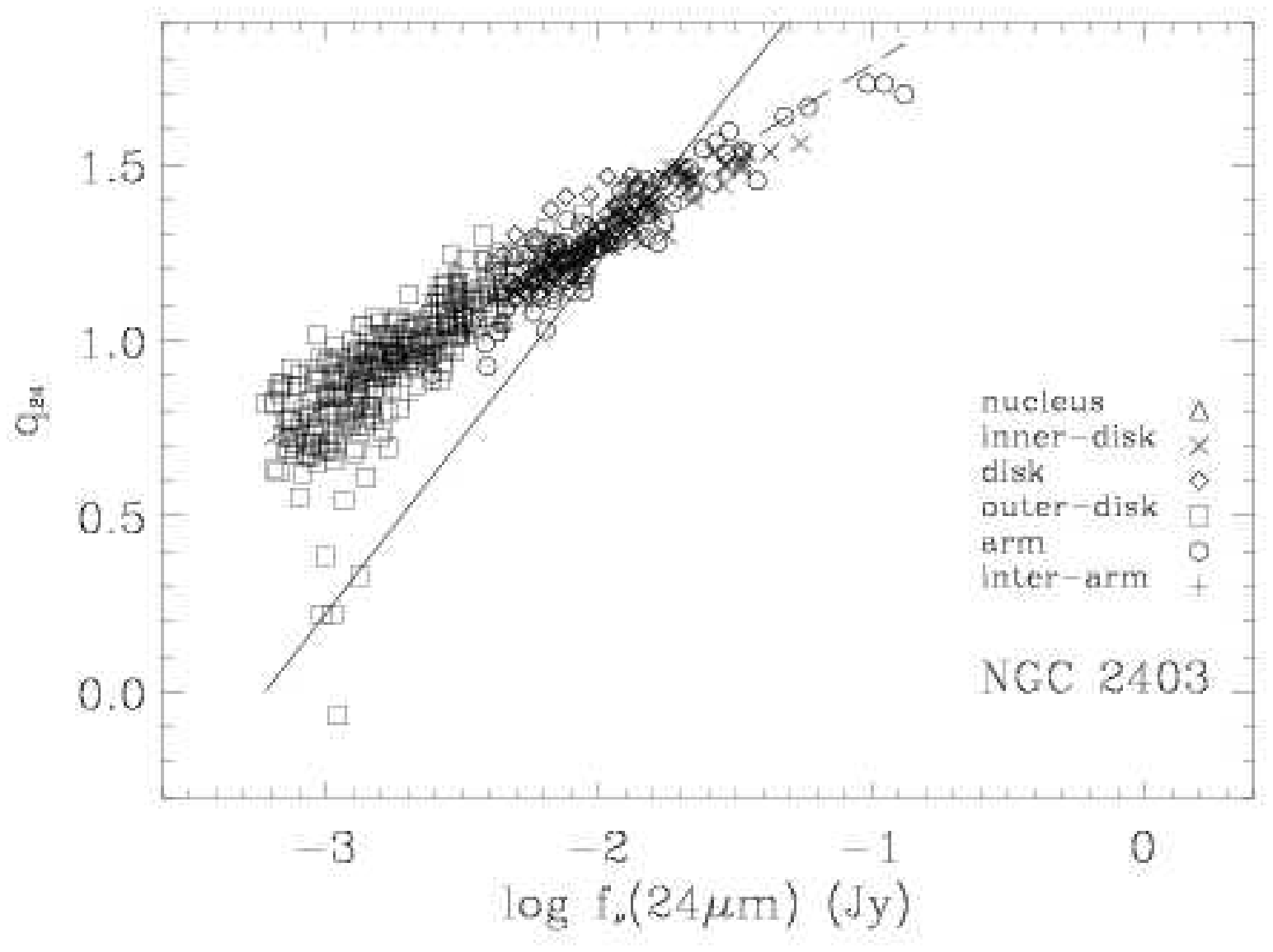}
      \plotone{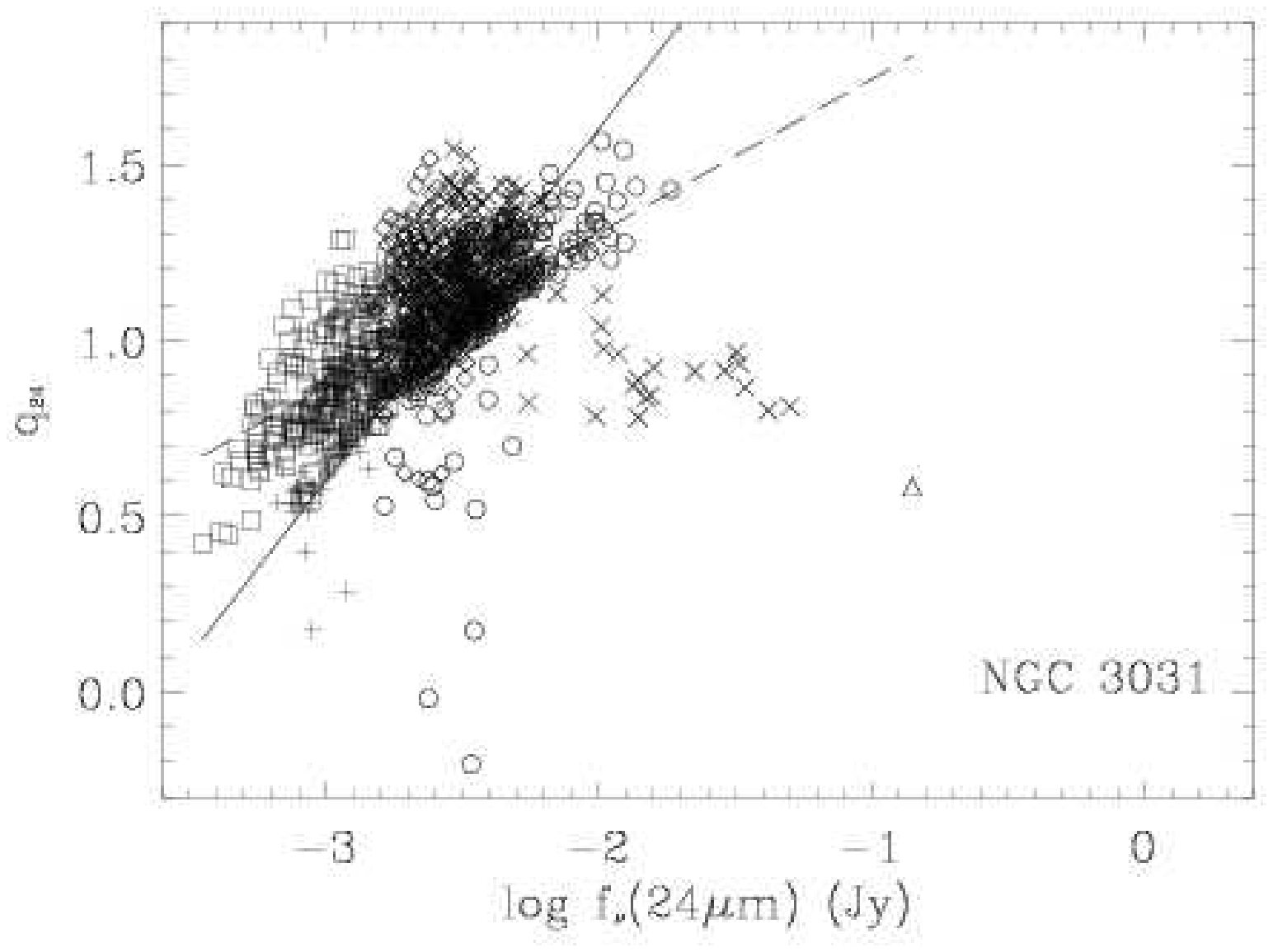}}}
  \vspace*{-.1cm}
  \resizebox{16.5cm}{!}{
    {\plotone{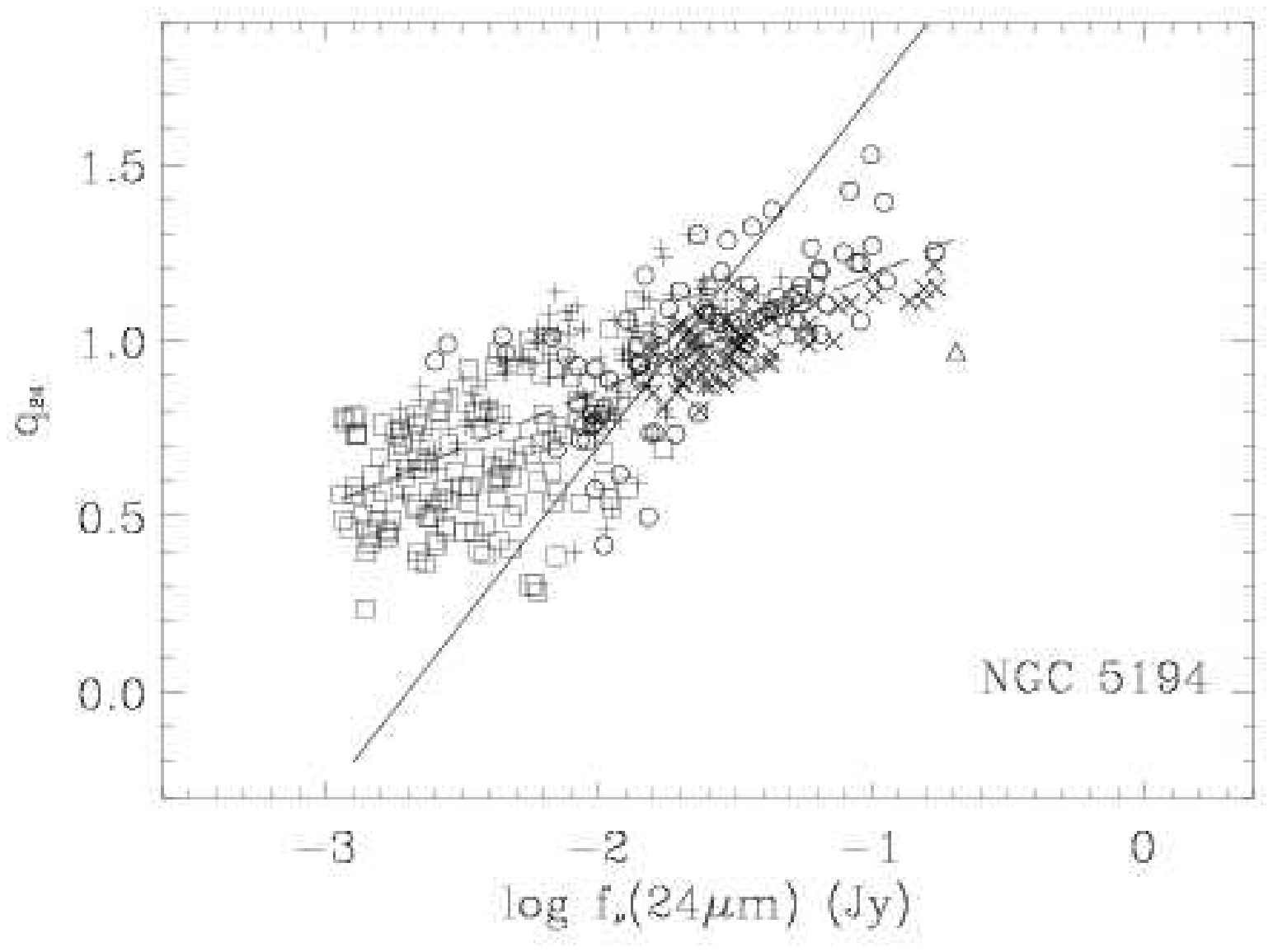}
      \plotone{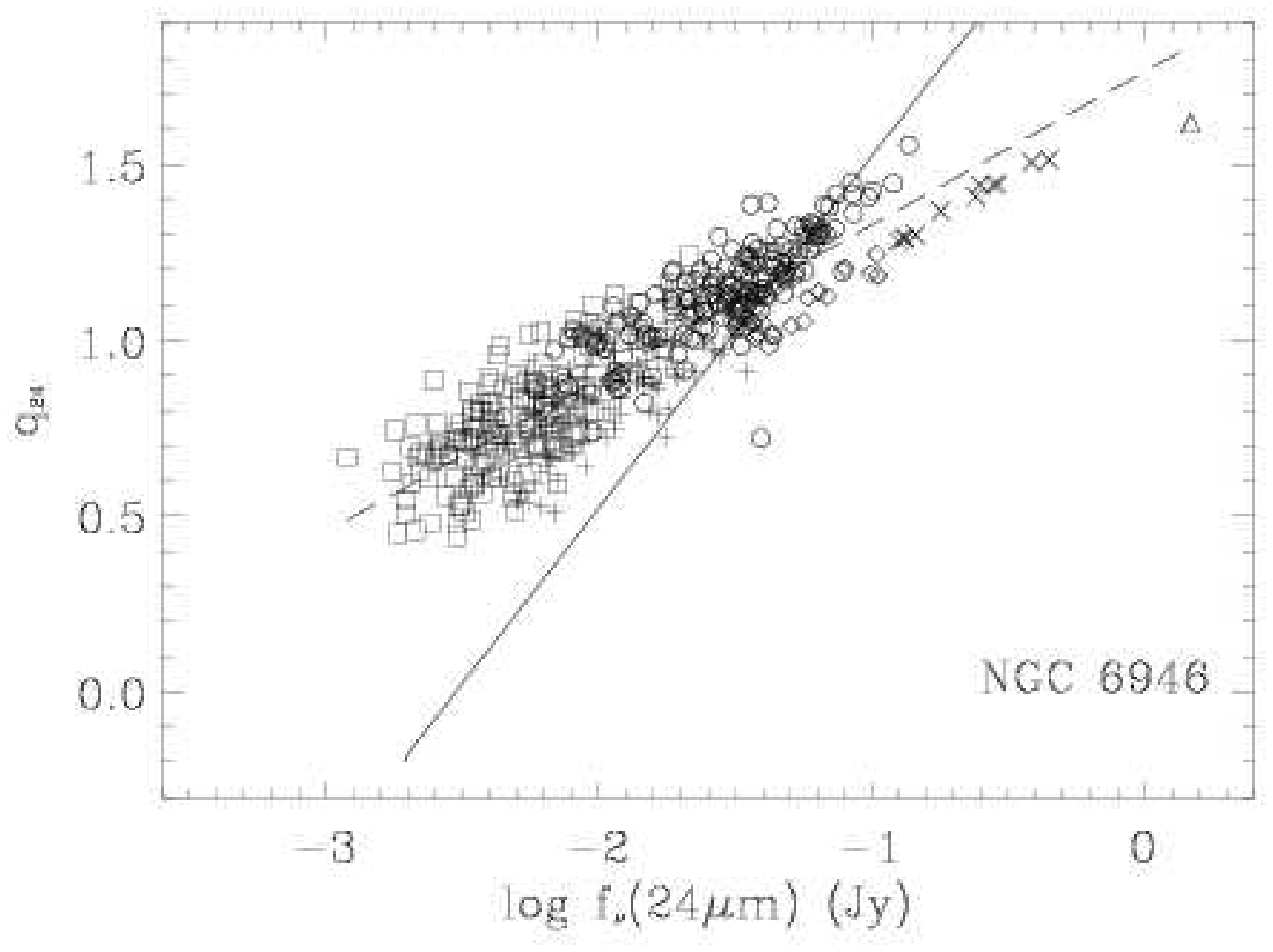}}}
  \caption{The same as Figure \ref{q70f70} except for $q_{24}$ as a
  function of the 24~$\micron$ flux density. \label{q24f24}}
\end{figure}

\clearpage
\begin{figure}[!ht]
  \resizebox{16.5cm}{!}{
    {\plotone{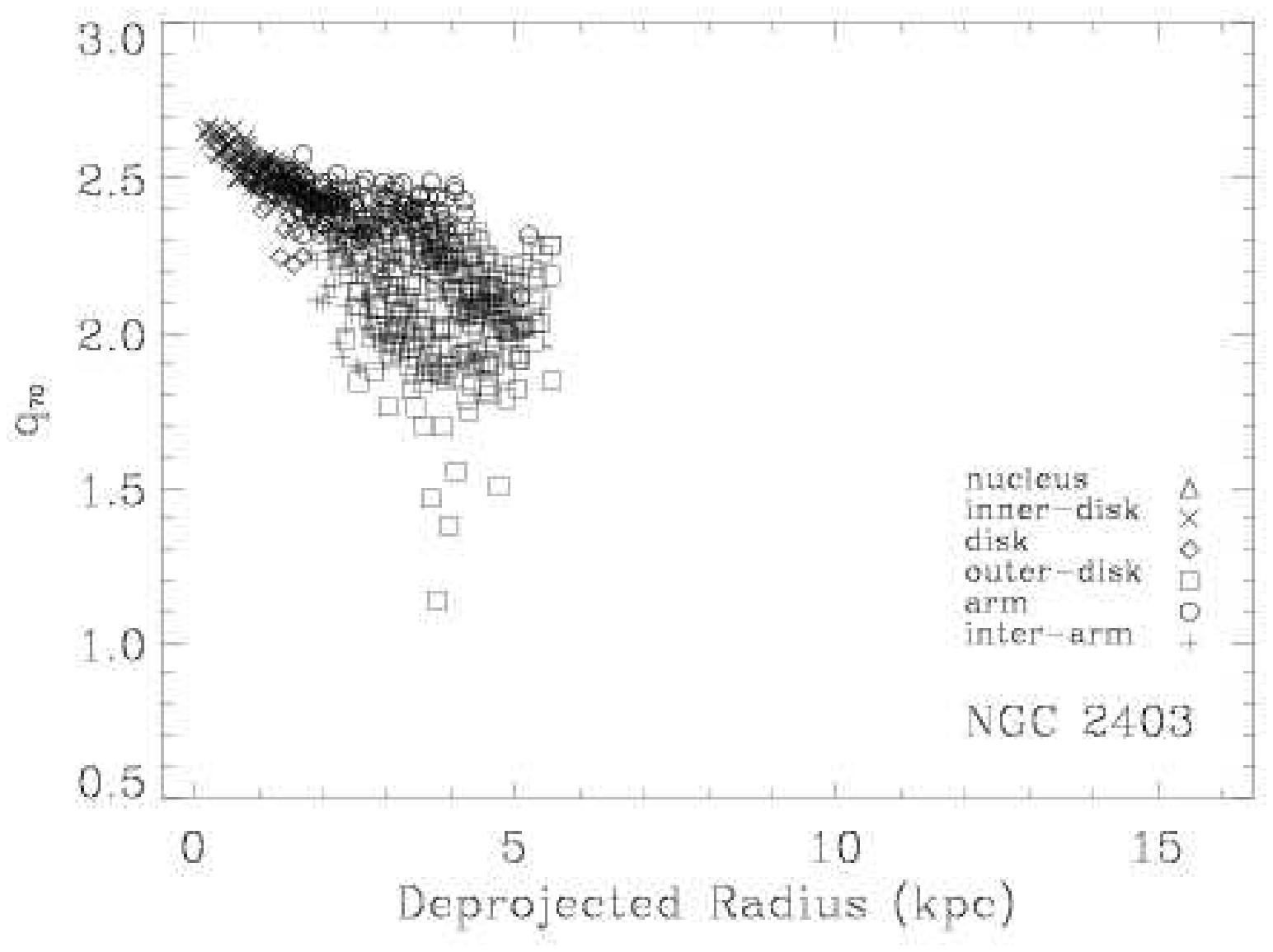}
      \plotone{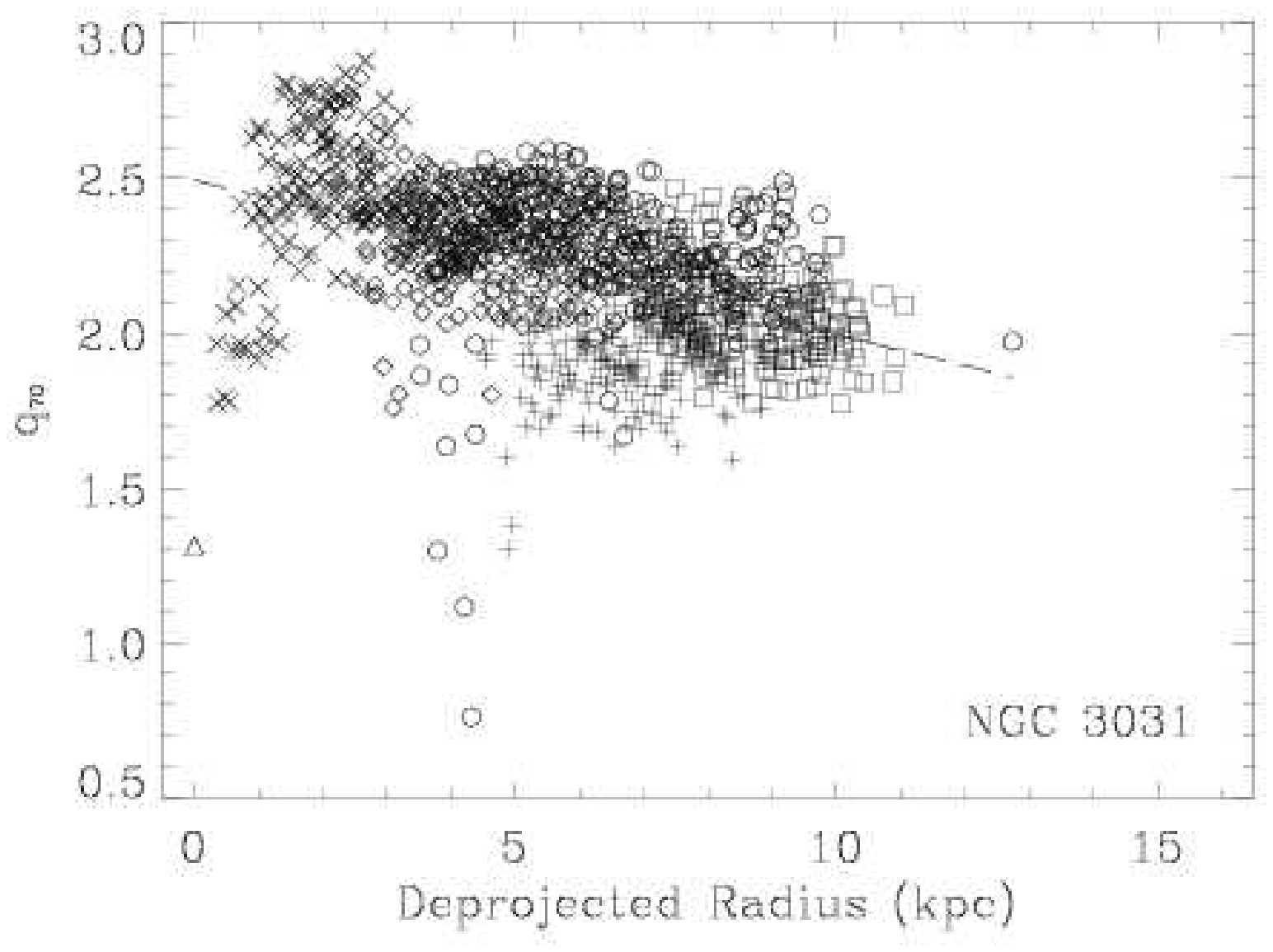}}}
  \vspace*{-.1cm}
  \resizebox{16.5cm}{!}{
    {\plotone{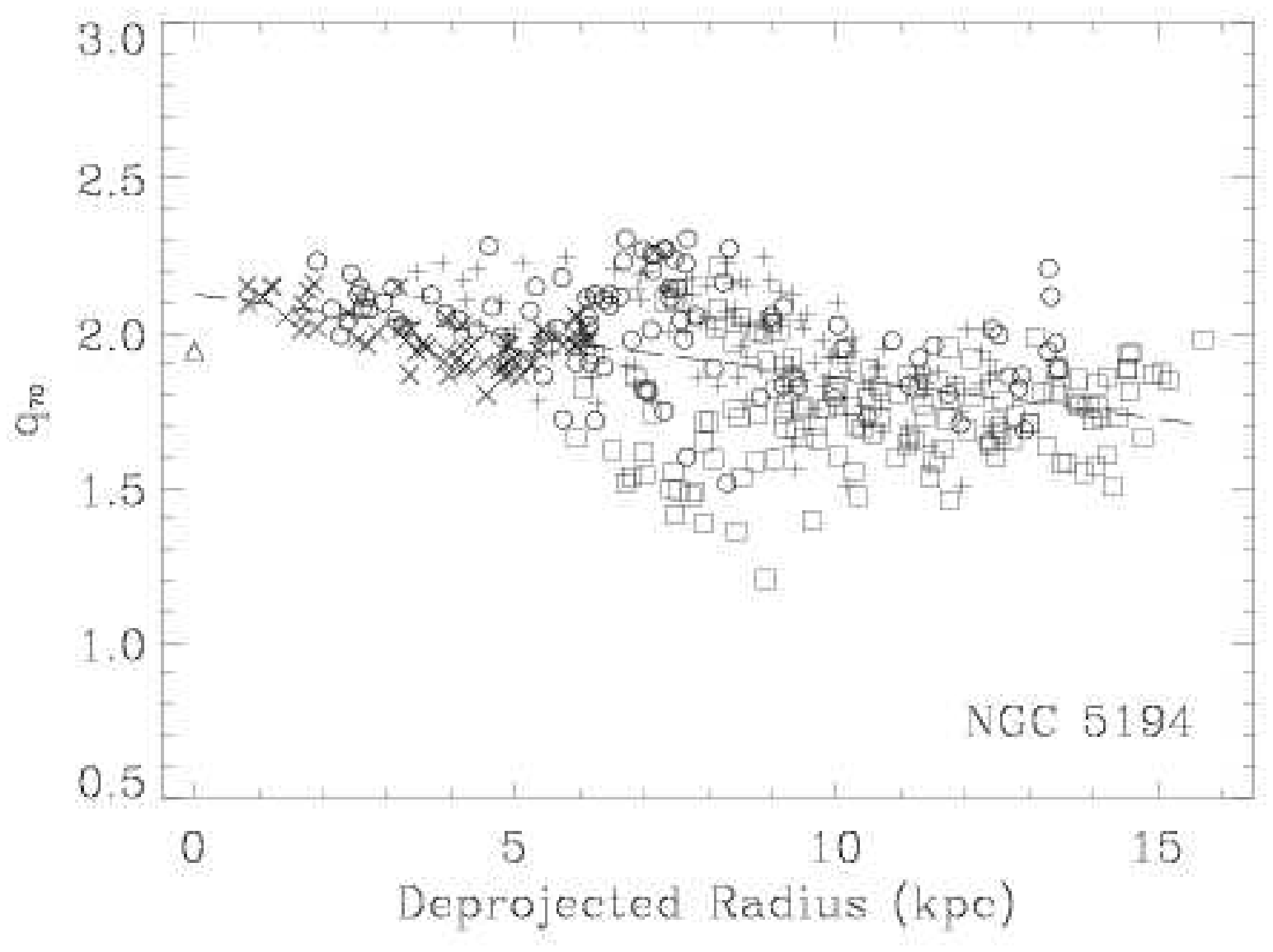}
      \plotone{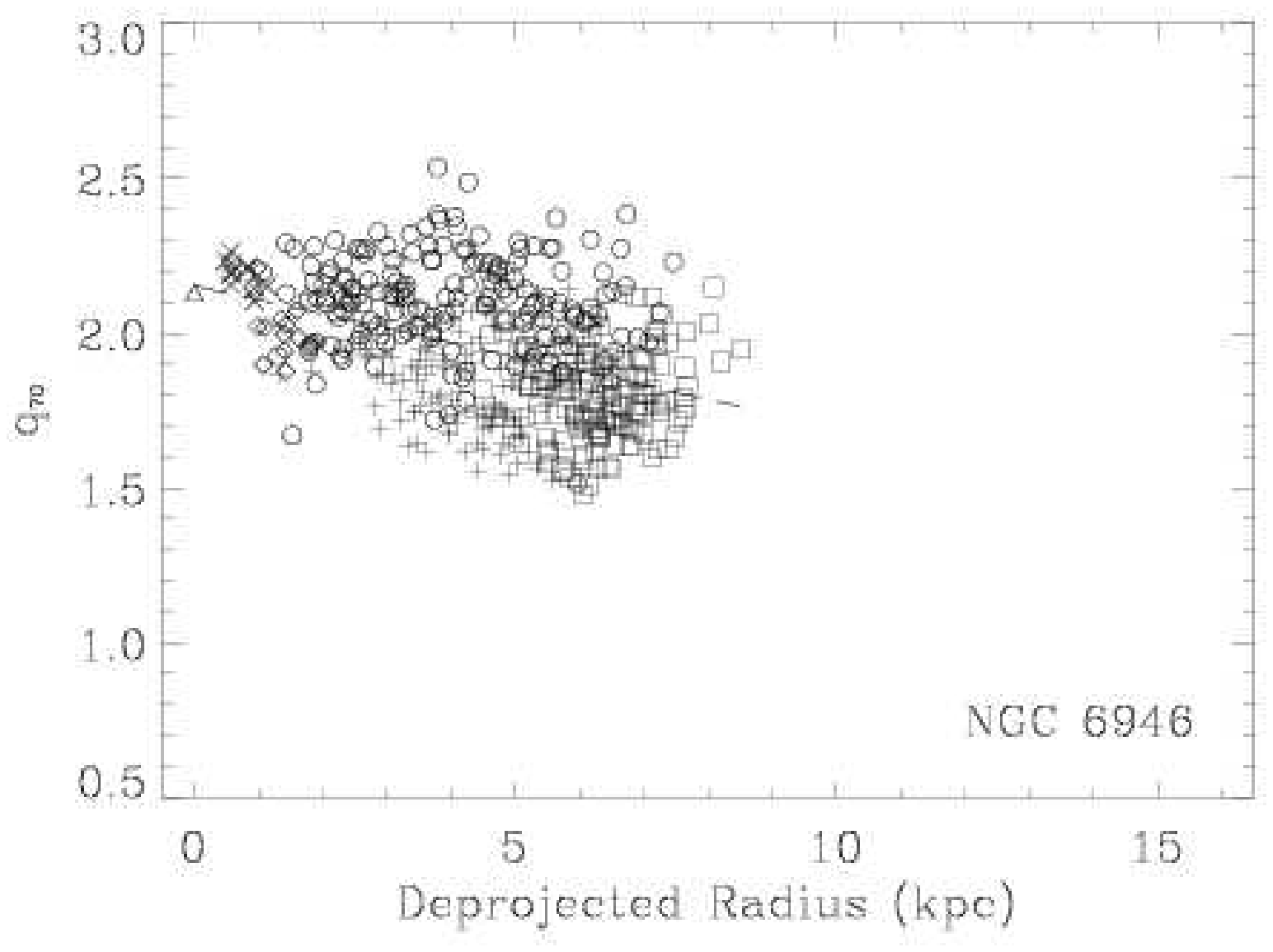}}}
  \caption{$q_{70}$ plotted as a function of galactocentric radius.
  The dashed line is the least-squares fit to the data.  \label {q70rad}}
\end{figure}

\clearpage
\begin{figure}[!ht]
  \resizebox{16.5cm}{!}{
    {\plotone{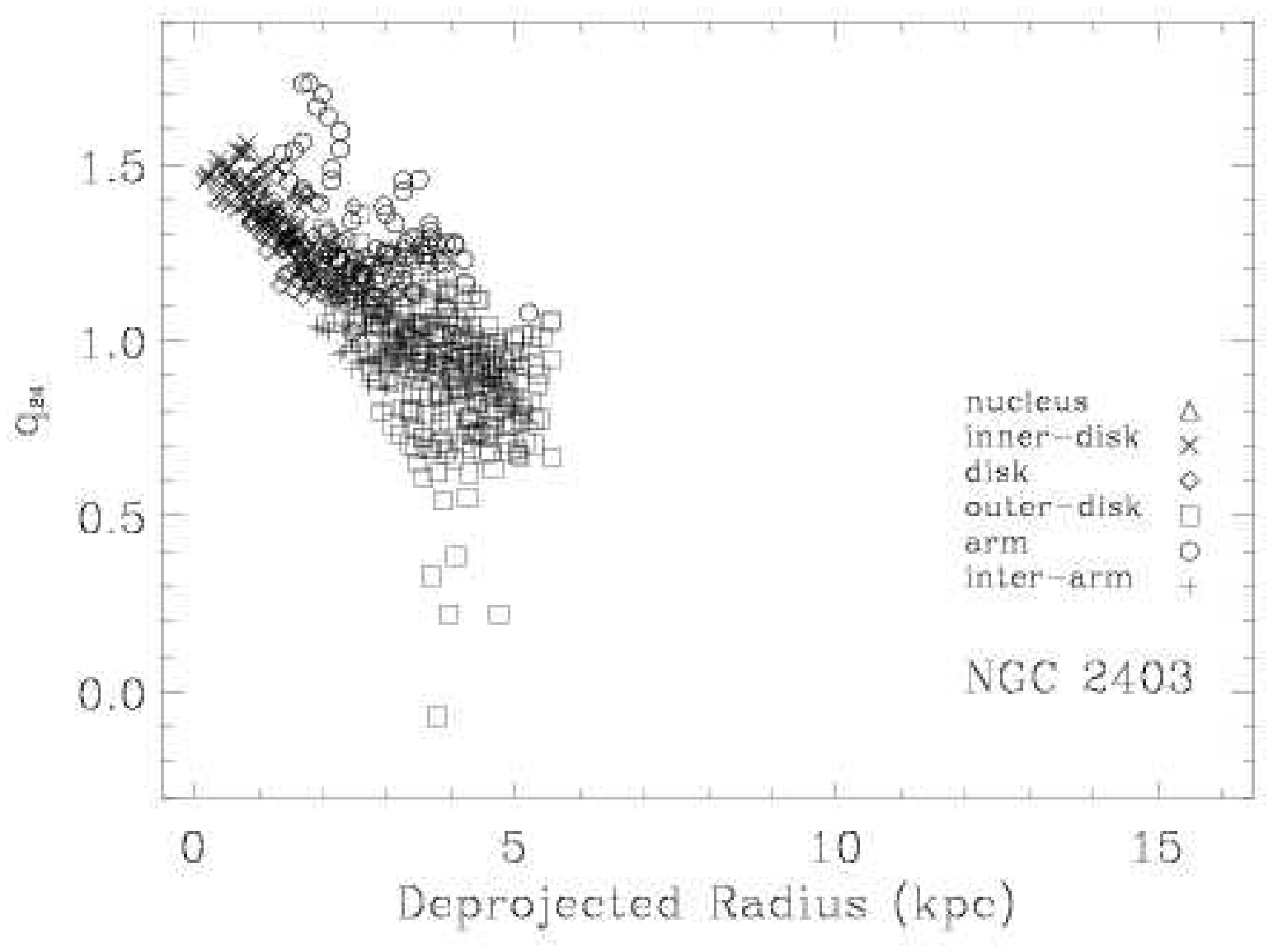}
      \plotone{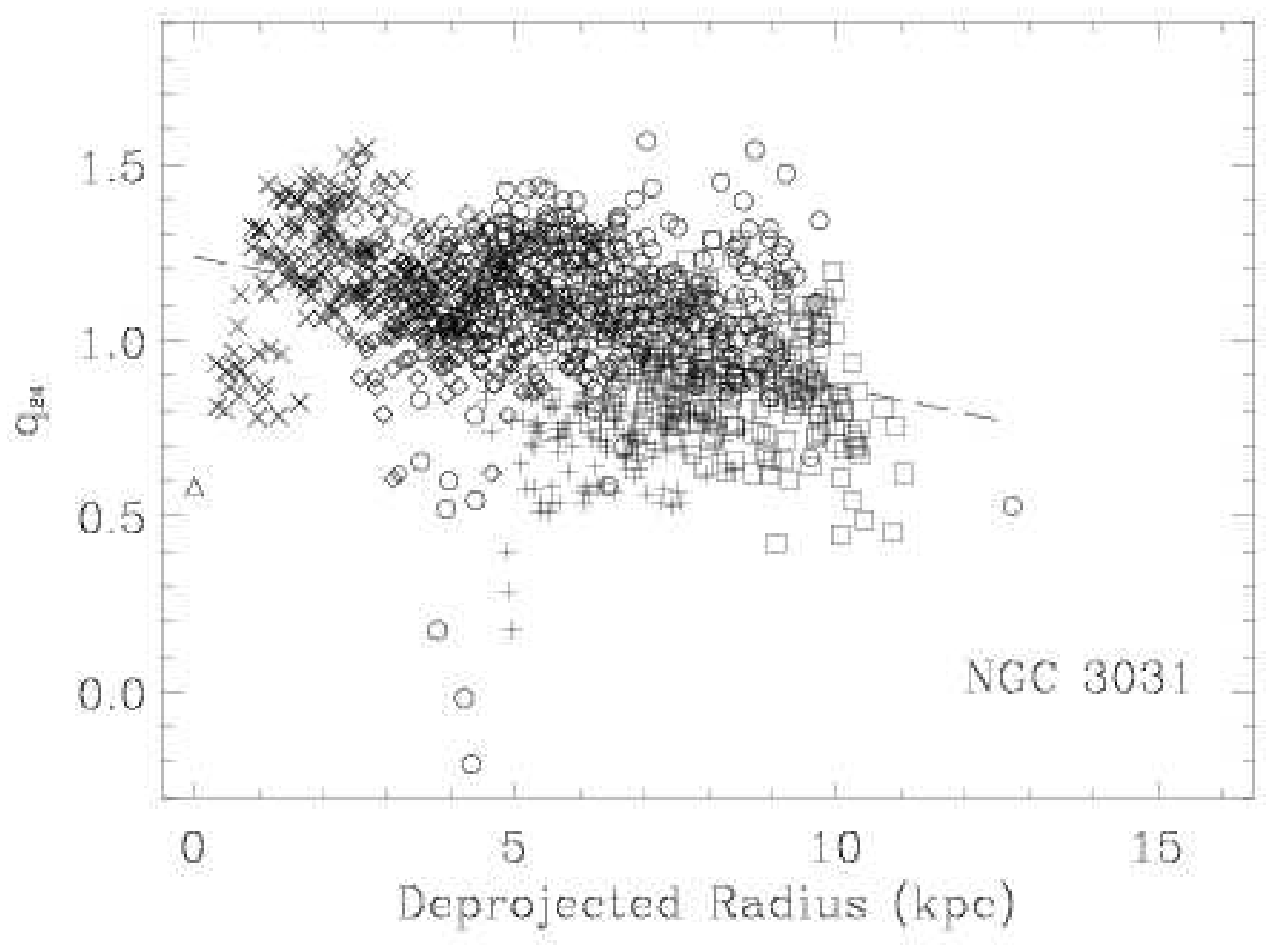}}}
  \resizebox{16.5cm}{!}{
    {\plotone{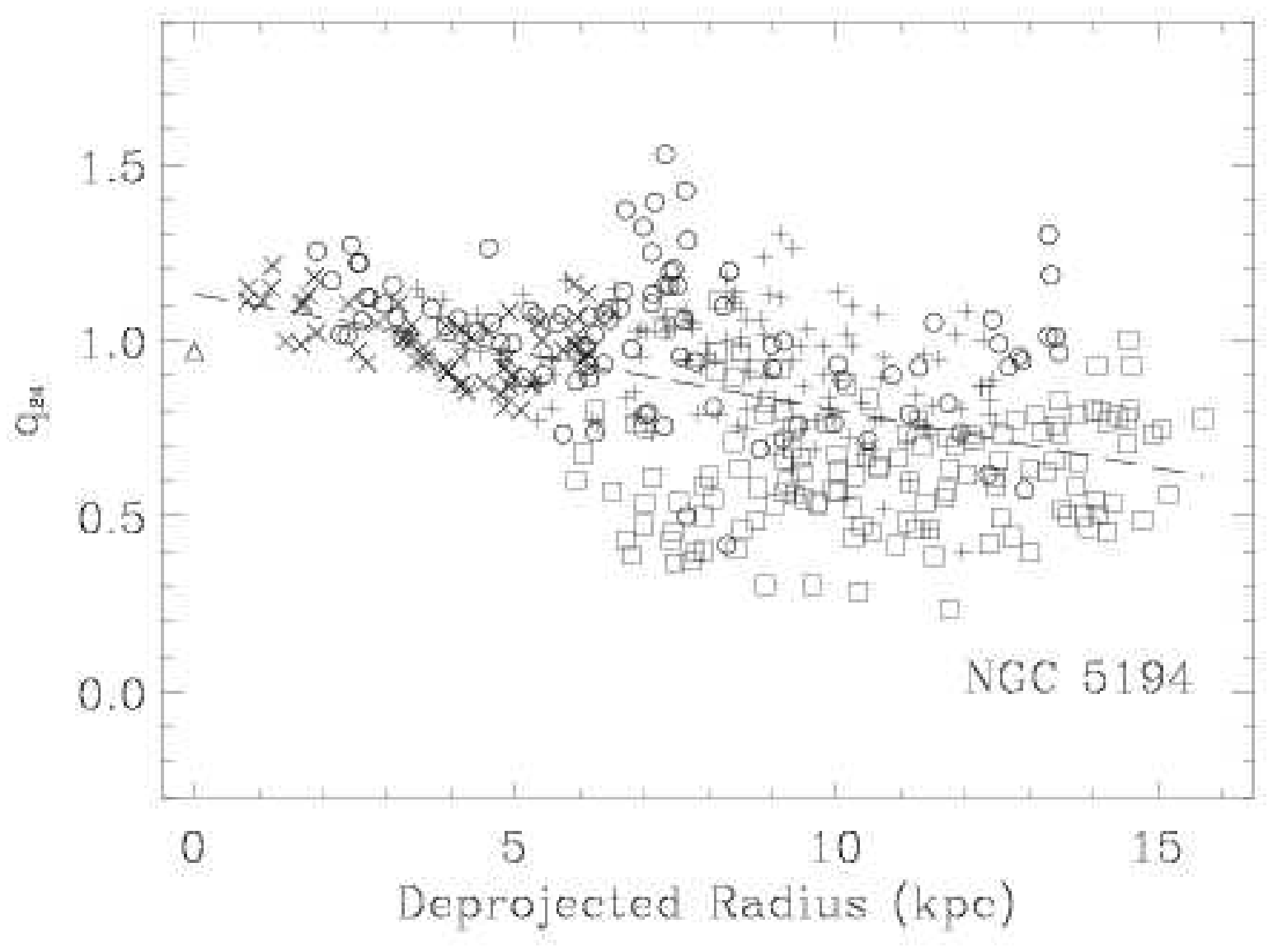}
      \plotone{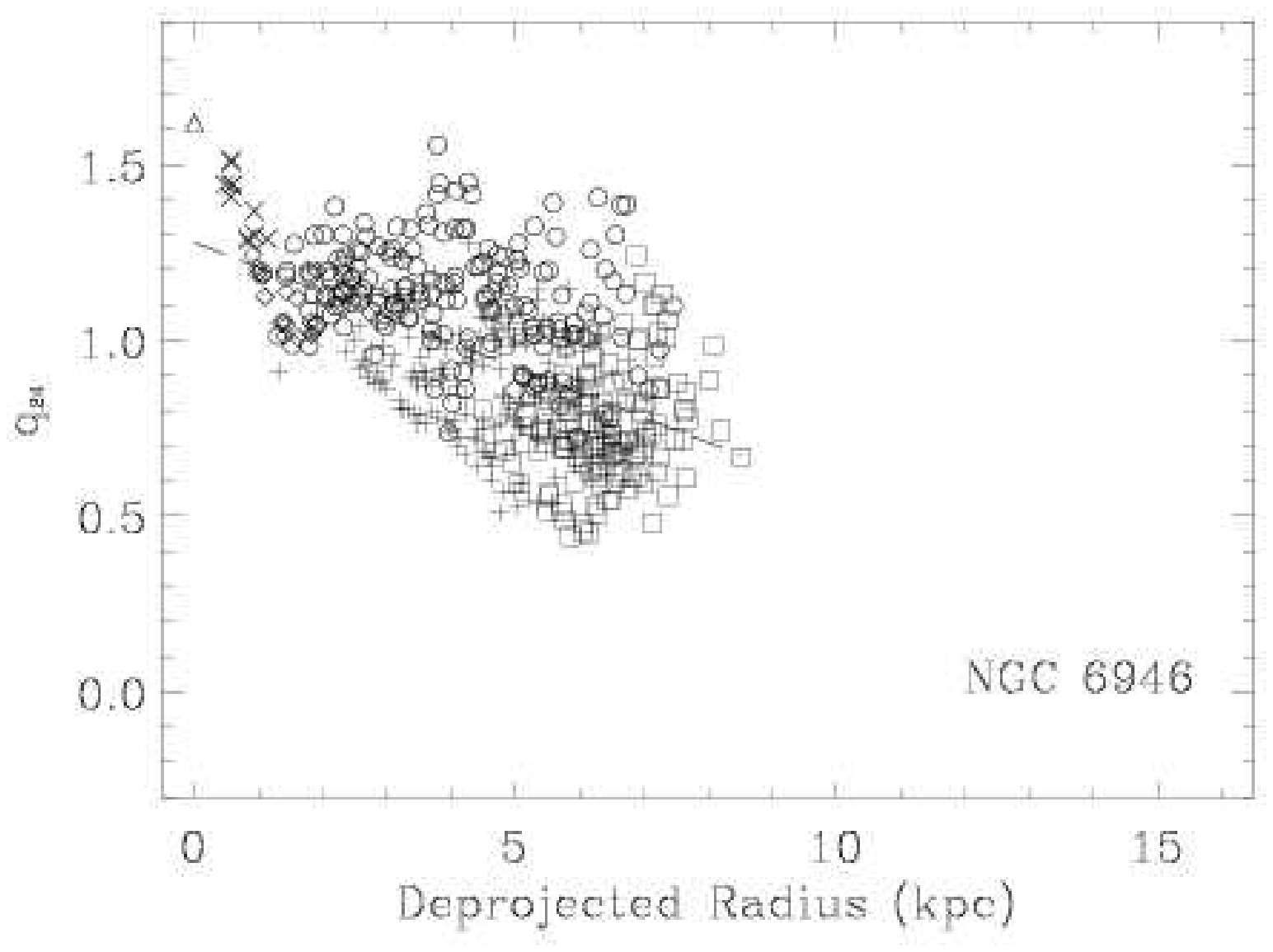}}}
  \caption{$q_{24}$ plotted as a function of galactocentric
  radius. The dashed line is the least-squares fit to the data.
  \label{q24rad}}
\end{figure}

\clearpage
\begin{figure}[!ht]
  \resizebox{15cm}{!}{
    \plotone{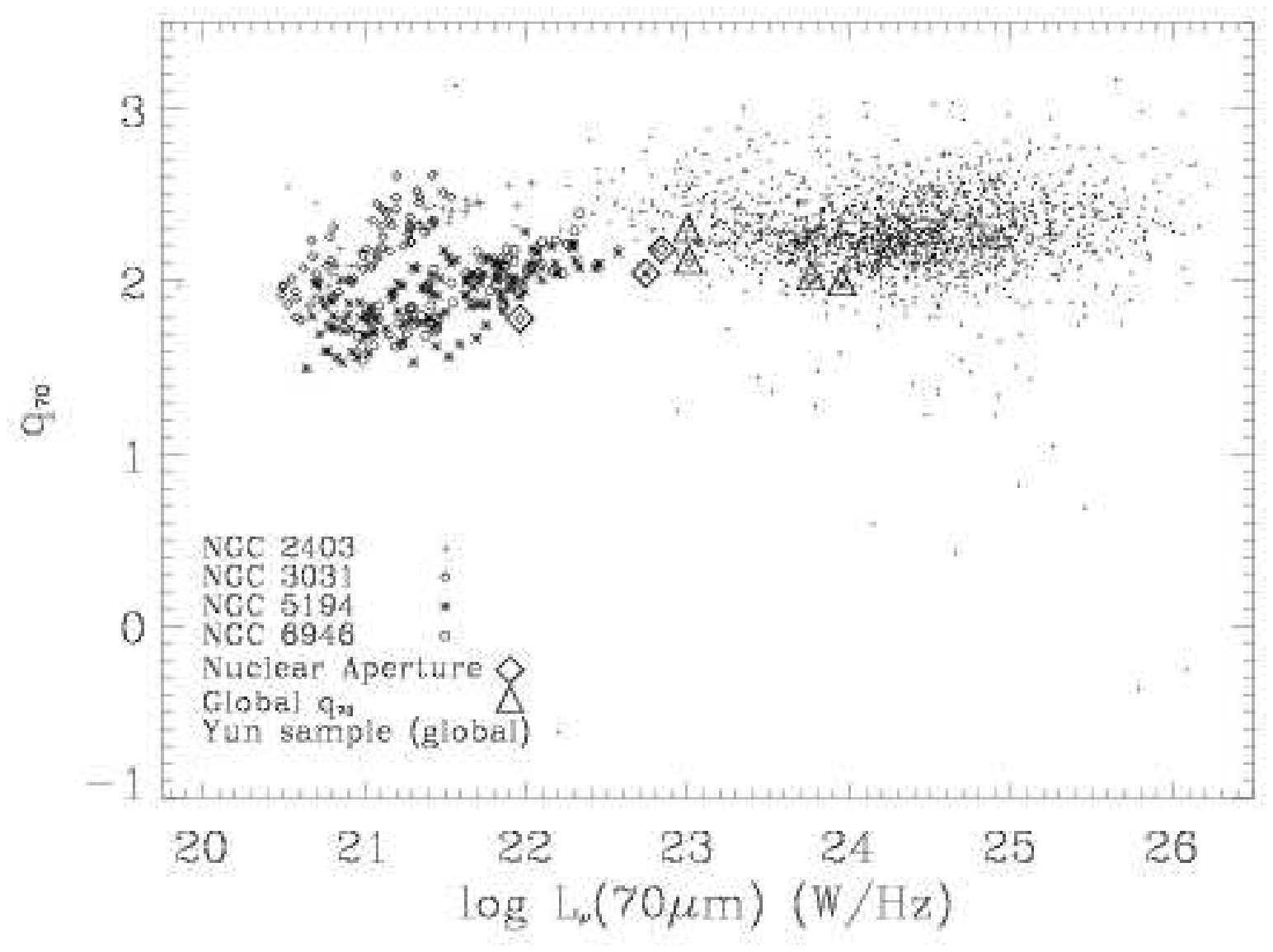}}
  \resizebox{15cm}{!}{
    \plotone{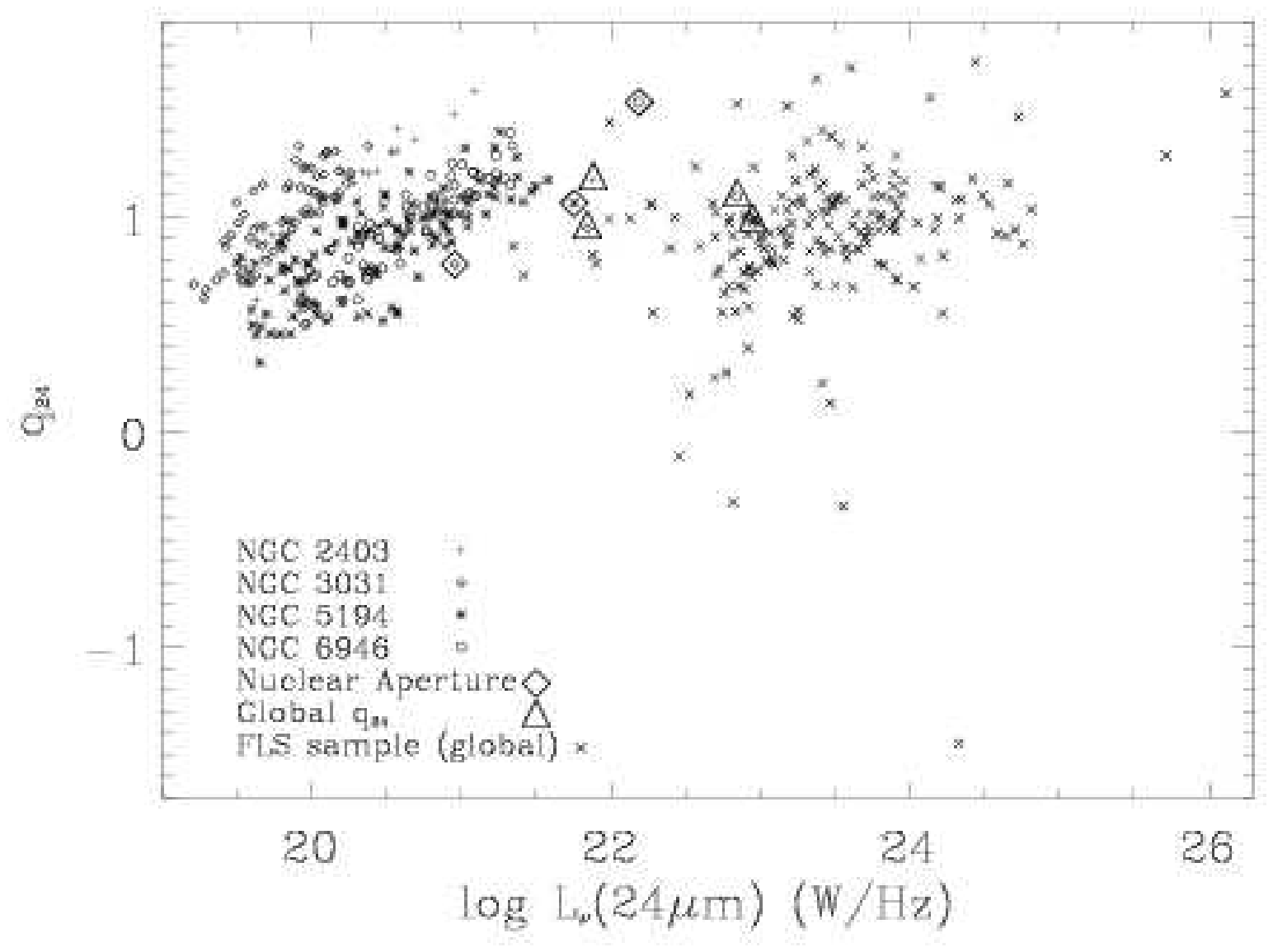}}
   \vspace*{-0.5cm}
 \caption{{\it Top}: 1.5~kpc aperture $q_{70}$ ratios for each sample
    galaxy plotted with global $q_{70}$ ratios estimated for the
    \citet{yun01} sample (see $\S$4.1.1). 
    {\it Bottom}: 1.5~kpc aperture $q_{24}$ ratios for each sample
    galaxy plotted with global $q_{24}$ ratios for the {\it Spitzer}
    FLS sample \citep{pa04}. 
    In both panels the nuclear and global $q_{\lambda}$ values for
    each of the four sample galaxies are identified by large diamonds
    and triangles, respectively.  \label{globloc}}  
\end{figure}

\clearpage
\begin{figure}[!ht]
  \resizebox{16.5cm}{!}{
    {\plotone{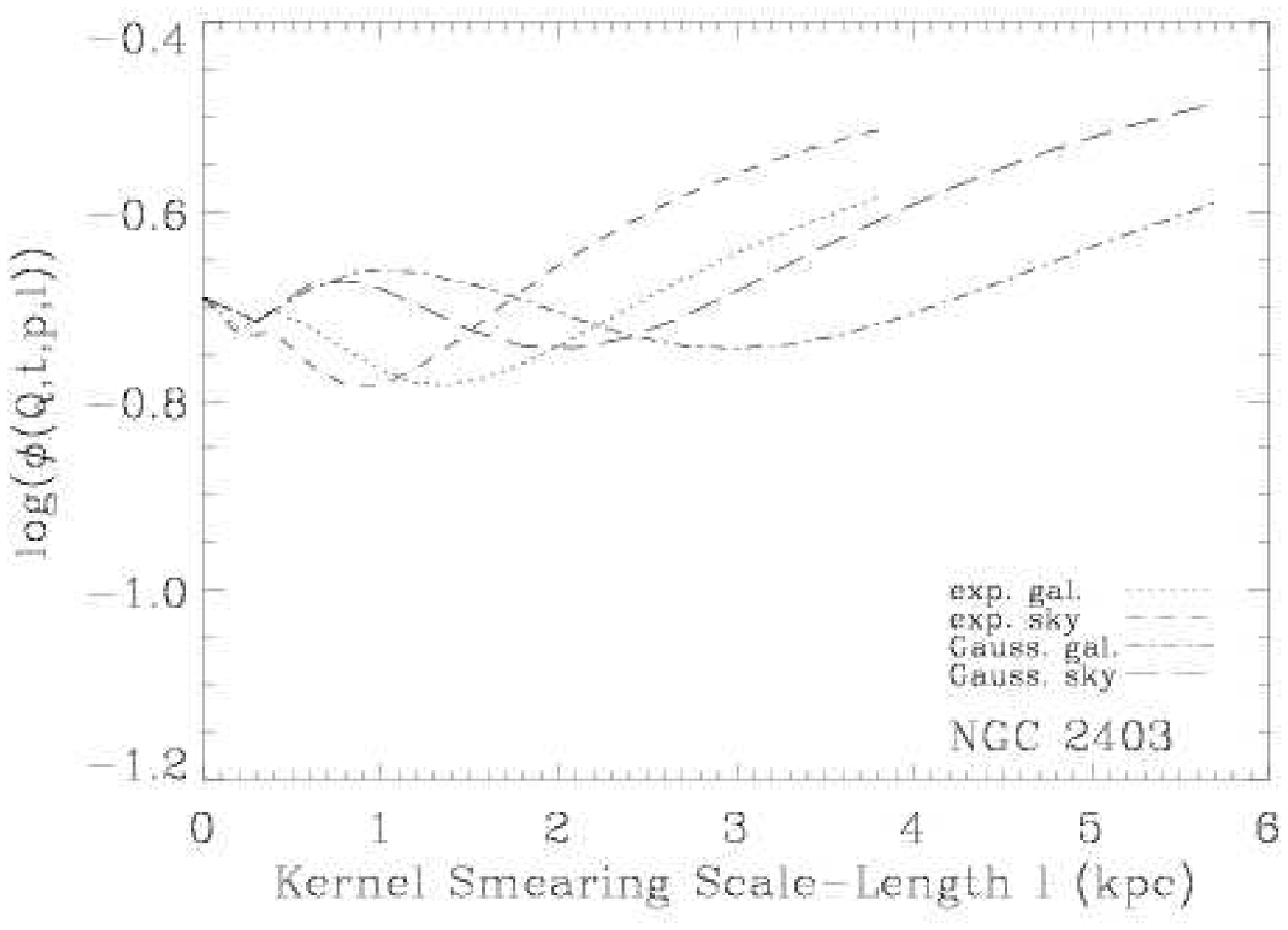}
      \plotone{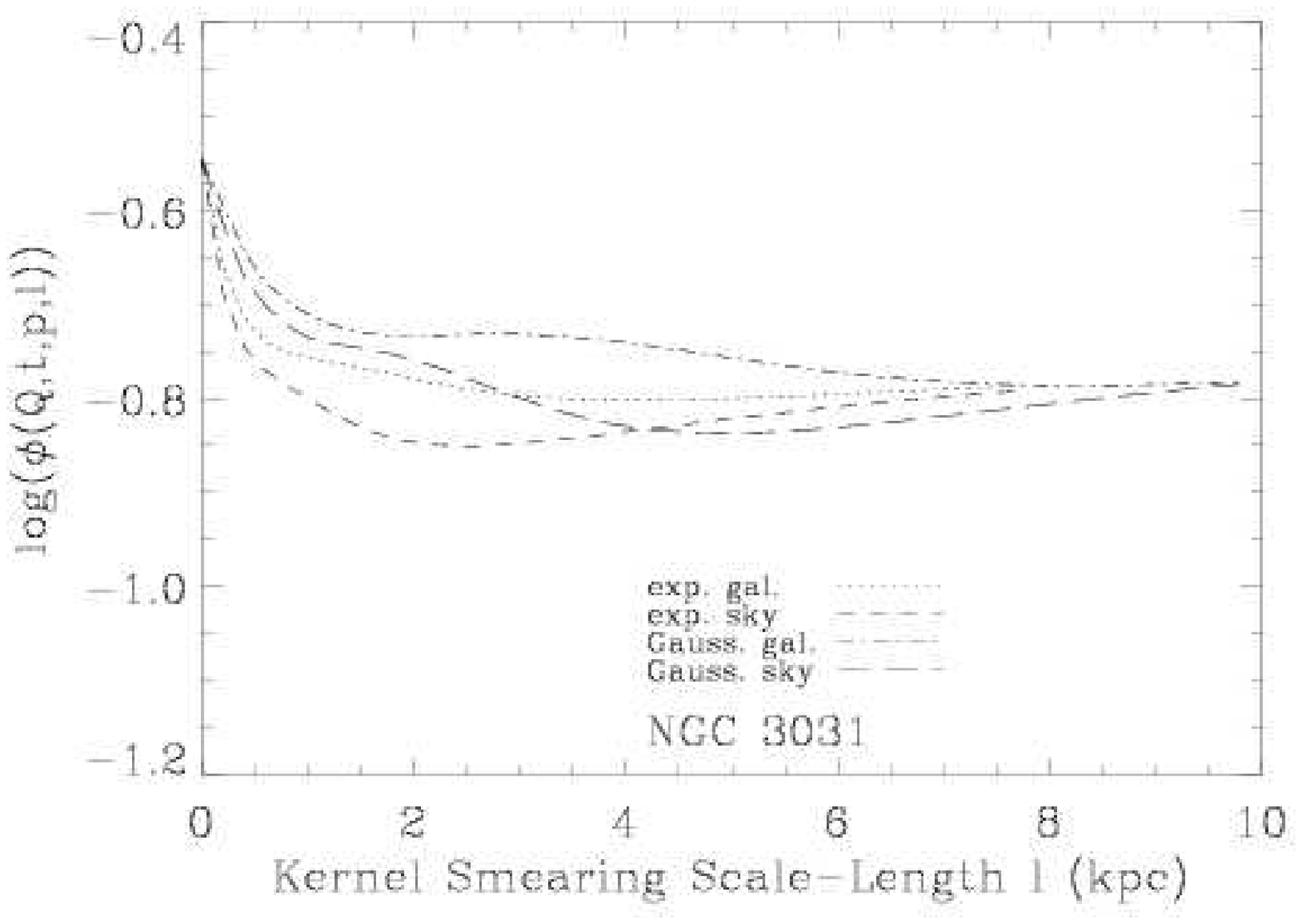}}}
  \resizebox{16.5cm}{!}{
    {\plotone{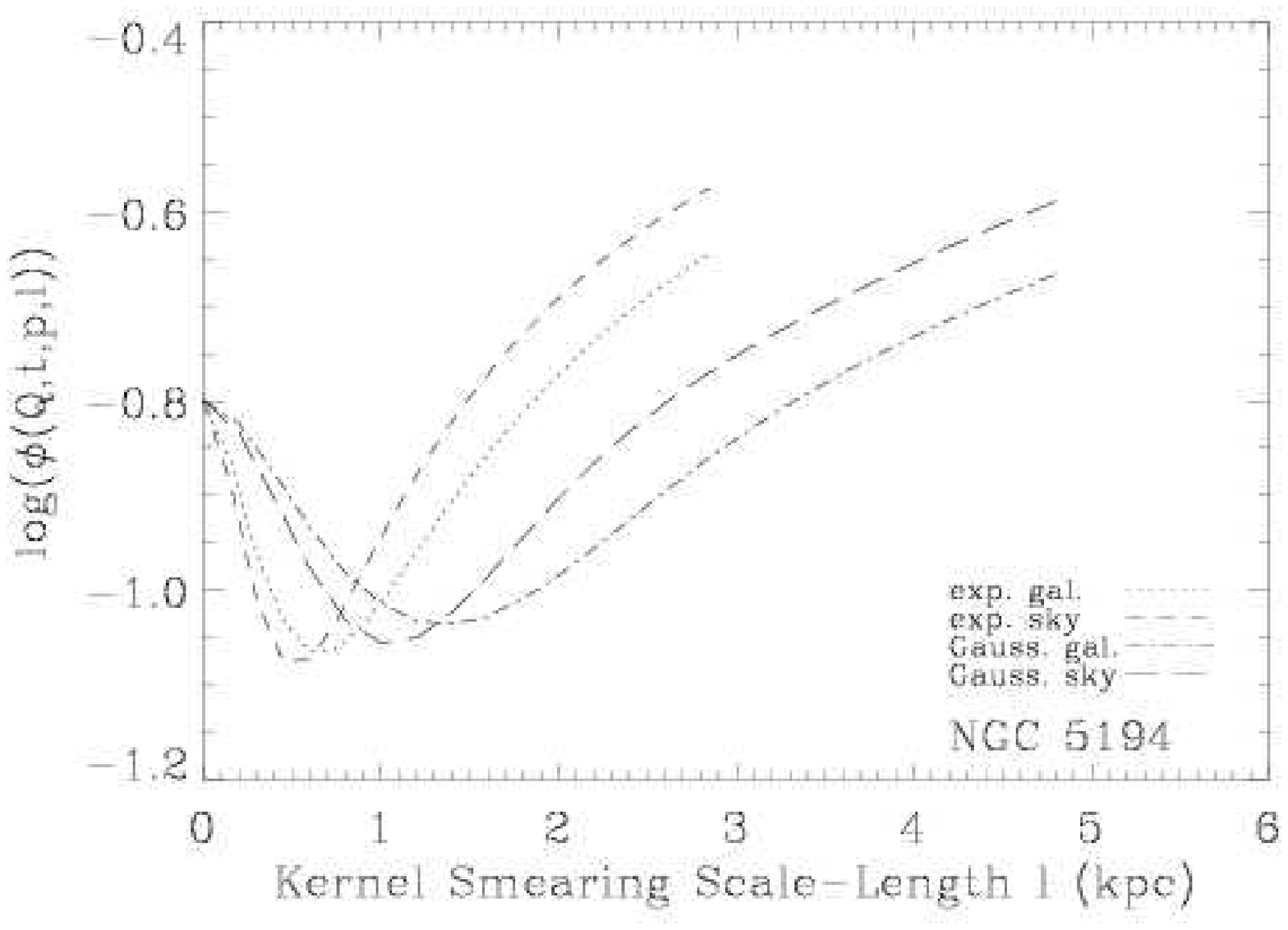}
      \plotone{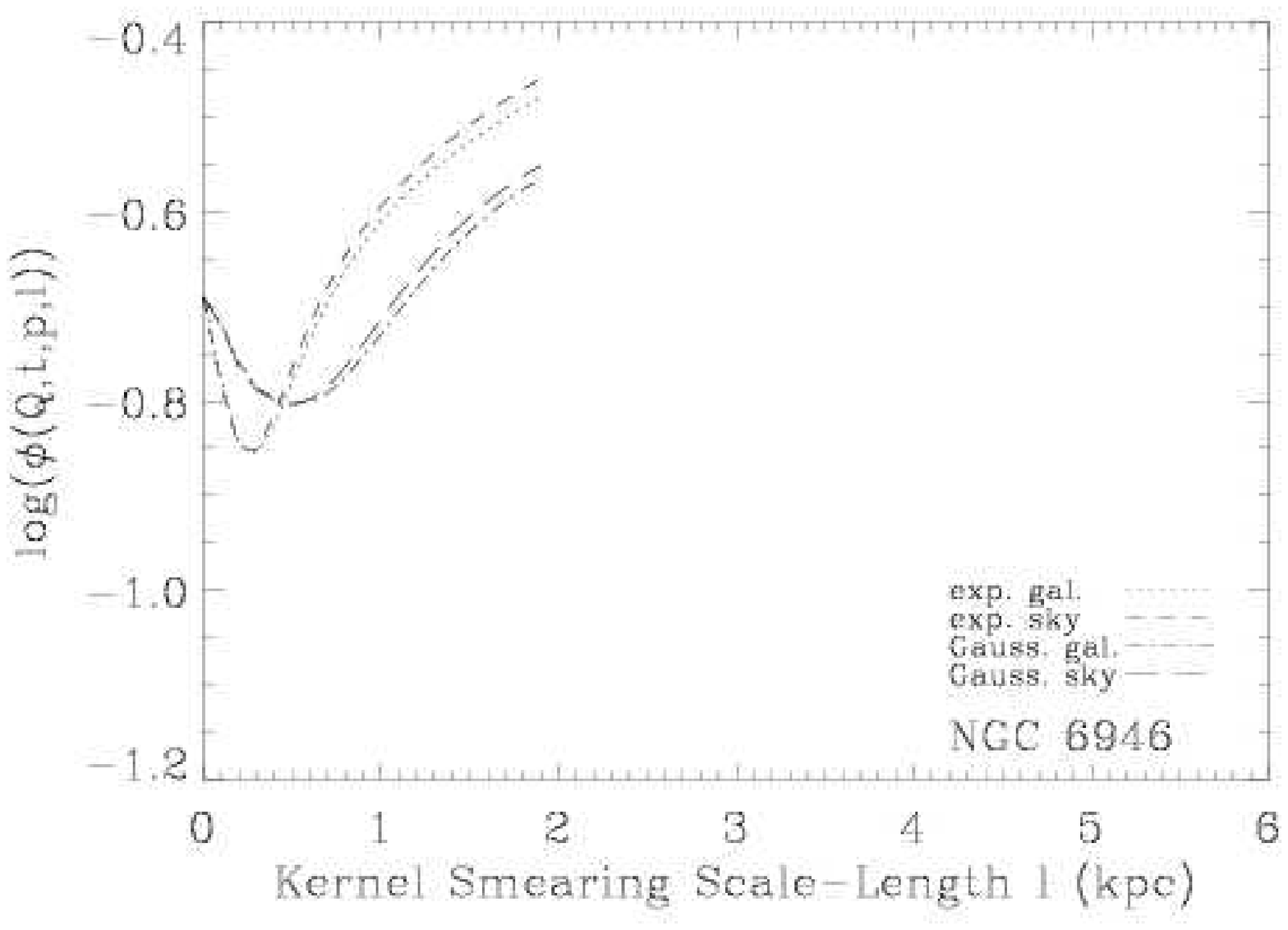}}}
    \caption{Residuals of observed radio maps with smeared 70~$\micron$
    images (as defined in $\S$4.2.1) as a function of smearing
    scale-length.  Results are shown for each kernel and
    galaxy.\label{res70}}
\end{figure}

\clearpage
\begin{figure}[!ht]
\resizebox{16.5cm}{!}{
    {\plotone{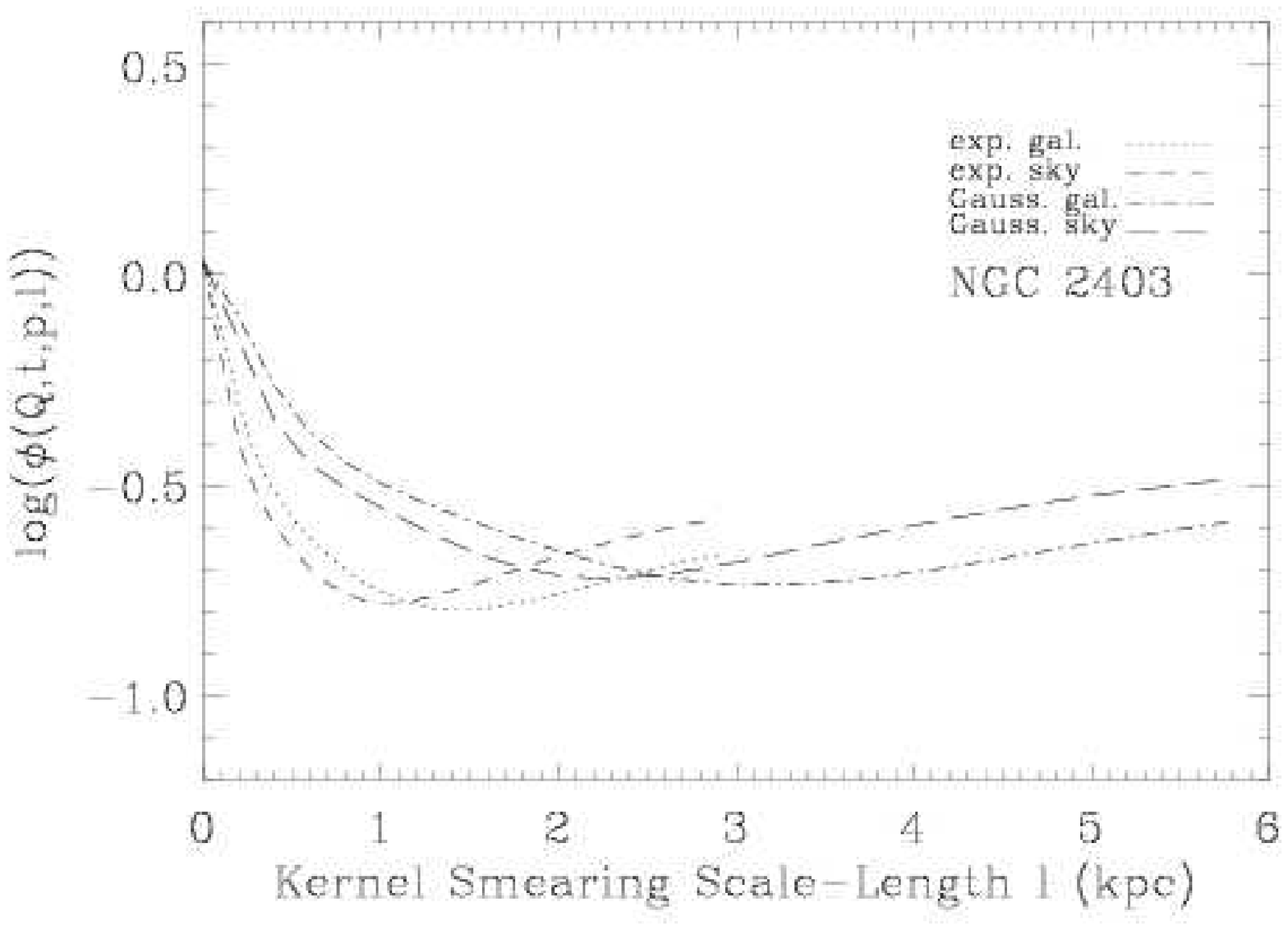}
      \plotone{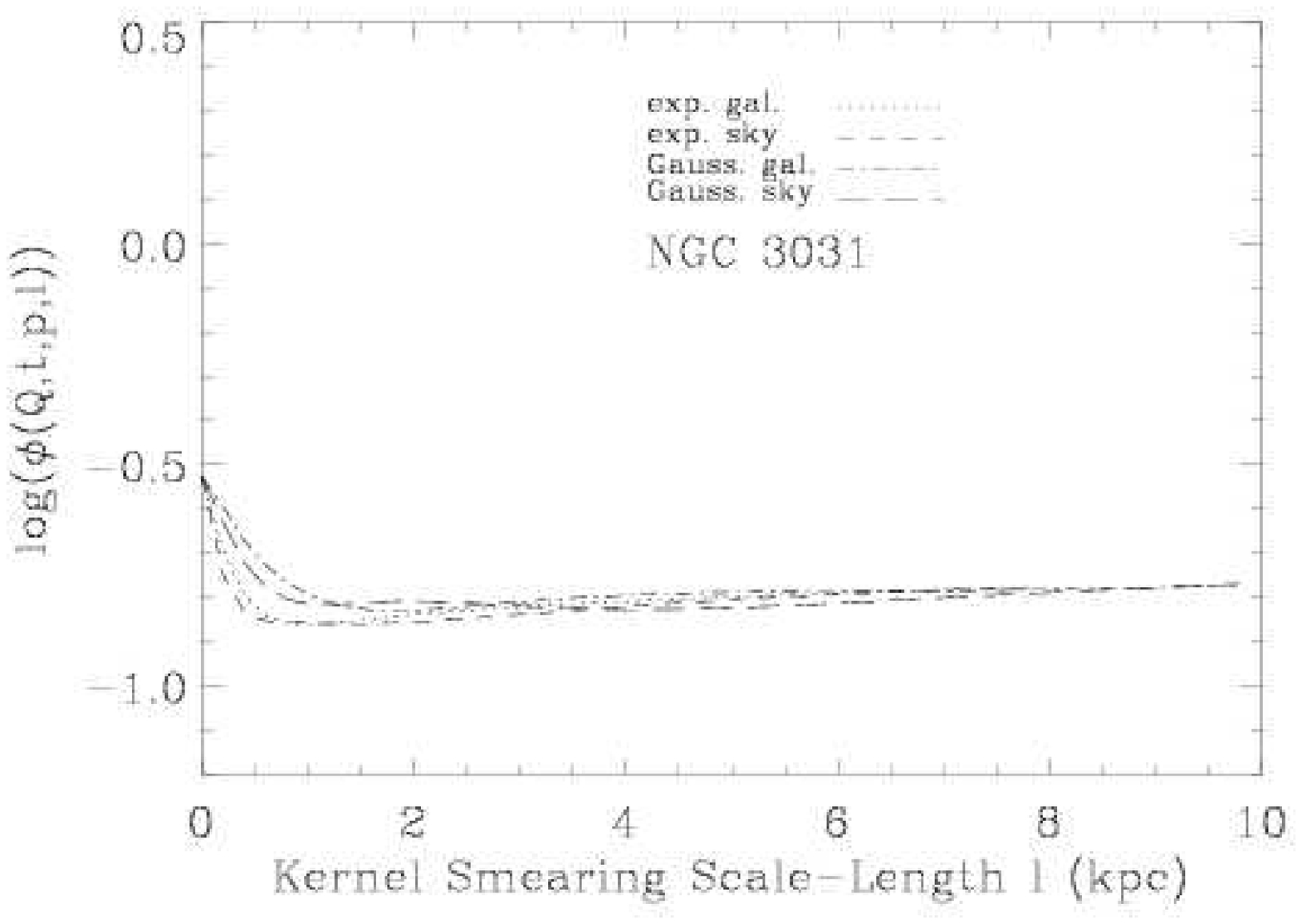}}}
  \vspace*{-.1cm}
  \resizebox{16.5cm}{!}{
    {\plotone{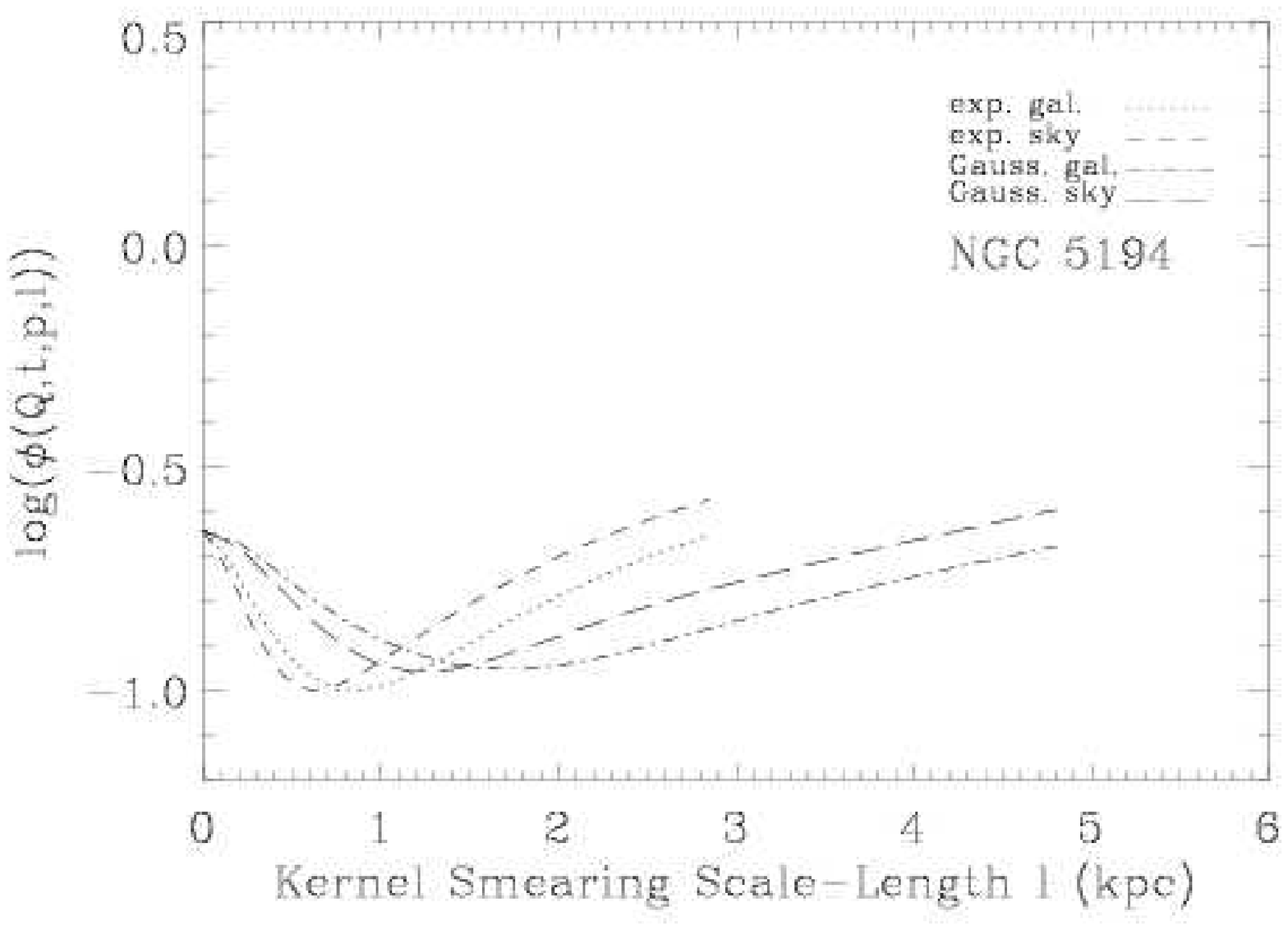}
      \plotone{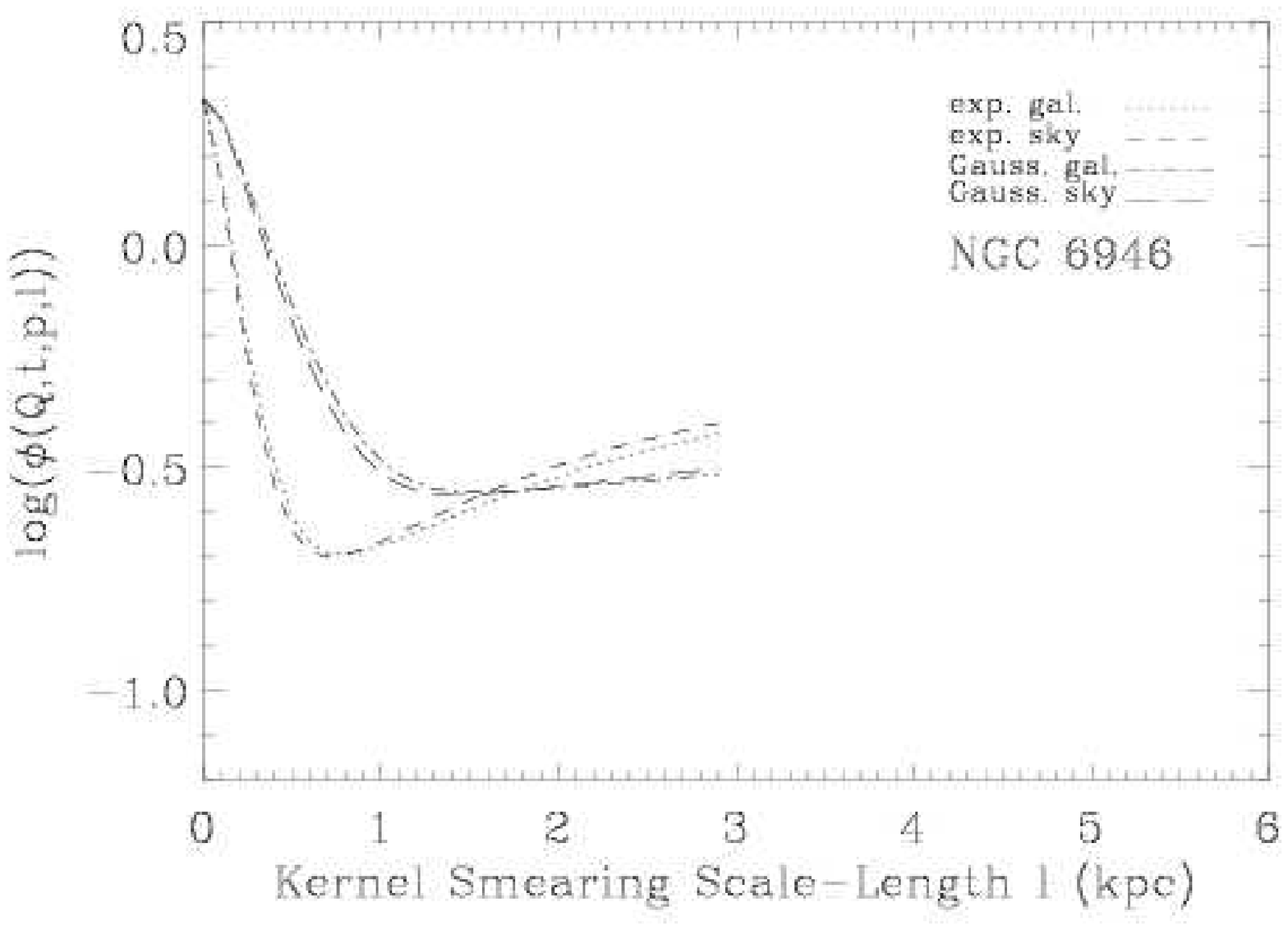}}}
  \caption{Same as Figure \ref{res70} except using
  smeared 24~$\micron$ images matched to the resolution of the
  70~$\micron$ beam.  \label{res24}} 
\end{figure}

\clearpage
\begin{figure}[!ht]
  \resizebox{16.5cm}{!}{
    {\plotone{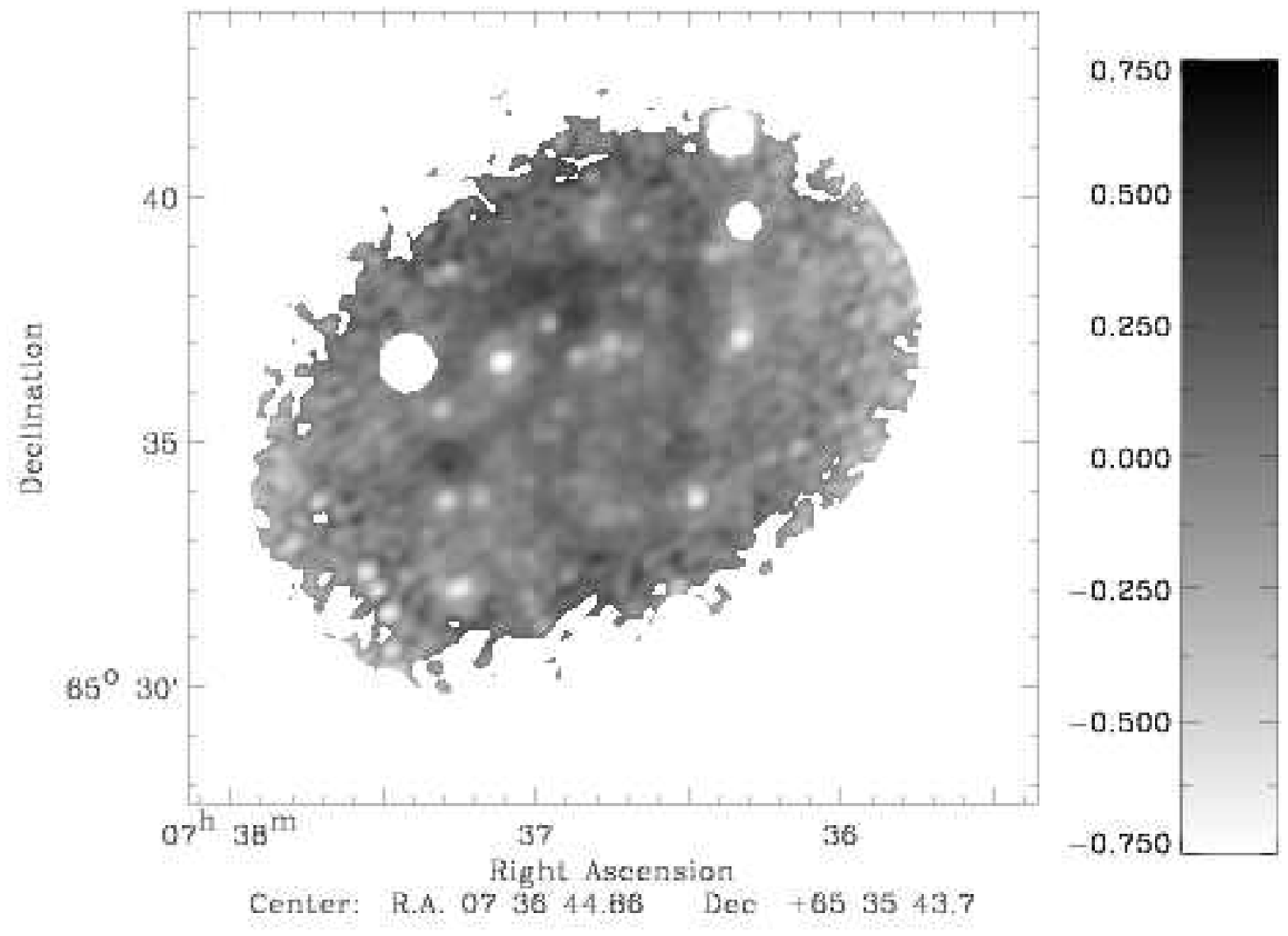}
      \plotone{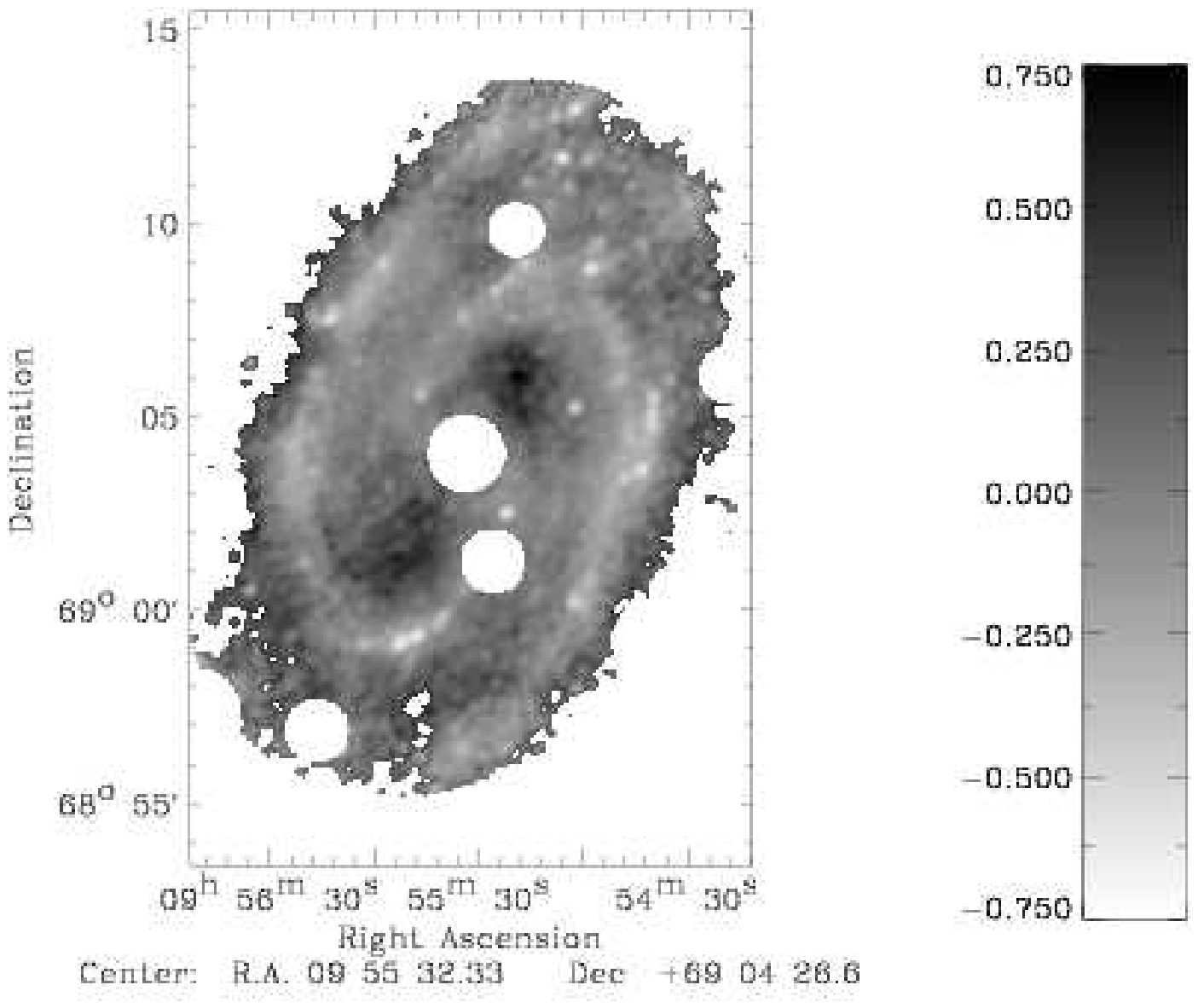}}}
  \vskip 1cm 
  \resizebox{16.5cm}{!}{
    {\plotone{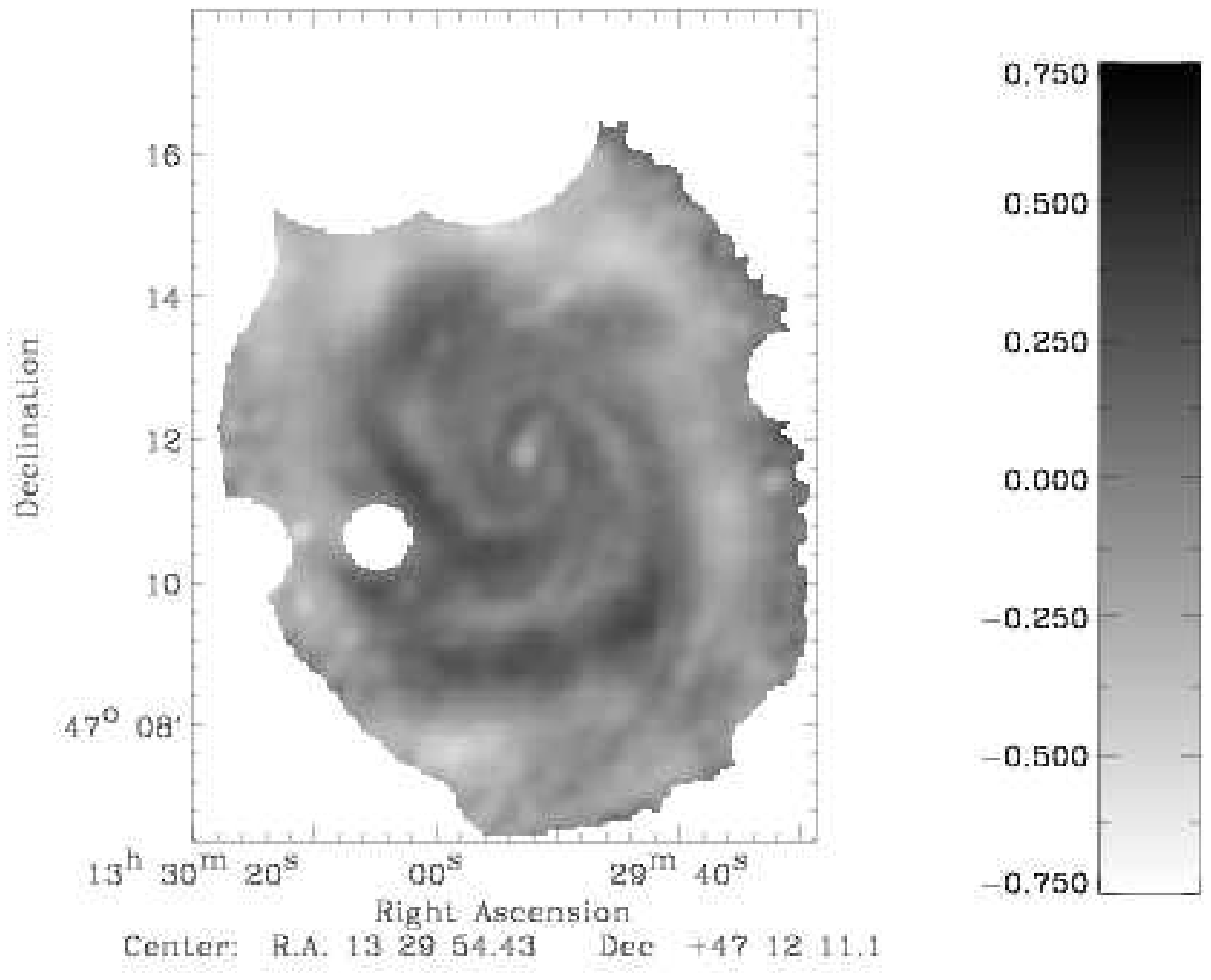}
      \plotone{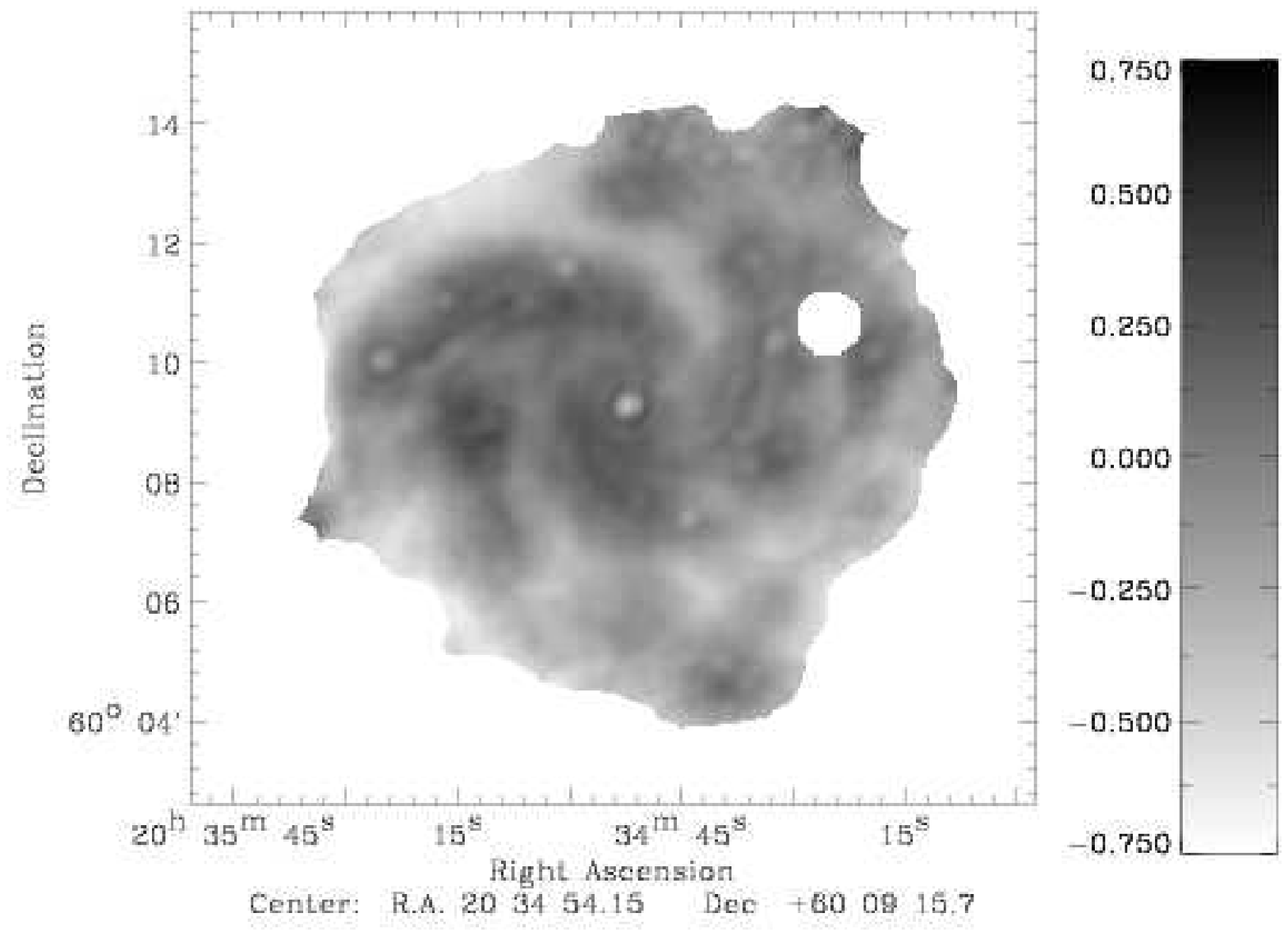}}}
  \vskip .5cm
  \caption{Residual images after subtracting the observed radio maps
    from the smeared 70~$\micron$ images (as defined in  $\S$4.2.1) for
    each galaxy using the best fit exponential kernel oriented in
    the plane of the sky. 
    \label{smear70}}
\end{figure} 

\clearpage
\begin{figure}[!ht]
  \resizebox{16.5cm}{!}{
    {\plotone{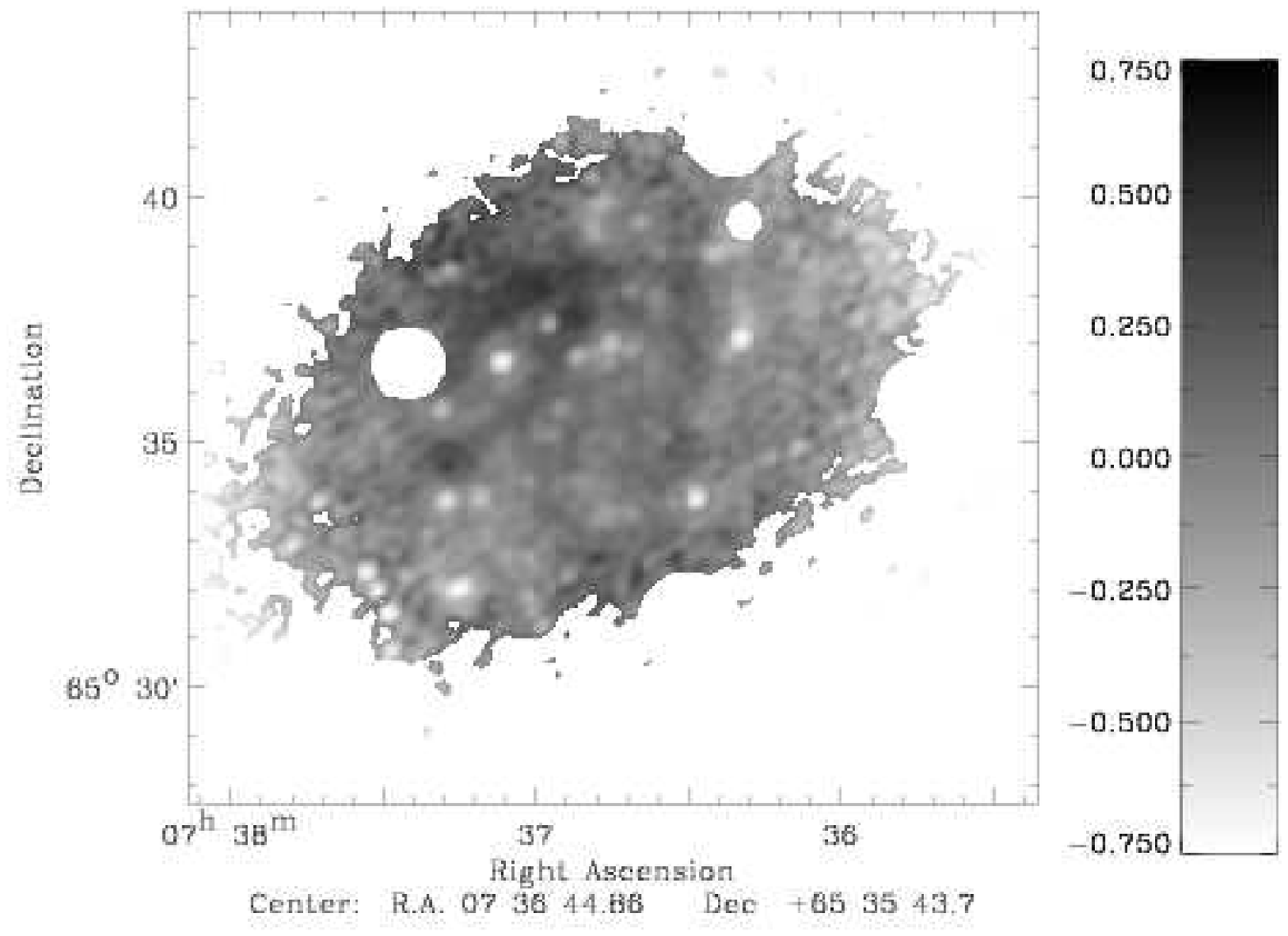}
      \plotone{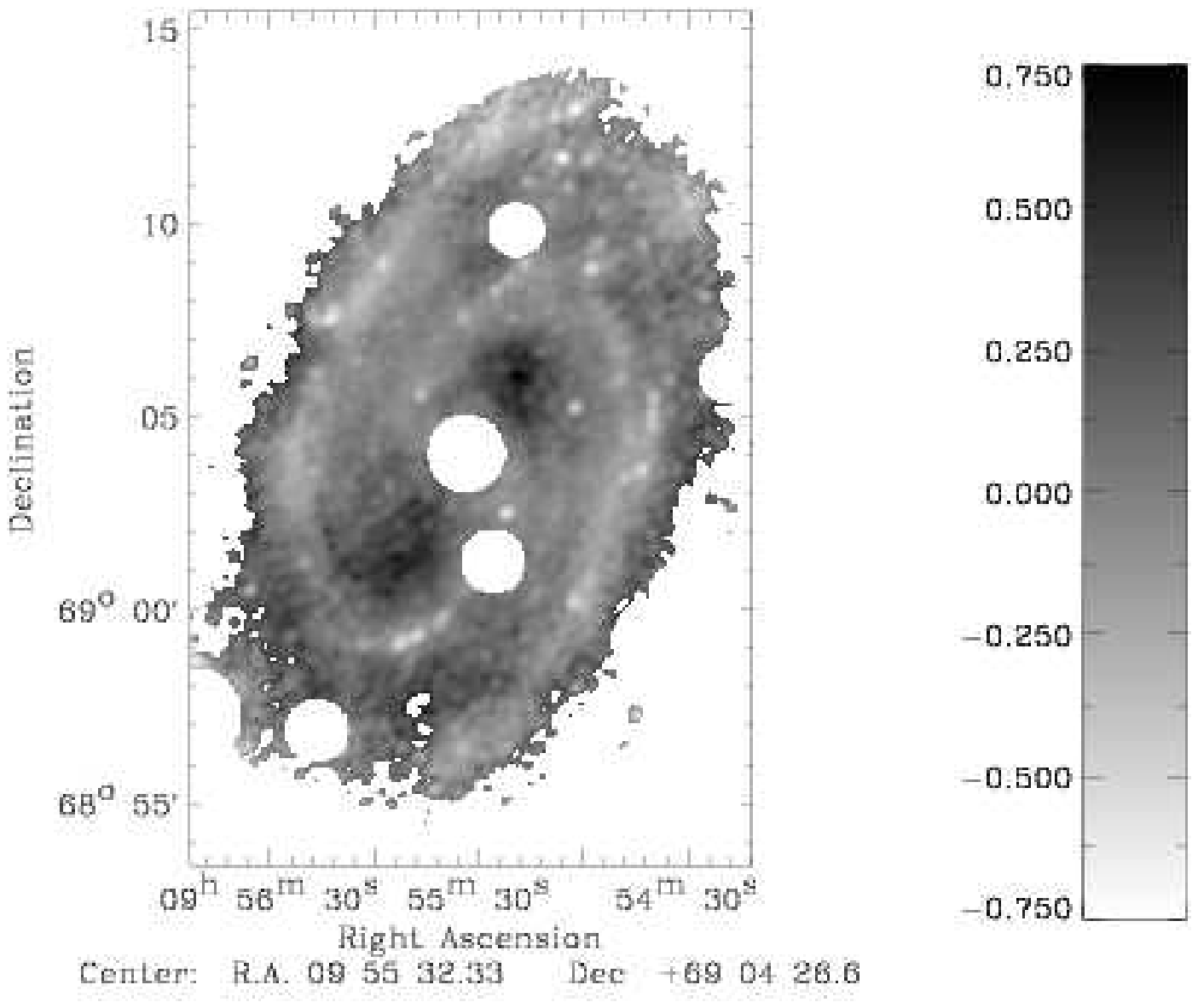}}}
  \vskip 1cm 
  \resizebox{16.5cm}{!}{
    {\plotone{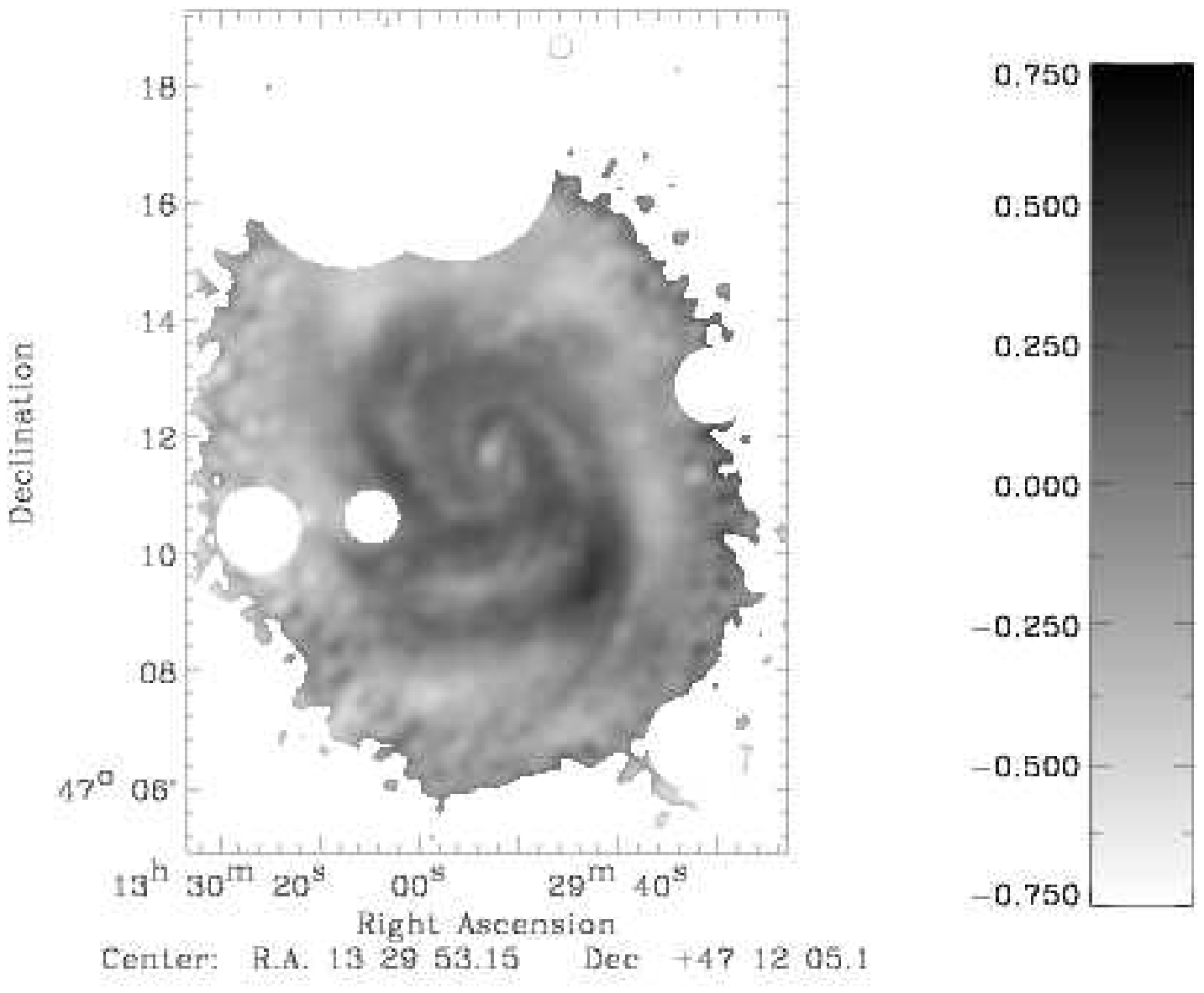}
      \plotone{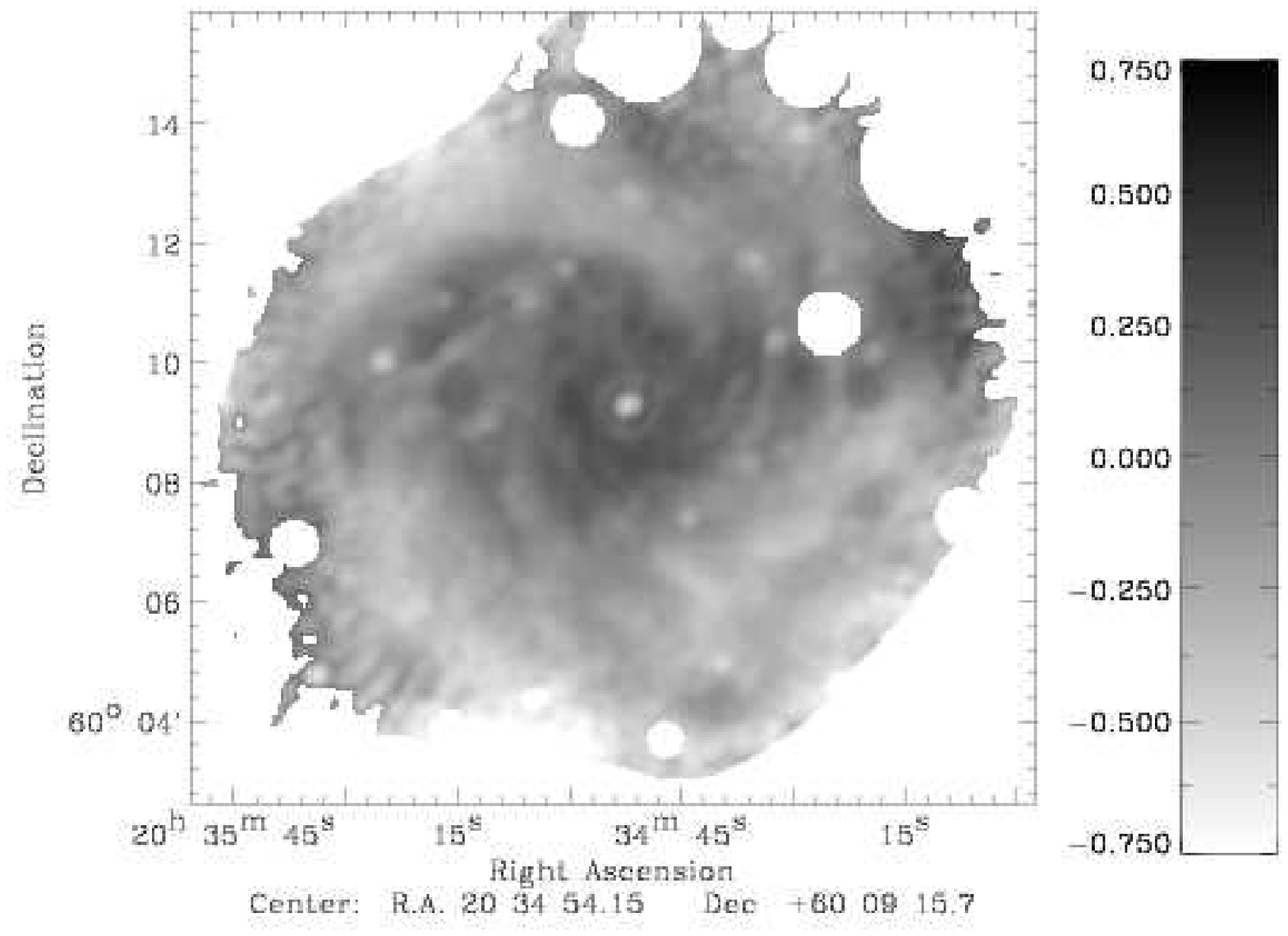}}}
  \vskip .5cm
  \caption{Residual images after subtracting the observed radio maps
      from the  smeared 24~$\micron$ images (as defined in $\S$4.2.1)
      for each galaxy using the best fit exponential kernel
      oriented in the plane of the sky. 
      The 24~$\micron$ maps were first convolved to match the
      70~$\micron$ PSF.  \label{smear24}}
\end{figure}

\end{document}